\documentclass[preprint]{aastex63}
\usepackage[utf8]{inputenc}
\usepackage{tikz}
\usetikzlibrary{decorations.pathreplacing}
\usepackage{graphicx}
\usepackage{capt-of}
\usepackage{xcolor}
\usepackage{amsmath,amssymb}
\usepackage{multirow}

\begin{document}

\title{Teaching neural networks to generate Fast Sunyaev Zel'dovich Maps}

\author{Leander Thiele}
\affiliation{Department of Physics, Princeton University, Jadwin Hall,
Princeton NJ 08544, USA}

\author{Francisco Villaescusa-Navarro}
\affiliation{Department of Astrophysical Sciences, Princeton University, Peyton Hall,
Princeton NJ 08544, USA}
\affiliation{Center for Computational Astrophysics, Flatiron Institute, 162 5th Avenue,
New York NY 10010, USA}

\author{David N. Spergel}
\affiliation{Center for Computational Astrophysics, Flatiron Institute, 162 5th Avenue,
New York NY 10010, USA}
\affiliation{Department of Astrophysical Sciences, Princeton University, Peyton Hall,
Princeton NJ 08544, USA}

\author{Dylan Nelson}
\affiliation{Max-Planck-Institut f\"ur Astrophysik, Karl-Schwarzschild-Str. 1,
85741 Garching, Germany}

\author{Annalisa Pillepich}
\affiliation{Max-Planck-Institut f\"ur Astronomie, K\"onigstuhl 17,
69117 Heidelberg, Germany}

\begin{abstract}
The thermal Sunyaev-Zel'dovich (tSZ) and the kinematic Sunyaev-Zel'dovich (kSZ) effects
trace the distribution of electron pressure and momentum in the hot Universe.
These observables depend on rich multi-scale physics,
thus, simulated maps should ideally be based on calculations
that capture baryonic feedback effects such as cooling, star formation, and other complex processes.
In this paper, we train deep convolutional neural networks with a U-Net architecture
to map from the three-dimensional distribution of dark matter
to electron density, momentum and pressure at $\sim 100\,\text{kpc}$ resolution.
These networks are trained on a combination of the
TNG300 volume and a set of cluster zoom-in simulations from the IllustrisTNG project.
The neural nets are able to reproduce the power spectrum, one-point probability distribution function, bispectrum,
and cross-correlation coefficients of the simulations more accurately than the
state-of-the-art semi-analytical models.
Our approach offers a route to capture the richness of a full cosmological hydrodynamical simulation
of galaxy formation with the speed of an analytical calculation.
\end{abstract}

\section{Introduction}

Over the past decade, observations of the cosmic microwave background (CMB)
have become one of the most powerful tools for studying the early Universe
and for determining its basic properties (age, density, composition).
Microwave background observations also have the potential
to provide detailed quantitative measurements of the properties of the Universe's recent evolution.
Through the thermal Sunyaev-Zel'dovich (tSZ) effect,
the microwave observations trace the large-scale distribution of electron pressure;
while the kinematic Sunyaev-Zel'dovich (kSZ) effect traces the large-scale distribution of electron momentum
\citep{ZS1969, SZ1970}.

High resolution microwave background experiments 
such as  ACT\footnote{\url{https://act.princeton.edu}}, SPT\footnote{\url{https://pole.uchicago.edu}},
and upcoming instruments like the Simons Observatory\footnote{\url{https://simonsobservatory.org}}
and CMB S4\footnote{\url{https://cmb-s4.org}},
provide ever improving measurements of small-scale fluctuations.
Accurate theoretical predictions are essential for extracting the full information from these rich observations.
Generating the simulations needed for these accurate predictions is particularly challenging
for the SZ effects as they depend not only on the underlying cosmology
but also on complex multi-scale processes
including feedback from star formation and Active Galactic Nuclei
and the rich plasma physics of cluster gas.
Hydrodynamical simulations of the Sunyaev-Zel'dovich effects began before the establishment
of precision cosmology
\citep{Scaramellaetal1993, daSilvaetal2000, daSilvaetal2001, SpringelWhiteHernquist2001}.
Later works investigated the effects of sub-grid physics
\citep{WhiteHernquistSpringel2002, Nagaietal2007, Pfrommeretal2007, Sijackietal2008,
       Battagliaetal2010}.
Large-scale simulation efforts have enabled direct comparison to microwave observations
\citep{Hallmanetal2007maps, Schaeferetal2006, Kayetal2012, Schayeetal2015, DolagKomatsuSunyaev2016,
       Spaceketal2018, Nelsonetal2019}.
Hydrodynamical uniform-resolution simulations
are computationally expensive and
must adopt a trade-off between large volumes, with robust statistics of high-mass objects,
and high numerical resolution to better capture the important physical processes.

Cosmologists have been taking a variety of approaches
to more efficiently generate predicted tSZ maps.
Most work builds on numerical simulations.
Early papers assumed that on large scales the gas pressure traces the dark matter distribution
\citep{Persietal1995, Refregieretal2000}.
In order to more accurately describe the non-linear evolution,
halo model approaches have been developed, starting from
\cite{KomatsuKitayama1999} and \cite{KomatsuSeljak2002}.
These assume that the halos have a characteristic profile \citep{KomatsuSeljak2001},
depending on mass and redshift and possibly a few other parameters.
A variety of other analytical approaches in the halo model framework have been developed
\citep{LeeSuto2003, OstrikerBodeBabul2005, Hallmanetal2007betamodel, Bodeetal2007, Bodeetal2009,
       Shawetal2010, EfstathiouMigliaccio2012, CapeloCoppiNatarajan2012, ShiKomatsu2014}.
These have some free parameters that are calibrated off observations of low-redshift objects.
A comparison of different analytical halo model approaches is performed in
\cite{TracBodeOstriker2011}.
Recent work has calibrated these halo electron pressure profiles off of numerical simulations
\citep{Allisonetal2011, Battagliaetal2012, NelsonLauNagai2014, LeBrunetal2017,
       Guptaetal2017, Planellesetal2017, Meadetal2020},
or observations
\citep{Afshordietal2007, AtrioBarandelaetal2008, Arnaudetal2010, ChaudhuriMajumdar2011,
       Planck2013TSZProfiles, ZandanelPfrommerPrada2014, LeBrunetal2014, RamosCejaetal2015}.
A common theme of these semi-analytical models is the assumption of symmetries (often spherical)
and the restriction to a small set of variables describing a given halo.
Especially the neglect of halo sub-structure leads to errors in summary statistics
\citep{Battagliaetal2012}.

Treatments of the kSZ effect begin with linear theory
\citep{OstrikerVishniac1986, Vishniac1987, JaffeKamionkowski1998},
and then were extended to non-linear scales
\citep{ValageasBalbiSilk2001, MaFry2002, ZhangPenTrac2004, Parketal2016}.
More recent works use numerical simulations to calibrate the analytic treatment
\citep{ShawRuddNagai2012, Alvarez2016, Parketal2018}.

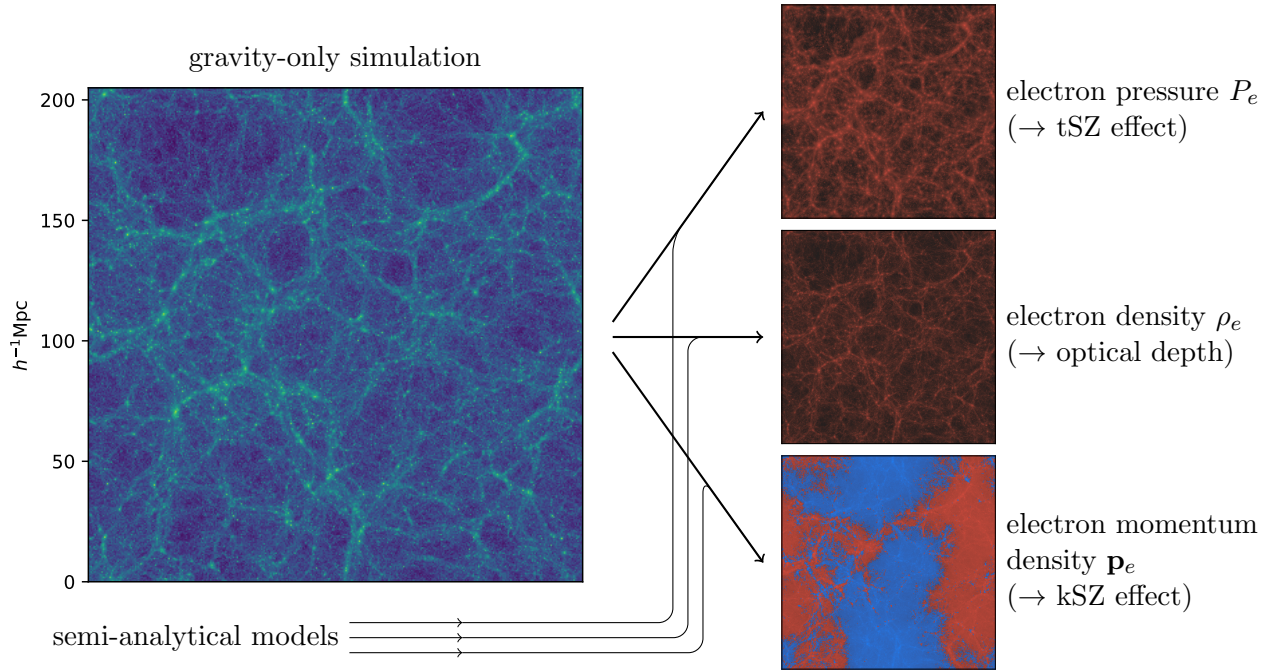
\begin{figure}
\begin{center}
\begin{tikzpicture}
	\node at (0,0) [left] {\includegraphics[scale=0.7]{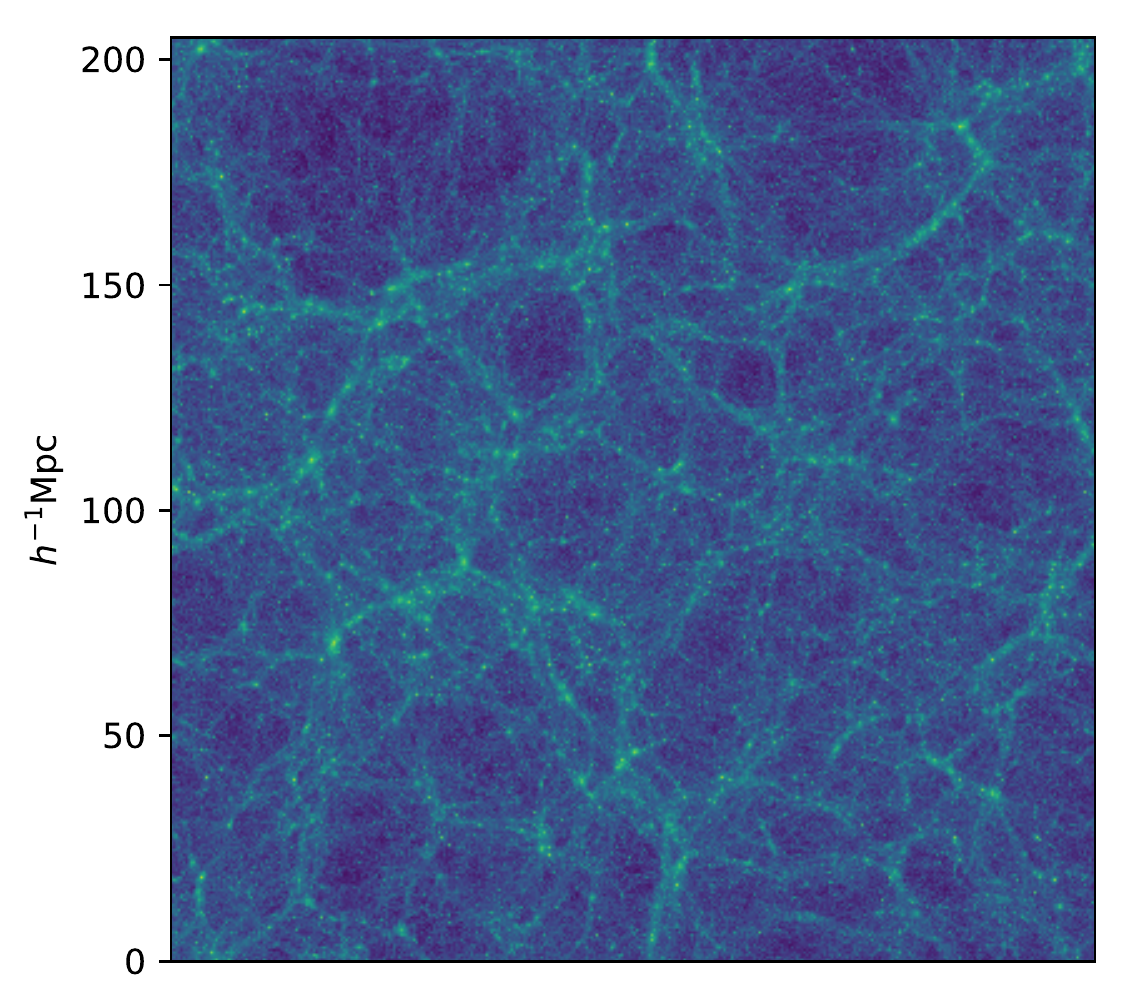}};
	\node at (-3.69,3.7) [centered] {gravity-only simulation};

	\node at (2,3) [right] {\includegraphics[width=0.17\textwidth]{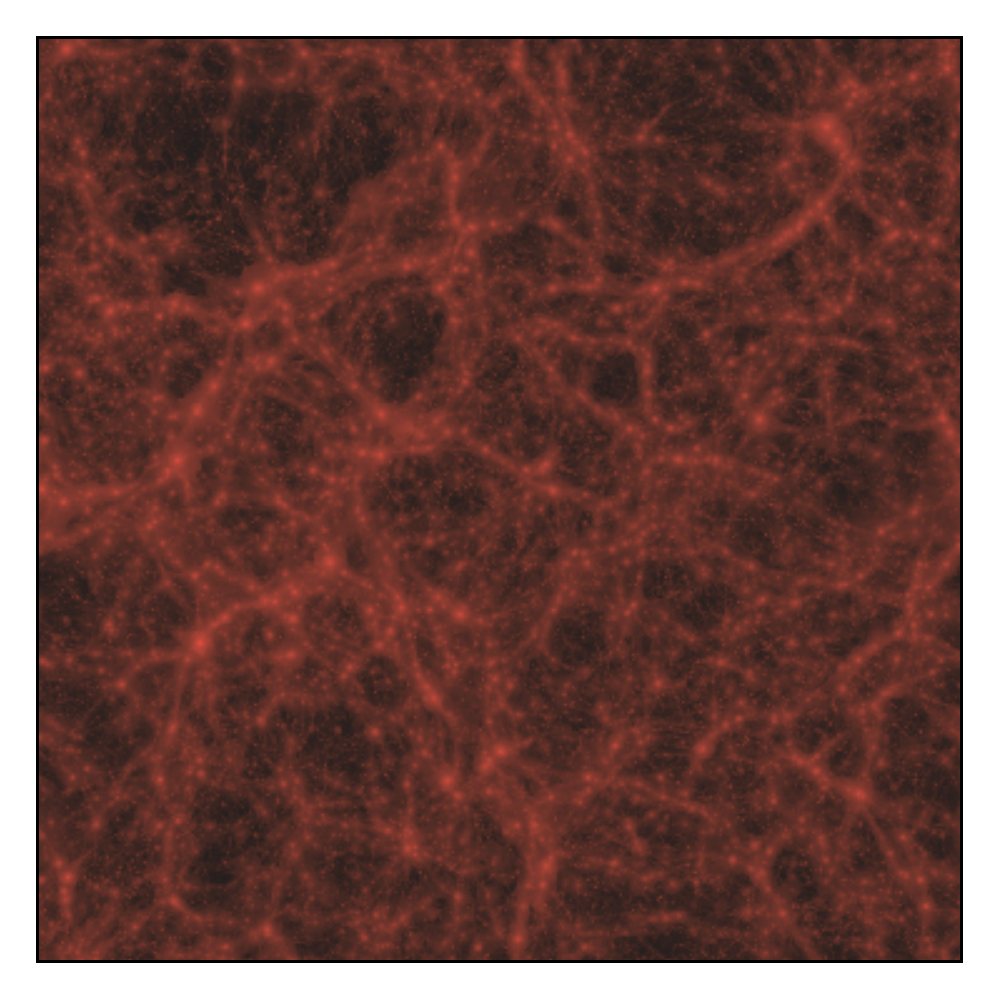}};
	\node at (5.1,3) [right,align=left] {electron pressure $P_e$\\($\rightarrow$ tSZ effect)};
	\node at (2,0) [right] {\includegraphics[width=0.17\textwidth]{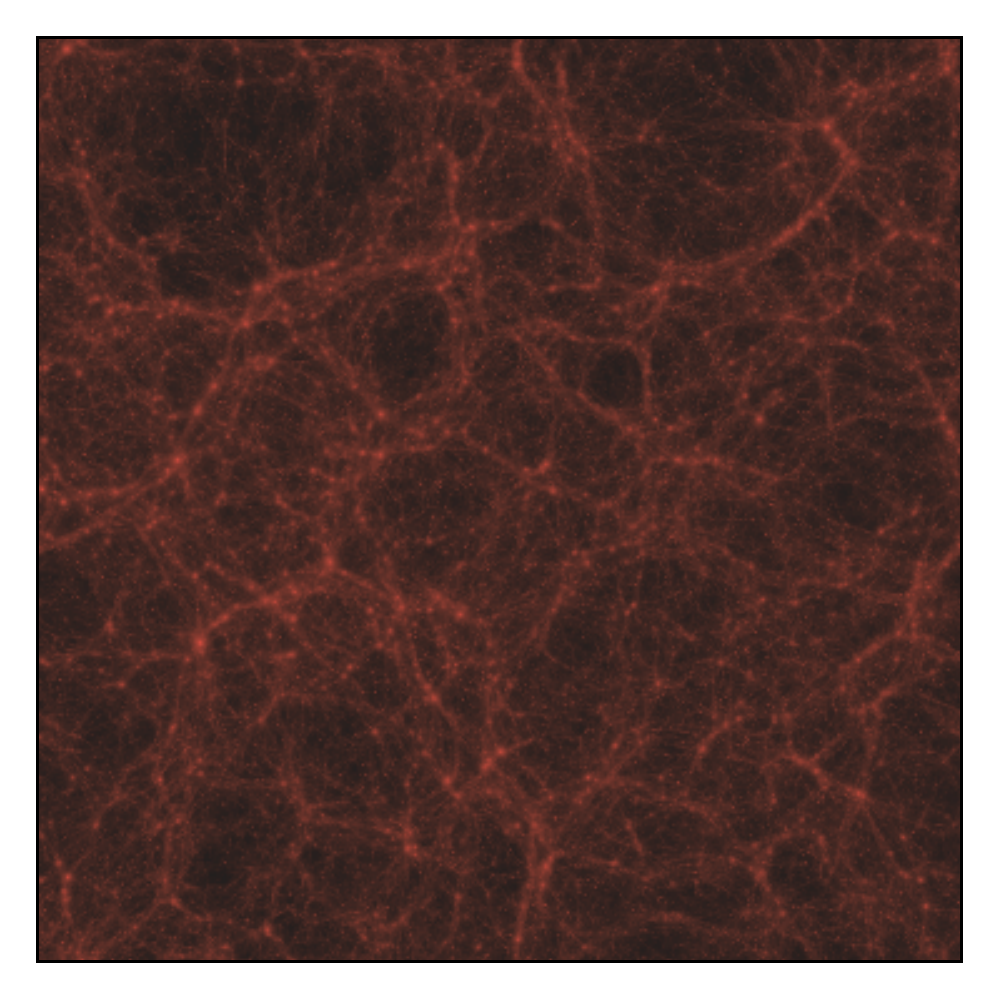}};
	\node at (5.1,0) [right,align=left] {electron density $\rho_e$\\($\rightarrow$ optical depth)};
	\node at (2,-3) [right] {\includegraphics[width=0.17\textwidth]{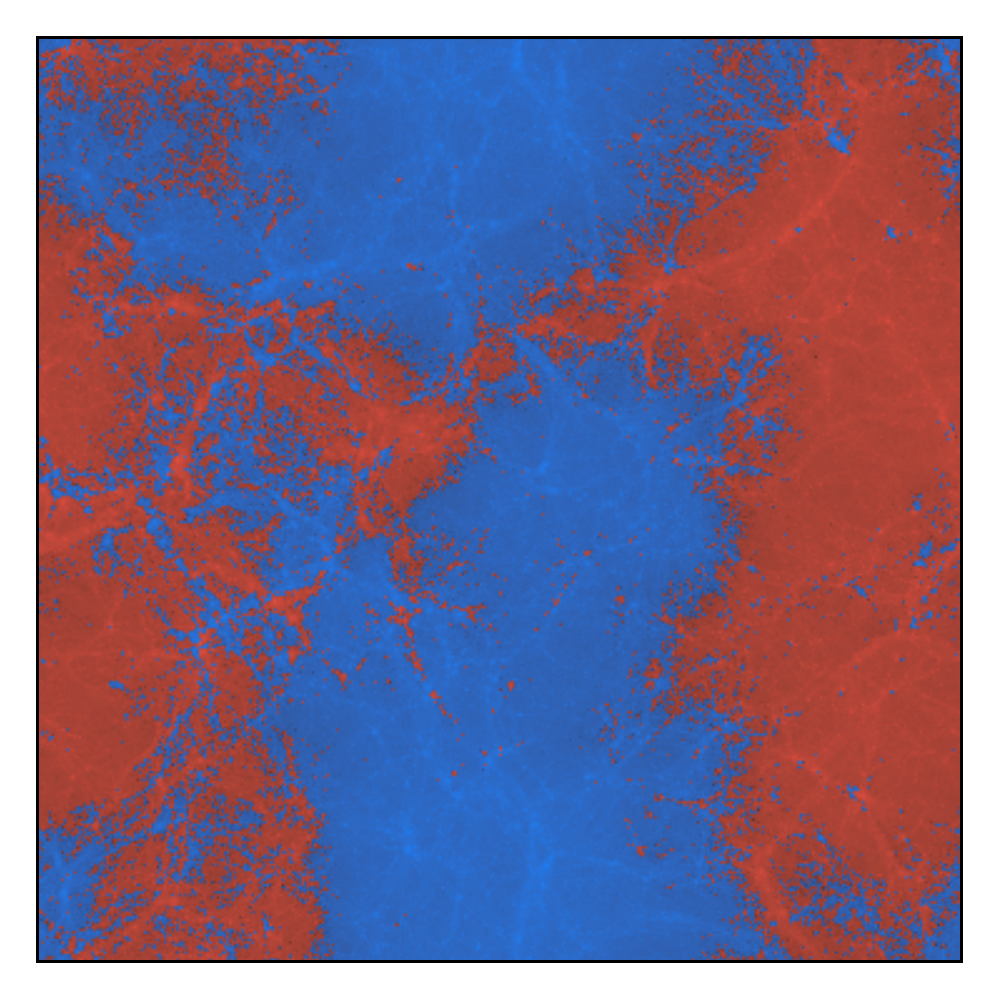}};
	\node at (5.1,-3) [right,align=left] {electron momentum\\density $\mathbf{p}_e$\\($\rightarrow$ kSZ effect)};
	
	\draw [thick,->] (0,0.2) -- (2,3);
	\draw [thick,->] (0,0) -- (2,0);
	\draw [thick,->] (0,-0.2) -- (2,-3);

	\draw [thin,->] (-3.5,-3.8) -- (-2,-3.8);
	\draw [thin,rounded corners=0.5em] (-2,-3.8) -- (0.8,-3.8) -- (0.8,1.32) -- (2,3);
	\draw [thin,->] (-3.5,-4) -- (-2,-4);
	\draw [thin,rounded corners=0.5em] (-2,-4) -- (1,-4) -- (1,0) -- (2,0);
	\draw [thin,->] (-3.5,-4.2) -- (-2,-4.2);
	\draw [thin,rounded corners=0.5em] (-2,-4.2) -- (1.2,-4.2) -- (1.2,-1.88) -- (2,-3);
	\node at (-3.5,-4) [left] {semi-analytical models};
\end{tikzpicture}
\caption{Schematic illustration of the approach taken in this work.
         We use properties of the matter field from gravity-only cosmological simulations, on the
	 left, as input into convolutional neural networks
	 which predict the electron gas properties
	 via training based on the corresponding full-physics hydrodynamical realization,
	 on the right.
	 The networks utilize simulation-calibrated semi-analytical models for the targets,
	 so that only residuals with respect to the hydrodynamical simulation output are required to be learned.}
\label{fig:intro}
\end{center}
\end{figure}

In this work we use a different approach, schematically illustrated in Fig.~\ref{fig:intro}.
We employ deep learning techniques to find the mapping
between the spatial distribution of dark matter\footnote{For brevity, here and in the remainder of
this paper we use the phrase ``dark matter" to mean the matter field obtained in a gravity-only
simulation.}
from gravity-only cosmological simulations and
(1) electron pressure, (2) electron density, and (3) electron momentum
from the corresponding state-of-the-art full-physics hydrodynamical realizations.
In particular, we use simulations from the IllustrisTNG project
\citep{IllustrisTNG1, IllustrisTNG2, IllustrisTNG3, IllustrisTNG4, IllustrisTNG5,
       Nelsonetal2019}. 
The advantage of this method with respect to the
semi-analytical models is that gravity-only simulations are accurate enough to model the clustering
of matter and halos down to rather small scales,
and deep learning can transform their output to account for baryonic effects,
capturing non-gravitational processes while avoiding oversimplifications such as spherical symmetry.

Machine learning has previously been used for similar tasks,
for example to generate two-dimensional maps of the tSZ effect \citep{Troesteretal2019},
and populating gravity-only simulations with galaxies
\citep{Xuetal2013, Zhangetal2019, Agarwaletal2018, JoKim2019, Mosteretal2020}.

We approach this work as part of a broader program
consisting of predicting baryonic observables from gravity-only simulations,
or even directly from the initial conditions \citep{Heetal2019}.
One application of particular interest would be the prediction of cross-correlation
statistics between different observables.
Because of our focus on cross-correlations,
we are primarily interested in generating the low-redshift kSZ signal
and defer the discussion of the high redshift signal to future work.
Our current approach assumes that there is an injective map between the matter field and the
observables of interest;
we discuss stochasticity (which is not captured in our approach) in Section~\ref{sec:conclusions}.


This approach will allow us to quickly generate tSZ and kSZ maps,
over large areas of the sky, with the astrophysical model of the IllustrisTNG simulation
\citep{Weinbergeretal2017, Pillepichetal2018},
but in comparatively negligible time (see the figures quoted in Section~\ref{sec:conclusions}).
Those maps can be used to extract cosmological and astrophysical information from CMB observations alone,
but also allow the modeling of cross-correlations
such as tSZ-weak lensing or kSZ with spectroscopic surveys.

This paper is organized as follows.
We describe the physics of the tSZ and kSZ effects in Section~\ref{sec:physics}.
Section~\ref{sec:methods} outlines the methods used in the analysis.
Section~\ref{sec:sparsity} describes the challenge introduced by sparsity:
rare massive clusters contribute a significant fraction of the tSZ signal
and dominate the higher moment statistics of the maps.
The section describes the use of zoom-in simulations of high density regions
to augment the data set and ameliorate this issue.
We show the results of our work in Section~\ref{sec:results}.
Finally, in Section~\ref{sec:conclusions} we draw the conclusions. 
Some technical details are collected in the Appendices.

\section{Physical background}
\label{sec:physics}


At low frequencies, clusters cast shadows against the microwave sky:
in the 90 and 150 GHz Planck, ACT and SPT maps, the tSZ effect makes clusters appear as cold spots.
This effect arises from random motions of thermal electrons which scatter CMB
photons into higher energy states.
It produces a non-blackbody distortion in the primary CMB's Planckian spectrum,
yielding a brightness temperature decrement (increment) at low (high) observation frequencies $\nu$.
It is usually parameterized in terms of the dimensionless Compton-$y$ parameter, given by
\begin{equation}
y = \sigma_T \int_\text{LOS} dr\,\frac{P_e}{m_e}\,,
\label{eq:Comptony}
\end{equation}
where $\sigma_T = (8\pi e^4/3)m_e^{-2}$ is the Thompson cross-section for electron-photon scattering, $r$ is physical
distance along the line of sight (LOS), $P_e = n_e T_e$ is electron pressure, and $m_e$ is the
electron mass. The dependence on the choice of line of sight $\mathbf{n}$ has been suppressed.
Thus, the tSZ effect is a measure of the line-of-sight integrated thermal energy density in electrons.

The kinematic SZ (kSZ) effect, on the other hand, is sourced by coherent motion of the electrons.
To leading order it preserves the blackbody energy distribution, yielding a shift in the local CMB
temperature.
We can write down a parameterization similar to the tSZ Compton-$y$,
referred to as the Doppler-$b$ parameter:
\begin{equation}
\mathbf{b} = \sigma_T \int_\text{LOS} dr\, \frac{\mathbf{p}_e}{m_e}e^{-\tau(r)}\,,
\end{equation}
where
\begin{equation}
\tau(r) = \sigma_T \int_0^r dr'\, \frac{\rho_e}{m_e}
\end{equation}
is the optical depth (average number of Compton scattering events up to $r$) and
$\mathbf{p}_e$ is the electron momentum density (momentum per unit volume).

For reference, the dimensionless parameters introduced above are related to the observed shifts in
the CMB temperature by
\begin{align}
\frac{\Delta T_\text{tSZ}(\mathbf{n},\nu)}{T_\text{CMB}} &= y(x\coth{x/2}-4)\,, \hspace{1cm} x\equiv\nu/T_\text{CMB}\,;\\
\frac{\Delta T_\text{kSZ}(\mathbf{n})}{T_\text{CMB}} &= -\mathbf{b}\mathbf{.n}\,.
\end{align}

In the late-time universe,
both SZ effects are predominantly sourced by electrons bound to dark matter halos.
The tSZ effect is mostly sourced by the hot electron gas in the cluster center (due to the
dependence on electron temperature), and the cluster integrated signal scales roughly as $Y\propto M_\text{halo}^{5/3}$
\citep{SZ1970}.
In contrast, the kSZ effect does not receive this temperature bias and is thus
more sensitive to the cool electron gas in cluster outskirts, lower mass halos, and the IGM.

As we explain in Section~\ref{subsec:dataprep}, our procedure begins with 
semi-analytical models for the desired electron gas properties.
For the electron pressure, we make use of the simulation-calibrated model from
\cite{Battagliaetal2012}, hereafter B12,
which gives a simple fitting function for the electron pressure as a function of halo mass and
redshift:
\begin{equation}
P_e^\text{model}(\mathbf{x}) = \sum_{\text{halos}~h}\text{B12}(M_h, |\mathbf{x}-\mathbf{x}_h|)\,.
\label{eq:Pemodel}
\end{equation}
The B12 semi-analytical model is calibrated on a particular cosmology with given assumptions on sub-grid baryonic
effects.

In contrast to the electron pressure,
we do not use a halo model fitting function \citep{Battaglia2016} for the electron density,
as we find that a simple linear fit relating dark matter and electron density with an additional
Gaussian smoothing is good enough for the purposes of this work:
\begin{equation}
\rho_e^\text{model}(\mathbf{x}) = A\times[\text{Gaussian}(\sigma=80\,h^{-1}\text{kpc})
                                  \circledast\rho_m](\mathbf{x})\,,
\label{eq:rhoemodel}
\end{equation}
with $A$ a scalar that was fitted to the TNG300 simulation.
The convolution kernel was chosen by visually inspecting some halos,
its precise shape and width are of little importance.

From this, we finally construct our semi-analytical model for the electron momentum density:
\begin{equation}
\mathbf{p}_e^\text{model}(\mathbf{x}) = \rho_e^\text{model}(\mathbf{x})\mathbf{v}_m(\mathbf{x}),
\label{eq:pemodel}
\end{equation}
where $\mathbf{v}_m$ is the dark matter velocity field.
In particular, we do not include any velocity bias between dark matter and baryon velocities.

\section{Methods}
\label{sec:methods}

In this section, we describe the methods used to prepare the data, as well as the general set-up for
the machine learning procedure (input and target data, network architecture).
We defer the solutions we developed for specific challenges arising from the properties of the data
to the next section.

\subsection{Data preparation}
\label{subsec:dataprep}

For the majority of this work, we use the TNG300-1 (hydrodynamical) and TNG300-1\_DM
(gravity-only) $z=0$ snapshots
\citep{IllustrisTNG1, IllustrisTNG2, IllustrisTNG3, IllustrisTNG4, IllustrisTNG5,
       Nelsonetal2019}.
We also make use of additional zoom-in hydrodynamical simulations,
as discussed in Section~\ref{subsec:fewinterestingvoxels};
using an available subset of halos from the TNG-Cluster sample \citep{TNGClusters}.
Details on the pixelization of the various fields are given in Appendix~\ref{app:pixelization}.

In addition to the dark matter properties, we also use simple semi-analytical models for the output as
input data.
The reason for this is that neural networks train faster on residuals than on the entire target.
We defer the reader to the previous section for a description
of how these semi-analytical models have been constructed for the different targets.

\subsection{Sampling}
\label{subsec:sampling}

We expect the mapping between the dark matter field and the target properties of the electron gas to be
dominated by operators whose spatial support does not exceed length-scales of a few Mpc.
Therefore, we choose our training samples as local boxes,
as described in more detail in Appendix~\ref{app:trainingboxes}.

\subsection{Neural network}
\label{subsec:nn}

\begin{figure}
\begin{center}
\tikzset{font={\fontsize{6pt}{12}{\selectfont}}}
\begin{tikzpicture}[yscale=0.39,xscale=0.39]
	\node [right] at (34.2,-21) {\footnotesize $p_\text{drop} = 0.7$};
	\draw [decoration={brace,amplitude=0.2em},decorate] (34,-20) -- (34,-22);
	\node [right] at (34.2,-18) {\footnotesize $p_\text{drop} = 0.5$};
	\draw [decoration={brace,amplitude=0.2em},decorate] (34,-17) -- (34,-19);
	\node [right] at (34.2,-15) {\footnotesize $p_\text{drop} = 0.4$};
	\draw [decoration={brace,amplitude=0.2em},decorate] (34,-14) -- (34,-16);
	\draw [thick,->] (1,-8) -- (15,-22)
		node [midway,below,rotate=-45] {\footnotesize resolution decreases, \# of feature channels increases};
	\node [above] at (21,-3) {\footnotesize skip connections};
	\draw [thin,<-] (41.7,-4) -- (42.5,-2)
		node [anchor=west,align=left] {\footnotesize apply\\sinh};
	\node [above,rotate=90] at (0.68,-3.00) {\footnotesize\textbf{Input}};
	\node [left, rotate=90] at (0.75,-4.50) {$1\times64^3$};
	\draw[->,thick] (0.79,-3.00) -- (1.43,-3.00);
	\draw (0.68,-4.65) -- (0.68,-1.65);
	\draw (0.75,-4.65) -- (0.75,-1.65);
	\draw (0.68,-4.65) -- (0.75,-4.65);
	\draw (0.68,-1.65) -- (0.75,-1.65);
	\draw (0.82,-4.35) -- (0.82,-1.35);
	\draw (0.75,-4.65) -- (0.82,-4.35);
	\draw (0.75,-1.65) -- (0.82,-1.35);
	\draw (0.68,-1.65) -- (0.75,-1.35);
	\draw (0.75,-1.35) -- (0.82,-1.35);
	\draw (1.43,-4.65) -- (1.43,-1.65);
	\draw (1.50,-4.65) -- (1.50,-1.65);
	\draw (1.43,-4.65) -- (1.50,-4.65);
	\draw (1.43,-1.65) -- (1.50,-1.65);
	\draw (1.57,-4.35) -- (1.57,-1.35);
	\draw (1.50,-4.65) -- (1.57,-4.35);
	\draw (1.50,-1.65) -- (1.57,-1.35);
	\draw (1.43,-1.65) -- (1.50,-1.35);
	\draw (1.50,-1.35) -- (1.57,-1.35);
	\node [left, rotate=90] at (1.50,-4.50) {$32\times64^3$ };
	\draw[->,thick] (1.54,-3.00) -- (2.17,-3.00);
	\draw (2.17,-4.65) -- (2.17,-1.65);
	\draw (2.25,-4.65) -- (2.25,-1.65);
	\draw (2.17,-4.65) -- (2.25,-4.65);
	\draw (2.17,-1.65) -- (2.25,-1.65);
	\draw (2.33,-4.35) -- (2.33,-1.35);
	\draw (2.25,-4.65) -- (2.33,-4.35);
	\draw (2.25,-1.65) -- (2.33,-1.35);
	\draw (2.17,-1.65) -- (2.25,-1.35);
	\draw (2.25,-1.35) -- (2.33,-1.35);
	\node [left, rotate=90] at (2.25,-4.50) {$32\times64^3$ };
	\draw[->,thick] (2.29,-3.00) -- (2.92,-3.00);
	\draw (2.92,-4.65) -- (2.92,-1.65);
	\draw (3.00,-4.65) -- (3.00,-1.65);
	\draw (2.92,-4.65) -- (3.00,-4.65);
	\draw (2.92,-1.65) -- (3.00,-1.65);
	\draw (3.08,-4.35) -- (3.08,-1.35);
	\draw (3.00,-4.65) -- (3.08,-4.35);
	\draw (3.00,-1.65) -- (3.08,-1.35);
	\draw (2.92,-1.65) -- (3.00,-1.35);
	\draw (3.00,-1.35) -- (3.08,-1.35);
	\node [left, rotate=90] at (3.00,-4.50) {$32\times64^3$ };
	\draw[->,thick] (3.04,-3.00) -- (3.67,-3.00);
	\draw (3.67,-4.65) -- (3.67,-1.65);
	\draw (3.75,-4.65) -- (3.75,-1.65);
	\draw (3.67,-4.65) -- (3.75,-4.65);
	\draw (3.67,-1.65) -- (3.75,-1.65);
	\draw (3.83,-4.35) -- (3.83,-1.35);
	\draw (3.75,-4.65) -- (3.83,-4.35);
	\draw (3.75,-1.65) -- (3.83,-1.35);
	\draw (3.67,-1.65) -- (3.75,-1.35);
	\draw (3.75,-1.35) -- (3.83,-1.35);
	\node [left, rotate=90] at (3.75,-4.50) {$32\times64^3$ };
	\draw[dashed] (3.79,-3.75) -- (4.11,-3.75);
	\draw[dashed] (4.11,-3.75) -- (4.11,-5.25);
	\draw [->,dashed] (4.11,-5.25) -- (4.42,-5.25);
	\draw (4.42,-7.65) -- (4.42,-4.65);
	\draw (4.50,-7.65) -- (4.50,-4.65);
	\draw (4.42,-7.65) -- (4.50,-7.65);
	\draw (4.42,-4.65) -- (4.50,-4.65);
	\draw (4.58,-7.35) -- (4.58,-4.35);
	\draw (4.50,-7.65) -- (4.58,-7.35);
	\draw (4.50,-4.65) -- (4.58,-4.35);
	\draw (4.42,-4.65) -- (4.50,-4.35);
	\draw (4.50,-4.35) -- (4.58,-4.35);
	\node [left, rotate=90] at (4.50,-7.50) {$32\times64^3$ };
	\draw[->,thick] (4.54,-6.00) -- (5.17,-6.00);
	\draw (5.17,-7.65) -- (5.17,-4.65);
	\draw (5.25,-7.65) -- (5.25,-4.65);
	\draw (5.17,-7.65) -- (5.25,-7.65);
	\draw (5.17,-4.65) -- (5.25,-4.65);
	\draw (5.33,-7.35) -- (5.33,-4.35);
	\draw (5.25,-7.65) -- (5.33,-7.35);
	\draw (5.25,-4.65) -- (5.33,-4.35);
	\draw (5.17,-4.65) -- (5.25,-4.35);
	\draw (5.25,-4.35) -- (5.33,-4.35);
	\node [left, rotate=90] at (5.25,-7.50) {$64\times32^3$ };
	\draw[->,thick] (5.29,-6.00) -- (5.92,-6.00);
	\draw (5.92,-7.65) -- (5.92,-4.65);
	\draw (6.00,-7.65) -- (6.00,-4.65);
	\draw (5.92,-7.65) -- (6.00,-7.65);
	\draw (5.92,-4.65) -- (6.00,-4.65);
	\draw (6.08,-7.35) -- (6.08,-4.35);
	\draw (6.00,-7.65) -- (6.08,-7.35);
	\draw (6.00,-4.65) -- (6.08,-4.35);
	\draw (5.92,-4.65) -- (6.00,-4.35);
	\draw (6.00,-4.35) -- (6.08,-4.35);
	\node [left, rotate=90] at (6.00,-7.50) {$64\times32^3$ };
	\draw[->,thick] (6.04,-6.00) -- (6.67,-6.00);
	\draw (6.67,-7.65) -- (6.67,-4.65);
	\draw (6.75,-7.65) -- (6.75,-4.65);
	\draw (6.67,-7.65) -- (6.75,-7.65);
	\draw (6.67,-4.65) -- (6.75,-4.65);
	\draw (6.83,-7.35) -- (6.83,-4.35);
	\draw (6.75,-7.65) -- (6.83,-7.35);
	\draw (6.75,-4.65) -- (6.83,-4.35);
	\draw (6.67,-4.65) -- (6.75,-4.35);
	\draw (6.75,-4.35) -- (6.83,-4.35);
	\node [left, rotate=90] at (6.75,-7.50) {$64\times32^3$ };
	\draw[dashed] (6.79,-6.75) -- (7.11,-6.75);
	\draw[dashed] (7.11,-6.75) -- (7.11,-8.25);
	\draw [->,dashed] (7.11,-8.25) -- (7.42,-8.25);
	\draw (7.42,-10.65) -- (7.42,-7.65);
	\draw (7.50,-10.65) -- (7.50,-7.65);
	\draw (7.42,-10.65) -- (7.50,-10.65);
	\draw (7.42,-7.65) -- (7.50,-7.65);
	\draw (7.58,-10.35) -- (7.58,-7.35);
	\draw (7.50,-10.65) -- (7.58,-10.35);
	\draw (7.50,-7.65) -- (7.58,-7.35);
	\draw (7.42,-7.65) -- (7.50,-7.35);
	\draw (7.50,-7.35) -- (7.58,-7.35);
	\node [left, rotate=90] at (7.50,-10.50) {$64\times32^3$ };
	\draw[->,thick] (7.54,-9.00) -- (8.18,-9.00);
	\draw (8.18,-10.65) -- (8.18,-7.65);
	\draw (8.25,-10.65) -- (8.25,-7.65);
	\draw (8.18,-10.65) -- (8.25,-10.65);
	\draw (8.18,-7.65) -- (8.25,-7.65);
	\draw (8.32,-10.35) -- (8.32,-7.35);
	\draw (8.25,-10.65) -- (8.32,-10.35);
	\draw (8.25,-7.65) -- (8.32,-7.35);
	\draw (8.18,-7.65) -- (8.25,-7.35);
	\draw (8.25,-7.35) -- (8.32,-7.35);
	\node [left, rotate=90] at (8.25,-10.50) {$128\times16^3$ };
	\draw[->,thick] (8.29,-9.00) -- (8.93,-9.00);
	\draw (8.93,-10.65) -- (8.93,-7.65);
	\draw (9.00,-10.65) -- (9.00,-7.65);
	\draw (8.93,-10.65) -- (9.00,-10.65);
	\draw (8.93,-7.65) -- (9.00,-7.65);
	\draw (9.07,-10.35) -- (9.07,-7.35);
	\draw (9.00,-10.65) -- (9.07,-10.35);
	\draw (9.00,-7.65) -- (9.07,-7.35);
	\draw (8.93,-7.65) -- (9.00,-7.35);
	\draw (9.00,-7.35) -- (9.07,-7.35);
	\node [left, rotate=90] at (9.00,-10.50) {$128\times16^3$ };
	\draw[->,thick] (9.04,-9.00) -- (9.68,-9.00);
	\draw (9.68,-10.65) -- (9.68,-7.65);
	\draw (9.75,-10.65) -- (9.75,-7.65);
	\draw (9.68,-10.65) -- (9.75,-10.65);
	\draw (9.68,-7.65) -- (9.75,-7.65);
	\draw (9.82,-10.35) -- (9.82,-7.35);
	\draw (9.75,-10.65) -- (9.82,-10.35);
	\draw (9.75,-7.65) -- (9.82,-7.35);
	\draw (9.68,-7.65) -- (9.75,-7.35);
	\draw (9.75,-7.35) -- (9.82,-7.35);
	\node [left, rotate=90] at (9.75,-10.50) {$128\times16^3$ };
	\draw[dashed] (9.79,-9.75) -- (10.11,-9.75);
	\draw[dashed] (10.11,-9.75) -- (10.11,-11.25);
	\draw [->,dashed] (10.11,-11.25) -- (10.43,-11.25);
	\draw (10.43,-13.65) -- (10.43,-10.65);
	\draw (10.50,-13.65) -- (10.50,-10.65);
	\draw (10.43,-13.65) -- (10.50,-13.65);
	\draw (10.43,-10.65) -- (10.50,-10.65);
	\draw (10.57,-13.35) -- (10.57,-10.35);
	\draw (10.50,-13.65) -- (10.57,-13.35);
	\draw (10.50,-10.65) -- (10.57,-10.35);
	\draw (10.43,-10.65) -- (10.50,-10.35);
	\draw (10.50,-10.35) -- (10.57,-10.35);
	\node [left, rotate=90] at (10.50,-13.50) {$128\times16^3$ };
	\draw[->,thick] (10.54,-12.00) -- (11.18,-12.00);
	\draw (11.18,-13.65) -- (11.18,-10.65);
	\draw (11.25,-13.65) -- (11.25,-10.65);
	\draw (11.18,-13.65) -- (11.25,-13.65);
	\draw (11.18,-10.65) -- (11.25,-10.65);
	\draw (11.32,-13.35) -- (11.32,-10.35);
	\draw (11.25,-13.65) -- (11.32,-13.35);
	\draw (11.25,-10.65) -- (11.32,-10.35);
	\draw (11.18,-10.65) -- (11.25,-10.35);
	\draw (11.25,-10.35) -- (11.32,-10.35);
	\node [left, rotate=90] at (11.25,-13.50) {$256\times8^3$ };
	\draw[->,thick] (11.29,-12.00) -- (11.93,-12.00);
	\draw (11.93,-13.65) -- (11.93,-10.65);
	\draw (12.00,-13.65) -- (12.00,-10.65);
	\draw (11.93,-13.65) -- (12.00,-13.65);
	\draw (11.93,-10.65) -- (12.00,-10.65);
	\draw (12.07,-13.35) -- (12.07,-10.35);
	\draw (12.00,-13.65) -- (12.07,-13.35);
	\draw (12.00,-10.65) -- (12.07,-10.35);
	\draw (11.93,-10.65) -- (12.00,-10.35);
	\draw (12.00,-10.35) -- (12.07,-10.35);
	\node [left, rotate=90] at (12.00,-13.50) {$256\times8^3$ };
	\draw[->,thick] (12.04,-12.00) -- (12.68,-12.00);
	\draw (12.68,-13.65) -- (12.68,-10.65);
	\draw (12.75,-13.65) -- (12.75,-10.65);
	\draw (12.68,-13.65) -- (12.75,-13.65);
	\draw (12.68,-10.65) -- (12.75,-10.65);
	\draw (12.82,-13.35) -- (12.82,-10.35);
	\draw (12.75,-13.65) -- (12.82,-13.35);
	\draw (12.75,-10.65) -- (12.82,-10.35);
	\draw (12.68,-10.65) -- (12.75,-10.35);
	\draw (12.75,-10.35) -- (12.82,-10.35);
	\node [left, rotate=90] at (12.75,-13.50) {$256\times8^3$ };
	\draw[dashed] (12.79,-12.75) -- (13.11,-12.75);
	\draw[dashed] (13.11,-12.75) -- (13.11,-14.25);
	\draw [->,dashed] (13.11,-14.25) -- (13.43,-14.25);
	\draw (13.43,-16.65) -- (13.43,-13.65);
	\draw (13.50,-16.65) -- (13.50,-13.65);
	\draw (13.43,-16.65) -- (13.50,-16.65);
	\draw (13.43,-13.65) -- (13.50,-13.65);
	\draw (13.57,-16.35) -- (13.57,-13.35);
	\draw (13.50,-16.65) -- (13.57,-16.35);
	\draw (13.50,-13.65) -- (13.57,-13.35);
	\draw (13.43,-13.65) -- (13.50,-13.35);
	\draw (13.50,-13.35) -- (13.57,-13.35);
	\node [left, rotate=90] at (13.50,-16.50) {$256\times8^3$ };
	\draw[->,thick] (13.54,-15.00) -- (14.18,-15.00);
	\draw (14.18,-16.65) -- (14.18,-13.65);
	\draw (14.25,-16.65) -- (14.25,-13.65);
	\draw (14.18,-16.65) -- (14.25,-16.65);
	\draw (14.18,-13.65) -- (14.25,-13.65);
	\draw (14.32,-16.35) -- (14.32,-13.35);
	\draw (14.25,-16.65) -- (14.32,-16.35);
	\draw (14.25,-13.65) -- (14.32,-13.35);
	\draw (14.18,-13.65) -- (14.25,-13.35);
	\draw (14.25,-13.35) -- (14.32,-13.35);
	\node [left, rotate=90] at (14.25,-16.50) {$512\times4^3$ };
	\draw[->,thick] (14.29,-15.00) -- (14.93,-15.00);
	\draw (14.93,-16.65) -- (14.93,-13.65);
	\draw (15.00,-16.65) -- (15.00,-13.65);
	\draw (14.93,-16.65) -- (15.00,-16.65);
	\draw (14.93,-13.65) -- (15.00,-13.65);
	\draw (15.07,-16.35) -- (15.07,-13.35);
	\draw (15.00,-16.65) -- (15.07,-16.35);
	\draw (15.00,-13.65) -- (15.07,-13.35);
	\draw (14.93,-13.65) -- (15.00,-13.35);
	\draw (15.00,-13.35) -- (15.07,-13.35);
	\node [left, rotate=90] at (15.00,-16.50) {$512\times4^3$ };
	\draw[->,thick] (15.04,-15.00) -- (15.68,-15.00);
	\draw (15.68,-16.65) -- (15.68,-13.65);
	\draw (15.75,-16.65) -- (15.75,-13.65);
	\draw (15.68,-16.65) -- (15.75,-16.65);
	\draw (15.68,-13.65) -- (15.75,-13.65);
	\draw (15.82,-16.35) -- (15.82,-13.35);
	\draw (15.75,-16.65) -- (15.82,-16.35);
	\draw (15.75,-13.65) -- (15.82,-13.35);
	\draw (15.68,-13.65) -- (15.75,-13.35);
	\draw (15.75,-13.35) -- (15.82,-13.35);
	\node [left, rotate=90] at (15.75,-16.50) {$512\times4^3$ };
	\draw[dashed] (15.79,-15.75) -- (16.11,-15.75);
	\draw[dashed] (16.11,-15.75) -- (16.11,-17.25);
	\draw [->,dashed] (16.11,-17.25) -- (16.43,-17.25);
	\draw (16.43,-19.65) -- (16.43,-16.65);
	\draw (16.50,-19.65) -- (16.50,-16.65);
	\draw (16.43,-19.65) -- (16.50,-19.65);
	\draw (16.43,-16.65) -- (16.50,-16.65);
	\draw (16.57,-19.35) -- (16.57,-16.35);
	\draw (16.50,-19.65) -- (16.57,-19.35);
	\draw (16.50,-16.65) -- (16.57,-16.35);
	\draw (16.43,-16.65) -- (16.50,-16.35);
	\draw (16.50,-16.35) -- (16.57,-16.35);
	\node [left, rotate=90] at (16.50,-19.50) {$512\times4^3$ };
	\draw[->,thick] (16.54,-18.00) -- (17.18,-18.00);
	\draw (17.18,-19.65) -- (17.18,-16.65);
	\draw (17.25,-19.65) -- (17.25,-16.65);
	\draw (17.18,-19.65) -- (17.25,-19.65);
	\draw (17.18,-16.65) -- (17.25,-16.65);
	\draw (17.32,-19.35) -- (17.32,-16.35);
	\draw (17.25,-19.65) -- (17.32,-19.35);
	\draw (17.25,-16.65) -- (17.32,-16.35);
	\draw (17.18,-16.65) -- (17.25,-16.35);
	\draw (17.25,-16.35) -- (17.32,-16.35);
	\node [left, rotate=90] at (17.25,-19.50) {$1024\times2^3$ };
	\draw[->,thick] (17.29,-18.00) -- (17.93,-18.00);
	\draw (17.93,-19.65) -- (17.93,-16.65);
	\draw (18.00,-19.65) -- (18.00,-16.65);
	\draw (17.93,-19.65) -- (18.00,-19.65);
	\draw (17.93,-16.65) -- (18.00,-16.65);
	\draw (18.07,-19.35) -- (18.07,-16.35);
	\draw (18.00,-19.65) -- (18.07,-19.35);
	\draw (18.00,-16.65) -- (18.07,-16.35);
	\draw (17.93,-16.65) -- (18.00,-16.35);
	\draw (18.00,-16.35) -- (18.07,-16.35);
	\node [left, rotate=90] at (18.00,-19.50) {$1024\times2^3$ };
	\draw[->,thick] (18.04,-18.00) -- (18.68,-18.00);
	\draw (18.68,-19.65) -- (18.68,-16.65);
	\draw (18.75,-19.65) -- (18.75,-16.65);
	\draw (18.68,-19.65) -- (18.75,-19.65);
	\draw (18.68,-16.65) -- (18.75,-16.65);
	\draw (18.82,-19.35) -- (18.82,-16.35);
	\draw (18.75,-19.65) -- (18.82,-19.35);
	\draw (18.75,-16.65) -- (18.82,-16.35);
	\draw (18.68,-16.65) -- (18.75,-16.35);
	\draw (18.75,-16.35) -- (18.82,-16.35);
	\node [left, rotate=90] at (18.75,-19.50) {$1024\times2^3$ };
	\draw[dashed] (18.79,-18.75) -- (19.11,-18.75);
	\draw[dashed] (19.11,-18.75) -- (19.11,-20.25);
	\draw [->,dashed] (19.11,-20.25) -- (19.43,-20.25);
	\draw (19.43,-22.65) -- (19.43,-19.65);
	\draw (19.50,-22.65) -- (19.50,-19.65);
	\draw (19.43,-22.65) -- (19.50,-22.65);
	\draw (19.43,-19.65) -- (19.50,-19.65);
	\draw (19.57,-22.35) -- (19.57,-19.35);
	\draw (19.50,-22.65) -- (19.57,-22.35);
	\draw (19.50,-19.65) -- (19.57,-19.35);
	\draw (19.43,-19.65) -- (19.50,-19.35);
	\draw (19.50,-19.35) -- (19.57,-19.35);
	\node [left, rotate=90] at (19.50,-22.50) {$1024\times2^3$ };
	\draw[->,thick] (19.54,-21.00) -- (20.18,-21.00);
	\draw (20.18,-22.65) -- (20.18,-19.65);
	\draw (20.25,-22.65) -- (20.25,-19.65);
	\draw (20.18,-22.65) -- (20.25,-22.65);
	\draw (20.18,-19.65) -- (20.25,-19.65);
	\draw (20.32,-22.35) -- (20.32,-19.35);
	\draw (20.25,-22.65) -- (20.32,-22.35);
	\draw (20.25,-19.65) -- (20.32,-19.35);
	\draw (20.18,-19.65) -- (20.25,-19.35);
	\draw (20.25,-19.35) -- (20.32,-19.35);
	\node [left, rotate=90] at (20.25,-22.50) {$2048\times1^3$ };
	\draw[->,thick] (20.29,-21.00) -- (20.93,-21.00);
	\draw (20.93,-22.65) -- (20.93,-19.65);
	\draw (21.00,-22.65) -- (21.00,-19.65);
	\draw (20.93,-22.65) -- (21.00,-22.65);
	\draw (20.93,-19.65) -- (21.00,-19.65);
	\draw (21.07,-22.35) -- (21.07,-19.35);
	\draw (21.00,-22.65) -- (21.07,-22.35);
	\draw (21.00,-19.65) -- (21.07,-19.35);
	\draw (20.93,-19.65) -- (21.00,-19.35);
	\draw (21.00,-19.35) -- (21.07,-19.35);
	\node [left, rotate=90] at (21.00,-22.50) {$2048\times1^3$ };
	\draw[->,thick] (21.04,-21.00) -- (21.68,-21.00);
	\draw (21.68,-22.65) -- (21.68,-19.65);
	\draw (21.75,-22.65) -- (21.75,-19.65);
	\draw (21.68,-22.65) -- (21.75,-22.65);
	\draw (21.68,-19.65) -- (21.75,-19.65);
	\draw (21.82,-22.35) -- (21.82,-19.35);
	\draw (21.75,-22.65) -- (21.82,-22.35);
	\draw (21.75,-19.65) -- (21.82,-19.35);
	\draw (21.68,-19.65) -- (21.75,-19.35);
	\draw (21.75,-19.35) -- (21.82,-19.35);
	\node [left, rotate=90] at (21.75,-22.50) {$2048\times1^3$ };
	\draw[->,thick] (21.79,-21.00) -- (22.43,-21.00);
	\draw (22.43,-22.65) -- (22.43,-19.65);
	\draw (22.50,-22.65) -- (22.50,-19.65);
	\draw (22.43,-22.65) -- (22.50,-22.65);
	\draw (22.43,-19.65) -- (22.50,-19.65);
	\draw (22.57,-22.35) -- (22.57,-19.35);
	\draw (22.50,-22.65) -- (22.57,-22.35);
	\draw (22.50,-19.65) -- (22.57,-19.35);
	\draw (22.43,-19.65) -- (22.50,-19.35);
	\draw (22.50,-19.35) -- (22.57,-19.35);
	\node [left, rotate=90] at (22.50,-22.50) {$1024\times2^3$ };
	\draw[dashed] (22.54,-20.25) -- (22.86,-20.25);
	\draw[dashed] (22.86,-20.25) -- (22.86,-18.75);
	\draw [->,dashed] (22.86,-18.75) -- (23.18,-18.75);
	\draw[->,dashed] (18.79,-18.00) -- (23.18,-18.00);
	\draw (23.18,-19.65) -- (23.18,-16.65);
	\draw (23.25,-19.65) -- (23.25,-16.65);
	\draw (23.18,-19.65) -- (23.25,-19.65);
	\draw (23.18,-16.65) -- (23.25,-16.65);
	\draw (23.32,-19.35) -- (23.32,-16.35);
	\draw (23.25,-19.65) -- (23.32,-19.35);
	\draw (23.25,-16.65) -- (23.32,-16.35);
	\draw (23.18,-16.65) -- (23.25,-16.35);
	\draw (23.25,-16.35) -- (23.32,-16.35);
	\node [left, rotate=90] at (23.25,-19.50) {$2048\times2^3$ };
	\draw[->,thick] (23.29,-18.00) -- (23.93,-18.00);
	\draw (23.93,-19.65) -- (23.93,-16.65);
	\draw (24.00,-19.65) -- (24.00,-16.65);
	\draw (23.93,-19.65) -- (24.00,-19.65);
	\draw (23.93,-16.65) -- (24.00,-16.65);
	\draw (24.07,-19.35) -- (24.07,-16.35);
	\draw (24.00,-19.65) -- (24.07,-19.35);
	\draw (24.00,-16.65) -- (24.07,-16.35);
	\draw (23.93,-16.65) -- (24.00,-16.35);
	\draw (24.00,-16.35) -- (24.07,-16.35);
	\node [left, rotate=90] at (24.00,-19.50) {$1024\times2^3$ };
	\draw[->,thick] (24.04,-18.00) -- (24.68,-18.00);
	\draw (24.68,-19.65) -- (24.68,-16.65);
	\draw (24.75,-19.65) -- (24.75,-16.65);
	\draw (24.68,-19.65) -- (24.75,-19.65);
	\draw (24.68,-16.65) -- (24.75,-16.65);
	\draw (24.82,-19.35) -- (24.82,-16.35);
	\draw (24.75,-19.65) -- (24.82,-19.35);
	\draw (24.75,-16.65) -- (24.82,-16.35);
	\draw (24.68,-16.65) -- (24.75,-16.35);
	\draw (24.75,-16.35) -- (24.82,-16.35);
	\node [left, rotate=90] at (24.75,-19.50) {$1024\times2^3$ };
	\draw[->,thick] (24.79,-18.00) -- (25.43,-18.00);
	\draw (25.43,-19.65) -- (25.43,-16.65);
	\draw (25.50,-19.65) -- (25.50,-16.65);
	\draw (25.43,-19.65) -- (25.50,-19.65);
	\draw (25.43,-16.65) -- (25.50,-16.65);
	\draw (25.57,-19.35) -- (25.57,-16.35);
	\draw (25.50,-19.65) -- (25.57,-19.35);
	\draw (25.50,-16.65) -- (25.57,-16.35);
	\draw (25.43,-16.65) -- (25.50,-16.35);
	\draw (25.50,-16.35) -- (25.57,-16.35);
	\node [left, rotate=90] at (25.50,-19.50) {$512\times4^3$ };
	\draw[dashed] (25.54,-17.25) -- (25.86,-17.25);
	\draw[dashed] (25.86,-17.25) -- (25.86,-15.75);
	\draw [->,dashed] (25.86,-15.75) -- (26.18,-15.75);
	\draw[->,dashed] (15.79,-15.00) -- (26.18,-15.00);
	\draw (26.18,-16.65) -- (26.18,-13.65);
	\draw (26.25,-16.65) -- (26.25,-13.65);
	\draw (26.18,-16.65) -- (26.25,-16.65);
	\draw (26.18,-13.65) -- (26.25,-13.65);
	\draw (26.32,-16.35) -- (26.32,-13.35);
	\draw (26.25,-16.65) -- (26.32,-16.35);
	\draw (26.25,-13.65) -- (26.32,-13.35);
	\draw (26.18,-13.65) -- (26.25,-13.35);
	\draw (26.25,-13.35) -- (26.32,-13.35);
	\node [left, rotate=90] at (26.25,-16.50) {$1024\times4^3$ };
	\draw[->,thick] (26.29,-15.00) -- (26.93,-15.00);
	\draw (26.93,-16.65) -- (26.93,-13.65);
	\draw (27.00,-16.65) -- (27.00,-13.65);
	\draw (26.93,-16.65) -- (27.00,-16.65);
	\draw (26.93,-13.65) -- (27.00,-13.65);
	\draw (27.07,-16.35) -- (27.07,-13.35);
	\draw (27.00,-16.65) -- (27.07,-16.35);
	\draw (27.00,-13.65) -- (27.07,-13.35);
	\draw (26.93,-13.65) -- (27.00,-13.35);
	\draw (27.00,-13.35) -- (27.07,-13.35);
	\node [left, rotate=90] at (27.00,-16.50) {$512\times4^3$ };
	\draw[->,thick] (27.04,-15.00) -- (27.68,-15.00);
	\draw (27.68,-16.65) -- (27.68,-13.65);
	\draw (27.75,-16.65) -- (27.75,-13.65);
	\draw (27.68,-16.65) -- (27.75,-16.65);
	\draw (27.68,-13.65) -- (27.75,-13.65);
	\draw (27.82,-16.35) -- (27.82,-13.35);
	\draw (27.75,-16.65) -- (27.82,-16.35);
	\draw (27.75,-13.65) -- (27.82,-13.35);
	\draw (27.68,-13.65) -- (27.75,-13.35);
	\draw (27.75,-13.35) -- (27.82,-13.35);
	\node [left, rotate=90] at (27.75,-16.50) {$512\times4^3$ };
	\draw[->,thick] (27.79,-15.00) -- (28.43,-15.00);
	\draw (28.43,-16.65) -- (28.43,-13.65);
	\draw (28.50,-16.65) -- (28.50,-13.65);
	\draw (28.43,-16.65) -- (28.50,-16.65);
	\draw (28.43,-13.65) -- (28.50,-13.65);
	\draw (28.57,-16.35) -- (28.57,-13.35);
	\draw (28.50,-16.65) -- (28.57,-16.35);
	\draw (28.50,-13.65) -- (28.57,-13.35);
	\draw (28.43,-13.65) -- (28.50,-13.35);
	\draw (28.50,-13.35) -- (28.57,-13.35);
	\node [left, rotate=90] at (28.50,-16.50) {$256\times8^3$ };
	\draw[dashed] (28.54,-14.25) -- (28.86,-14.25);
	\draw[dashed] (28.86,-14.25) -- (28.86,-12.75);
	\draw [->,dashed] (28.86,-12.75) -- (29.18,-12.75);
	\draw[->,dashed] (12.79,-12.00) -- (29.18,-12.00);
	\draw (29.18,-13.65) -- (29.18,-10.65);
	\draw (29.25,-13.65) -- (29.25,-10.65);
	\draw (29.18,-13.65) -- (29.25,-13.65);
	\draw (29.18,-10.65) -- (29.25,-10.65);
	\draw (29.32,-13.35) -- (29.32,-10.35);
	\draw (29.25,-13.65) -- (29.32,-13.35);
	\draw (29.25,-10.65) -- (29.32,-10.35);
	\draw (29.18,-10.65) -- (29.25,-10.35);
	\draw (29.25,-10.35) -- (29.32,-10.35);
	\node [left, rotate=90] at (29.25,-13.50) {$512\times8^3$ };
	\draw[->,thick] (29.29,-12.00) -- (29.93,-12.00);
	\draw (29.93,-13.65) -- (29.93,-10.65);
	\draw (30.00,-13.65) -- (30.00,-10.65);
	\draw (29.93,-13.65) -- (30.00,-13.65);
	\draw (29.93,-10.65) -- (30.00,-10.65);
	\draw (30.07,-13.35) -- (30.07,-10.35);
	\draw (30.00,-13.65) -- (30.07,-13.35);
	\draw (30.00,-10.65) -- (30.07,-10.35);
	\draw (29.93,-10.65) -- (30.00,-10.35);
	\draw (30.00,-10.35) -- (30.07,-10.35);
	\node [left, rotate=90] at (30.00,-13.50) {$256\times8^3$ };
	\draw[->,thick] (30.04,-12.00) -- (30.68,-12.00);
	\draw (30.68,-13.65) -- (30.68,-10.65);
	\draw (30.75,-13.65) -- (30.75,-10.65);
	\draw (30.68,-13.65) -- (30.75,-13.65);
	\draw (30.68,-10.65) -- (30.75,-10.65);
	\draw (30.82,-13.35) -- (30.82,-10.35);
	\draw (30.75,-13.65) -- (30.82,-13.35);
	\draw (30.75,-10.65) -- (30.82,-10.35);
	\draw (30.68,-10.65) -- (30.75,-10.35);
	\draw (30.75,-10.35) -- (30.82,-10.35);
	\node [left, rotate=90] at (30.75,-13.50) {$256\times8^3$ };
	\draw[->,thick] (30.79,-12.00) -- (31.43,-12.00);
	\draw (31.43,-13.65) -- (31.43,-10.65);
	\draw (31.50,-13.65) -- (31.50,-10.65);
	\draw (31.43,-13.65) -- (31.50,-13.65);
	\draw (31.43,-10.65) -- (31.50,-10.65);
	\draw (31.57,-13.35) -- (31.57,-10.35);
	\draw (31.50,-13.65) -- (31.57,-13.35);
	\draw (31.50,-10.65) -- (31.57,-10.35);
	\draw (31.43,-10.65) -- (31.50,-10.35);
	\draw (31.50,-10.35) -- (31.57,-10.35);
	\node [left, rotate=90] at (31.50,-13.50) {$128\times16^3$ };
	\draw[dashed] (31.54,-11.25) -- (31.86,-11.25);
	\draw[dashed] (31.86,-11.25) -- (31.86,-9.75);
	\draw [->,dashed] (31.86,-9.75) -- (32.17,-9.75);
	\draw[->,dashed] (9.79,-9.00) -- (32.17,-9.00);
	\draw (32.17,-10.65) -- (32.17,-7.65);
	\draw (32.25,-10.65) -- (32.25,-7.65);
	\draw (32.17,-10.65) -- (32.25,-10.65);
	\draw (32.17,-7.65) -- (32.25,-7.65);
	\draw (32.33,-10.35) -- (32.33,-7.35);
	\draw (32.25,-10.65) -- (32.33,-10.35);
	\draw (32.25,-7.65) -- (32.33,-7.35);
	\draw (32.17,-7.65) -- (32.25,-7.35);
	\draw (32.25,-7.35) -- (32.33,-7.35);
	\node [left, rotate=90] at (32.25,-10.50) {$256\times16^3$ };
	\draw[->,thick] (32.29,-9.00) -- (32.92,-9.00);
	\draw (32.92,-10.65) -- (32.92,-7.65);
	\draw (33.00,-10.65) -- (33.00,-7.65);
	\draw (32.92,-10.65) -- (33.00,-10.65);
	\draw (32.92,-7.65) -- (33.00,-7.65);
	\draw (33.08,-10.35) -- (33.08,-7.35);
	\draw (33.00,-10.65) -- (33.08,-10.35);
	\draw (33.00,-7.65) -- (33.08,-7.35);
	\draw (32.92,-7.65) -- (33.00,-7.35);
	\draw (33.00,-7.35) -- (33.08,-7.35);
	\node [left, rotate=90] at (33.00,-10.50) {$128\times16^3$ };
	\draw[->,thick] (33.04,-9.00) -- (33.67,-9.00);
	\draw (33.67,-10.65) -- (33.67,-7.65);
	\draw (33.75,-10.65) -- (33.75,-7.65);
	\draw (33.67,-10.65) -- (33.75,-10.65);
	\draw (33.67,-7.65) -- (33.75,-7.65);
	\draw (33.83,-10.35) -- (33.83,-7.35);
	\draw (33.75,-10.65) -- (33.83,-10.35);
	\draw (33.75,-7.65) -- (33.83,-7.35);
	\draw (33.67,-7.65) -- (33.75,-7.35);
	\draw (33.75,-7.35) -- (33.83,-7.35);
	\node [left, rotate=90] at (33.75,-10.50) {$128\times16^3$ };
	\draw[->,thick] (33.79,-9.00) -- (34.42,-9.00);
	\draw (34.42,-10.65) -- (34.42,-7.65);
	\draw (34.50,-10.65) -- (34.50,-7.65);
	\draw (34.42,-10.65) -- (34.50,-10.65);
	\draw (34.42,-7.65) -- (34.50,-7.65);
	\draw (34.58,-10.35) -- (34.58,-7.35);
	\draw (34.50,-10.65) -- (34.58,-10.35);
	\draw (34.50,-7.65) -- (34.58,-7.35);
	\draw (34.42,-7.65) -- (34.50,-7.35);
	\draw (34.50,-7.35) -- (34.58,-7.35);
	\node [left, rotate=90] at (34.50,-10.50) {$64\times32^3$ };
	\draw[dashed] (34.54,-8.25) -- (34.86,-8.25);
	\draw[dashed] (34.86,-8.25) -- (34.86,-6.75);
	\draw [->,dashed] (34.86,-6.75) -- (35.17,-6.75);
	\draw[->,dashed] (6.79,-6.00) -- (35.17,-6.00);
	\draw (35.17,-7.65) -- (35.17,-4.65);
	\draw (35.25,-7.65) -- (35.25,-4.65);
	\draw (35.17,-7.65) -- (35.25,-7.65);
	\draw (35.17,-4.65) -- (35.25,-4.65);
	\draw (35.33,-7.35) -- (35.33,-4.35);
	\draw (35.25,-7.65) -- (35.33,-7.35);
	\draw (35.25,-4.65) -- (35.33,-4.35);
	\draw (35.17,-4.65) -- (35.25,-4.35);
	\draw (35.25,-4.35) -- (35.33,-4.35);
	\node [left, rotate=90] at (35.25,-7.50) {$128\times32^3$ };
	\draw[->,thick] (35.29,-6.00) -- (35.92,-6.00);
	\draw (35.92,-7.65) -- (35.92,-4.65);
	\draw (36.00,-7.65) -- (36.00,-4.65);
	\draw (35.92,-7.65) -- (36.00,-7.65);
	\draw (35.92,-4.65) -- (36.00,-4.65);
	\draw (36.08,-7.35) -- (36.08,-4.35);
	\draw (36.00,-7.65) -- (36.08,-7.35);
	\draw (36.00,-4.65) -- (36.08,-4.35);
	\draw (35.92,-4.65) -- (36.00,-4.35);
	\draw (36.00,-4.35) -- (36.08,-4.35);
	\node [left, rotate=90] at (36.00,-7.50) {$64\times32^3$ };
	\draw[->,thick] (36.04,-6.00) -- (36.67,-6.00);
	\draw (36.67,-7.65) -- (36.67,-4.65);
	\draw (36.75,-7.65) -- (36.75,-4.65);
	\draw (36.67,-7.65) -- (36.75,-7.65);
	\draw (36.67,-4.65) -- (36.75,-4.65);
	\draw (36.83,-7.35) -- (36.83,-4.35);
	\draw (36.75,-7.65) -- (36.83,-7.35);
	\draw (36.75,-4.65) -- (36.83,-4.35);
	\draw (36.67,-4.65) -- (36.75,-4.35);
	\draw (36.75,-4.35) -- (36.83,-4.35);
	\node [left, rotate=90] at (36.75,-7.50) {$64\times32^3$ };
	\draw[->,thick] (36.79,-6.00) -- (37.42,-6.00);
	\draw (37.42,-7.65) -- (37.42,-4.65);
	\draw (37.50,-7.65) -- (37.50,-4.65);
	\draw (37.42,-7.65) -- (37.50,-7.65);
	\draw (37.42,-4.65) -- (37.50,-4.65);
	\draw (37.58,-7.35) -- (37.58,-4.35);
	\draw (37.50,-7.65) -- (37.58,-7.35);
	\draw (37.50,-4.65) -- (37.58,-4.35);
	\draw (37.42,-4.65) -- (37.50,-4.35);
	\draw (37.50,-4.35) -- (37.58,-4.35);
	\node [left, rotate=90] at (37.50,-7.50) {$32\times32^3$ };
	\draw[dashed] (37.54,-5.25) -- (37.86,-5.25);
	\draw[dashed] (37.86,-5.25) -- (37.86,-3.75);
	\draw [->,dashed] (37.86,-3.75) -- (38.17,-3.75);
	\draw[->,dashed] (3.79,-3.00) -- (38.17,-3.00);
	\draw (38.17,-4.65) -- (38.17,-1.65);
	\draw (38.25,-4.65) -- (38.25,-1.65);
	\draw (38.17,-4.65) -- (38.25,-4.65);
	\draw (38.17,-1.65) -- (38.25,-1.65);
	\draw (38.33,-4.35) -- (38.33,-1.35);
	\draw (38.25,-4.65) -- (38.33,-4.35);
	\draw (38.25,-1.65) -- (38.33,-1.35);
	\draw (38.17,-1.65) -- (38.25,-1.35);
	\draw (38.25,-1.35) -- (38.33,-1.35);
	\node [left, rotate=90] at (38.25,-4.50) {$64\times32^3$ };
	\draw[->,thick] (38.29,-3.00) -- (38.92,-3.00);
	\draw (38.92,-4.65) -- (38.92,-1.65);
	\draw (39.00,-4.65) -- (39.00,-1.65);
	\draw (38.92,-4.65) -- (39.00,-4.65);
	\draw (38.92,-1.65) -- (39.00,-1.65);
	\draw (39.08,-4.35) -- (39.08,-1.35);
	\draw (39.00,-4.65) -- (39.08,-4.35);
	\draw (39.00,-1.65) -- (39.08,-1.35);
	\draw (38.92,-1.65) -- (39.00,-1.35);
	\draw (39.00,-1.35) -- (39.08,-1.35);
	\node [left, rotate=90] at (39.00,-4.50) {$64\times32^3$ };
	\draw[->,thick] (39.04,-3.00) -- (39.67,-3.00);
	\draw (39.67,-4.65) -- (39.67,-1.65);
	\draw (39.75,-4.65) -- (39.75,-1.65);
	\draw (39.67,-4.65) -- (39.75,-4.65);
	\draw (39.67,-1.65) -- (39.75,-1.65);
	\draw (39.83,-4.35) -- (39.83,-1.35);
	\draw (39.75,-4.65) -- (39.83,-4.35);
	\draw (39.75,-1.65) -- (39.83,-1.35);
	\draw (39.67,-1.65) -- (39.75,-1.35);
	\draw (39.75,-1.35) -- (39.83,-1.35);
	\node [left, rotate=90] at (39.75,-4.50) {$64\times32^3$ };
	\draw[->,thick] (39.79,-3.00) -- (40.42,-3.00);
	\draw (40.42,-4.65) -- (40.42,-1.65);
	\draw (40.50,-4.65) -- (40.50,-1.65);
	\draw (40.42,-4.65) -- (40.50,-4.65);
	\draw (40.42,-1.65) -- (40.50,-1.65);
	\draw (40.58,-4.35) -- (40.58,-1.35);
	\draw (40.50,-4.65) -- (40.58,-4.35);
	\draw (40.50,-1.65) -- (40.58,-1.35);
	\draw (40.42,-1.65) -- (40.50,-1.35);
	\draw (40.50,-1.35) -- (40.58,-1.35);
	\node [left, rotate=90] at (40.50,-4.50) {$64\times32^3$ };
	\draw[->,thick] (40.54,-3.00) -- (41.17,-3.00);
	\draw (41.17,-4.65) -- (41.17,-1.65);
	\draw (41.25,-4.65) -- (41.25,-1.65);
	\draw (41.17,-4.65) -- (41.25,-4.65);
	\draw (41.17,-1.65) -- (41.25,-1.65);
	\draw (41.33,-4.35) -- (41.33,-1.35);
	\draw (41.25,-4.65) -- (41.33,-4.35);
	\draw (41.25,-1.65) -- (41.33,-1.35);
	\draw (41.17,-1.65) -- (41.25,-1.35);
	\draw (41.25,-1.35) -- (41.33,-1.35);
	\node [left, rotate=90] at (41.25,-4.50) {$63\times32^3$ };
	\draw[dashed] (41.29,-3.75) -- (41.61,-3.75);
	\draw[dashed] (41.61,-3.75) -- (41.61,-5.25);
	\draw [->,dashed] (41.61,-5.25) -- (41.92,-5.25);
	%
	\draw (41.17,-13.65) -- (41.17,-10.65);
	\draw (41.25,-13.65) -- (41.25,-10.65);
	\draw (41.17,-13.65) -- (41.25,-13.65);
	\draw (41.17,-10.65) -- (41.25,-10.65);
	\draw (41.33,-13.35) -- (41.33,-10.35);
	\draw (41.25,-13.65) -- (41.33,-13.35);
	\draw (41.25,-10.65) -- (41.33,-10.35);
	\draw (41.17,-10.65) -- (41.25,-10.35);
	\draw (41.25,-10.35) -- (41.33,-10.35);
	\node [left, rotate=90] at (41.25,-13.5) {$1\times32^3$};
	\node [above,rotate=90] at (41.17,-12.00) {\footnotesize\textbf{Model}};
	\draw[dashed] (41.29,-11.25) -- (41.61,-11.25);
	\draw[dashed] (41.61,-11.25) -- (41.61,-6.75);
	\draw [->,dashed] (41.61,-6.75) -- (41.92,-6.75);
	%
	\draw (41.92,-7.65) -- (41.92,-4.65);
	\draw (42.00,-7.65) -- (42.00,-4.65);
	\draw (41.92,-7.65) -- (42.00,-7.65);
	\draw (41.92,-4.65) -- (42.00,-4.65);
	\draw (42.08,-7.35) -- (42.08,-4.35);
	\draw (42.00,-7.65) -- (42.08,-7.35);
	\draw (42.00,-4.65) -- (42.08,-4.35);
	\draw (41.92,-4.65) -- (42.00,-4.35);
	\draw (42.00,-4.35) -- (42.08,-4.35);
	\node [left, rotate=90] at (42.00,-7.50) {$64\times32^3$ };
	\draw[->,thick] (42.04,-6.00) -- (42.67,-6.00);
	\draw (42.67,-7.65) -- (42.67,-4.65);
	\draw (42.75,-7.65) -- (42.75,-4.65);
	\draw (42.67,-7.65) -- (42.75,-7.65);
	\draw (42.67,-4.65) -- (42.75,-4.65);
	\draw (42.83,-7.35) -- (42.83,-4.35);
	\draw (42.75,-7.65) -- (42.83,-7.35);
	\draw (42.75,-4.65) -- (42.83,-4.35);
	\draw (42.67,-4.65) -- (42.75,-4.35);
	\draw (42.75,-4.35) -- (42.83,-4.35);
	\node [left, rotate=90] at (42.75,-7.50) {$1\times32^3$};
	\node [below,rotate=90] at (42.83,-6.00) {\footnotesize\textbf{Output}};
\end{tikzpicture}
\caption{U-Net architecture for the CNN used in predicting electron pressure and density
	 from dark matter distributions in 3D.
	 The layer dimensions are given in the format 
	 $(\text{\# of feature channels})\times\text{grid size}$.
         The dashed lines represent copies, while the solid arrows are convolutional operations
	 with $3^3$-sized filters (except for the very last operation, where the channels are simply
	 collapsed into the output).
	 If more than one dashed line ends at a single layer we concatenate along the feature
	 channel direction.
	 In the lowest three levels we apply dropout of entire feature channels with the
	 probabilities given.
	 We usually apply batch normalization, except wherever dropout is used.
	 All activation functions are ReLUs.
	 The network used to predict the electron momentum density is of similar structure,
	 with the following differences: the semi-analytical model is not concatenated but rather multiplied with
	 the first channel of the network representation; no $\sinh$ function is applied; since the resolution is a factor
	 2 smaller the network is smaller; in the two levels with highest resolution we apply
	 \texttt{Hardshrink} activation functions.
	 Most of these modifications are natural consequences of the negative-positive symmetry of the
	 momentum data.}
\label{fig:network}
\end{center}
\end{figure}
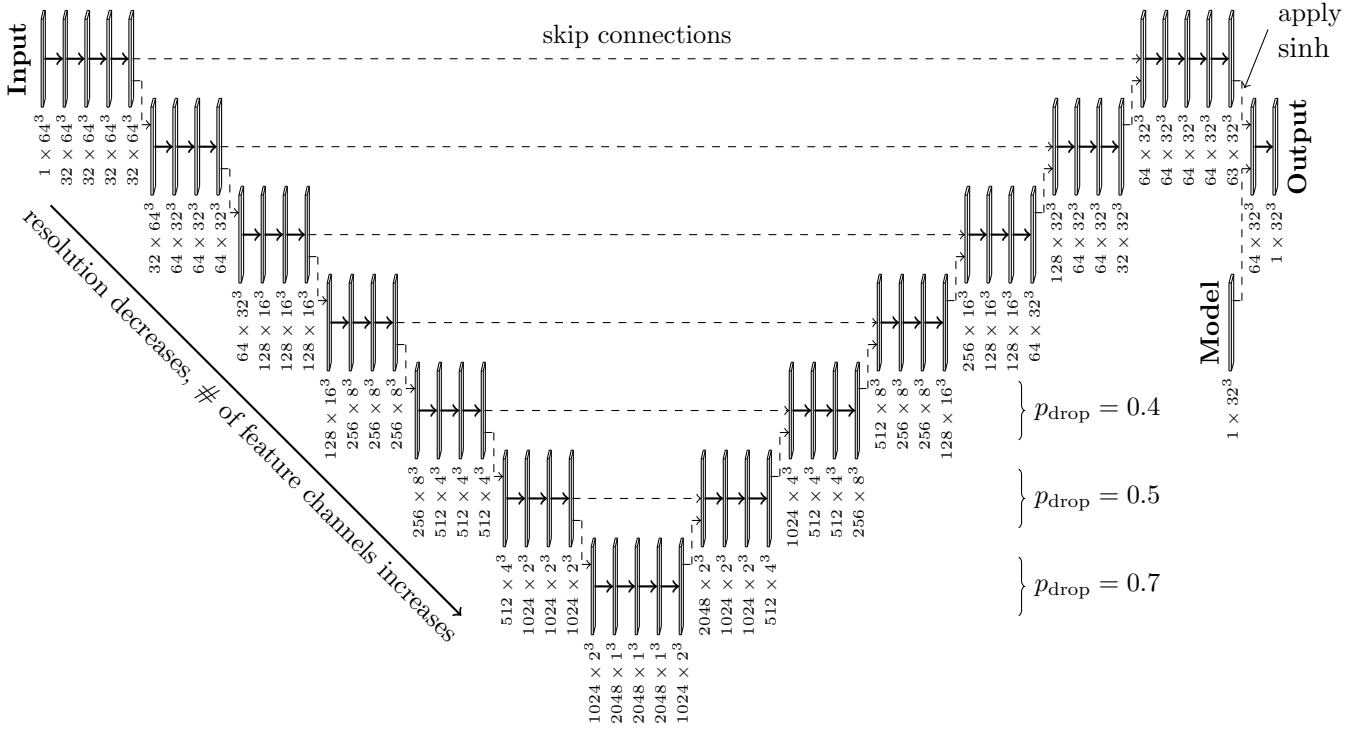

Having defined the input and target data, we now discuss the neural network architecture.
Given the locality of the problem as well as translational, rotational, and parity invariance, a
deep convolutional neural network architecture is the logical choice.
Motivated by previous successes in similar applications \citep[e.g.,][]{Heetal2019, Zhangetal2019}, we choose a U-Net
architecture, displayed in Fig.~\ref{fig:network} (this network was used for electron pressure and
density as target data, the architecture for the electron momentum density was slightly modified to
take the sign symmetry of the data as well as lower resolution into account).
The general idea of the U-Net is that while the receptive field increases in deeper layers,
the number of feature channels increases.
This allows the network to find more spatial correlations on larger scales.
Skip connections (horizontal dashed lines in the figure) make it easier for the network to retain
more local, high resolution features.
Note that the semi-analytical model is fed into the network at a rather late stage,
this makes it easier for the network to recognize
that the target is close to the semi-analytical model with small residuals
that need to be learned in the deeper layers.
More details on the network architecture may be found in the caption of Fig.~\ref{fig:network}.

\subsection{Training procedure}
\label{subsec:training}

To train the network, we use the Adam optimizer with $\vec\beta = \{0.9, 0.999\}$
and choose samples from the training set according
to a strongly biased selection function, as described in Section~\ref{subsec:fewinterestingvoxels}.
These samples consist of in- and output boxes which are described in
Appendix~\ref{app:trainingboxes} (there, we also specify the splitting into training,
validation, and testing data).
Although there is no natural definition of an epoch, for convenience,
we define an epoch as 8192 training samples.
The validation loss is evaluated on a random subset (chosen according to the same selection function
as for the training set) of 256 boxes from the validation set.
We find that a batch size of 32 yields the best performance.
We generally start training with a learning rate of $10^{-3}$, train for about 100 epochs, and then
resume training from the state of the network with the best validation loss.
We adjust the learning rate whenever we see the validation loss to become dominated by noise over
any perceptible downwards trend.
While during training only the loss function discussed in Section~\ref{subsec:fewinterestingvoxels}
is computed, after each training run (i.e. every $\sim 100$ epochs),
we evaluate the network on the whole validation box and compute power spectrum, cross-correlation
coefficient with the reference hydrodynamical simulation, and one-point PDF.
We base our decisions regarding the tuning of hyperparameters, changes in network architecture,
and the eventual stop of training principally on these summary statistics as well as the validation
loss curve.

We perform the usual data augmentation of rotations, reflections, and transpositions;
in addition, in order to avoid overfitting, we multiply input data and target by voxel-wise Gaussian
noise of $10\,\%$ standard deviation and unit mean.
Although it is impossible to explore all directions in the space of hyperparameters, we are
confident that with about 200 runs with different hyperparameters and network architectures we have
found a relatively well optimized training procedure and network architecture.

\section{Challenge: sparsity}
\label{sec:sparsity}

The main challenge
we need to overcome to train the neural network can be summarized as the sparsity of the dataset.
In this section, we will argue that this problem comes in two different, but related, aspects,
and explain the strategies we use to solve it.
We will concentrate on the case of electron pressure here, the arguments carry over to electron
density and electron momentum density.

\subsection{Few interesting voxels}
\label{subsec:fewinterestingvoxels}

\begin{figure}
\begin{center}
\begin{minipage}[t]{0.48\textwidth}
\includegraphics[width=\textwidth]{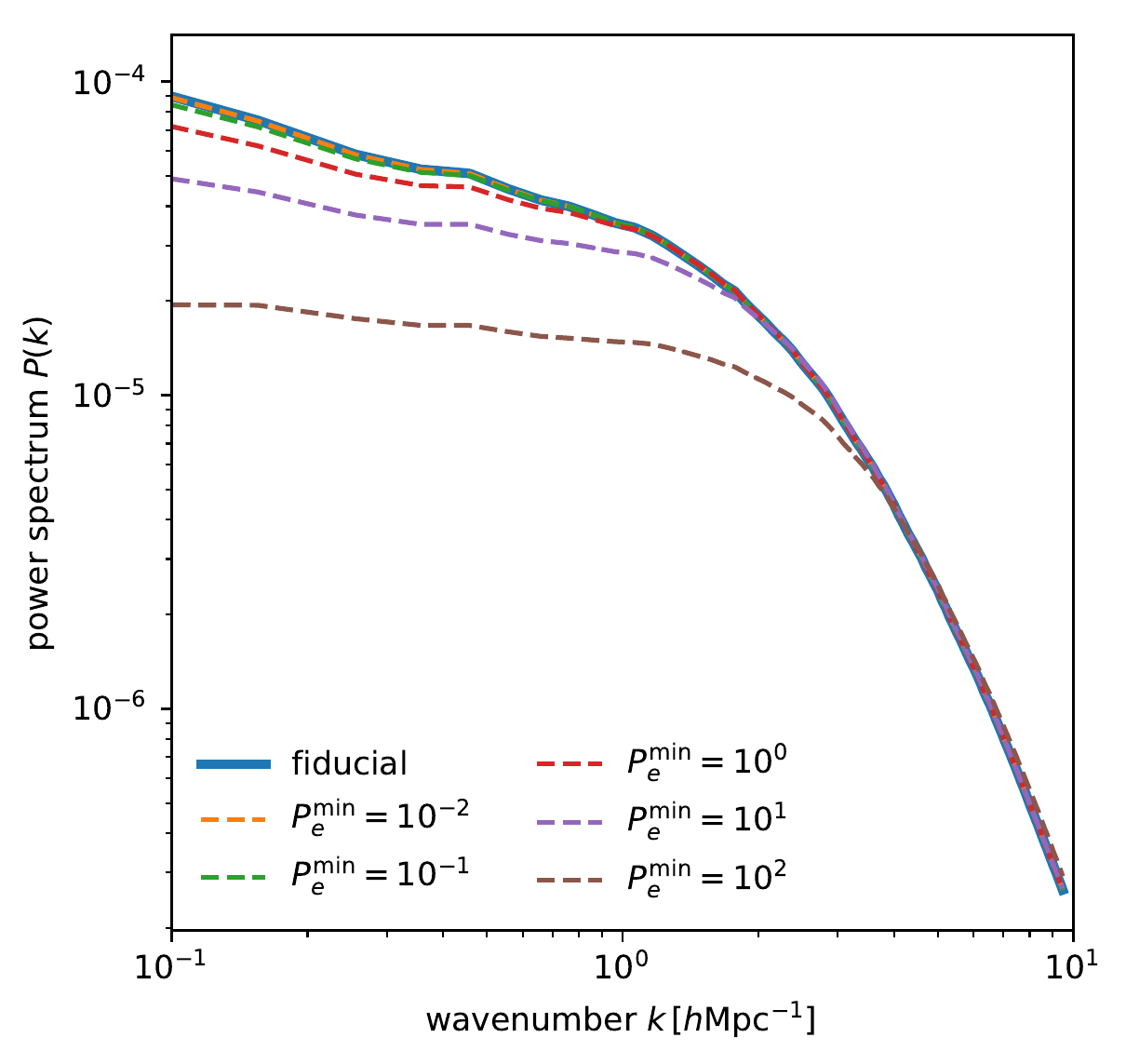}
\captionof{figure}
	{\emph{solid blue:} electron pressure power spectrum for 20\% of the TNG300
	 box. \emph{dashed:} the same with electron pressure values below
	 minimum values $P_e^\text{min}$ (given in units of $\sigma(P_e)$) set to zero.
	 It can be seen that the electron pressure power spectrum can be reproduced
	 using only voxels with $P_e\gtrsim 10^{-2}\sigma(P_e)$.
	 These voxels represent only a small fraction of the entire box,
	 as illustrated in Fig.~\ref{fig:PDFsparsity}.}
\label{fig:thresholds}
\end{minipage}\hspace{1em}%
\begin{minipage}[t]{0.48\textwidth}
\includegraphics[width=\textwidth]{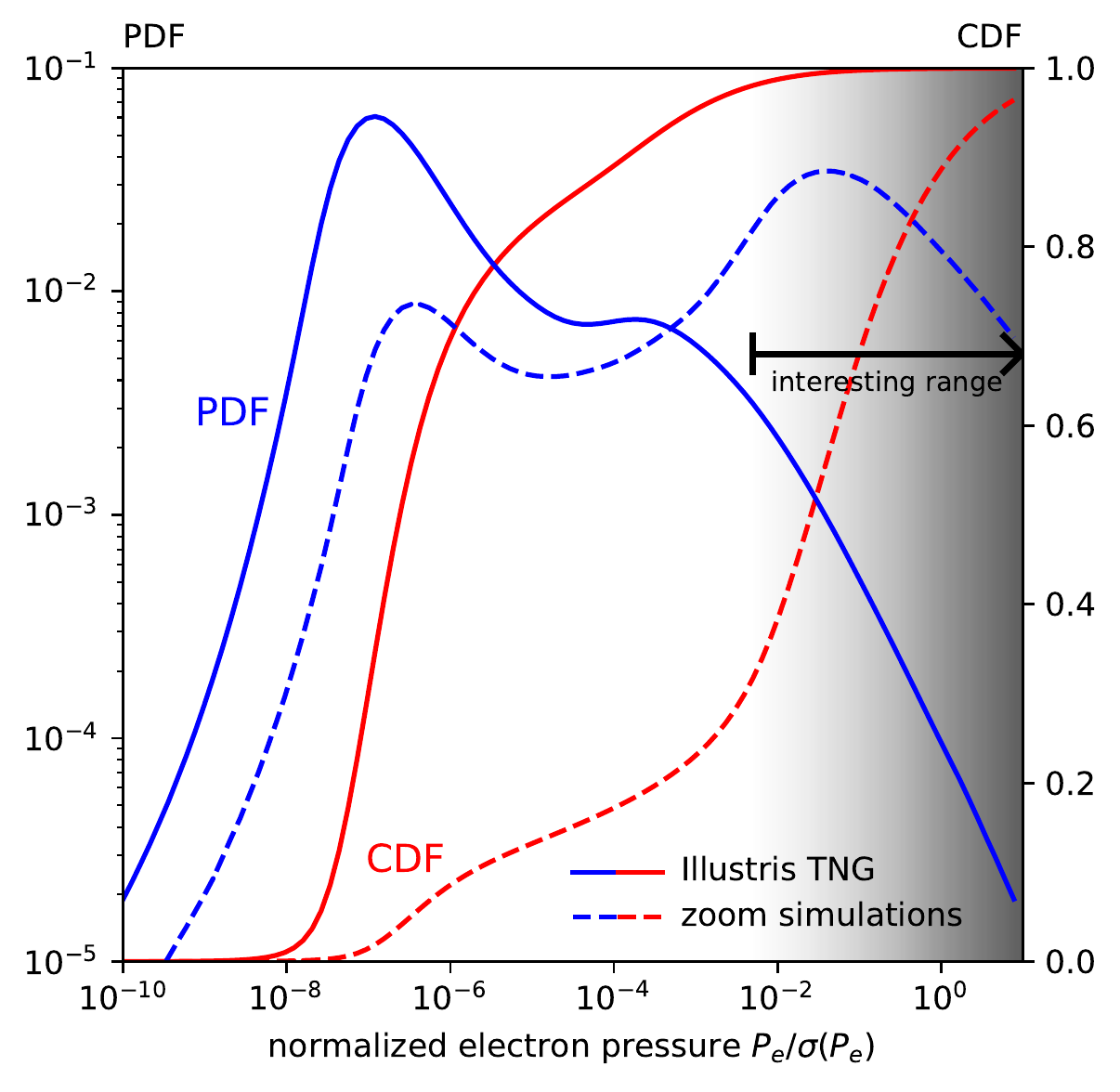}
\captionof{figure}
	{Illustration of the sparsity problem in the electron pressure distribution.
         Plotted are both probability distribution function (PDF; blue)
	 and cumulative distribution function (CDF; red)
	 for the TNG300 volume (solid) and the 181 TNG-Cluster zoom-in simulation boxes (dashed).
	 The shaded area labeled ``interesting range" was chosen according to the conclusions
	 from Fig.~\ref{fig:thresholds}.
	 In the reference uniform-volume simulation, TNG300, only of order $1\,\%$ of the voxels
	 fall into the ``interesting" range.}
\label{fig:PDFsparsity}
\end{minipage}
\end{center}
\end{figure}

The first aspect to consider is the fact that only very few voxels are actually ``interesting", in the sense
that they contribute to any commonly used summary statistics.
This point is illustrated in Fig.~\ref{fig:thresholds}, where we plot the electron pressure power
spectrum. While the solid blue line is the power spectrum for the target TNG300 hydrodynamical box, the dashed lines
result from setting voxels with electron pressure below varying thresholds to zero.
We observe that only voxels with $P_e \gtrsim 10^{-2}\sigma(P_e)$ are relevant (this argument is only rigorous
for the case of the power spectrum, but it is clear that for higher order statistics the situation
is even more severe).
Referring now to the solid lines in Fig.~\ref{fig:PDFsparsity}, in which we plot the probability
distribution function (PDF) and cumulative distribution function (CDF) of the electron pressure,
we observe that those relevant voxels constitute only a tiny fraction (of order $1\,\%$) of the simulation volume.

The small fraction of interesting voxels implies that a naive training procedure in which we would
randomly pick training samples from the simulation volume would be highly inefficient.
Thus, we solve this first aspect of the sparsity problem by biasing the training samples.
We know that the high electron pressure voxels are found in massive halos. Therefore, it seems
natural to bias the sample selection such that always at least one halo of a given
minimum mass is to be found in any training sample.

In the case of the electron density as target, we find that biasing according to the selection
function\footnote{From here onwards, wherever not otherwise stated, we use the following units:\\
                  \mbox{unit mass = $10^{10}\,h^{-1}M_\odot$};
		  \mbox{unit length = $1\,h^{-1}\text{kpc}$};
		  \mbox{unit speed = $1\,\text{km}/\text{s}$}.\label{fn:units}}
\begin{equation}
w_{\rho_e}(M_\text{500c}) \propto \Theta(M_\text{500c}-10^3) (\log_{10} M_\text{500c})^{0.25}
\label{eq:rhoesampling}
\end{equation}
with $\Theta$ the Heaviside function gives good results.

Note that the actual halo mass function of the training samples is \emph{not}
identical to Eq.~\eqref{eq:rhoesampling}
but rather equals $w(M)$ times the TNG300 mass function
(since $w(M)$ chooses from the available halos).\footnote{To be precise:
we make a list of TNG300 halos with masses above the argument in the $\Theta$-function,
assign weights to them according to $w(M)$ and then, in each training step,
randomly choose a halo according to these weights.
The input (gravity-only) sample box, whose geometry is described in
Appendix~\ref{app:trainingboxes},
is then chosen such that the halo's center falls into one of its voxels,
with equal probability for each voxel to contain the halo center.}

In the case of the electron momentum density, we find that a less aggressive selection function
works best:
\begin{equation}
w_{\mathbf{p}_e}(M_\text{500c}) \propto \Theta(M_\text{500c}-10^3)\,.
\label{eq:pesampling}
\end{equation}

In the case of electron pressure, the biasing schemes described above are not sufficient.
The reason is that the integrated
electron pressure in halos of mass $M_\text{halo}$ is not simply $\propto M_\text{halo}$,
but scales as $\propto M_\text{halo}^{5/3}$.
To overcome this problem, for electron pressure alone,
we do not use the original TNG300 simulation for training samples, but instead draw our
training samples from additional zoom-in simulations centered on massive halos.
These are performed with the same astrophysical galaxy-formation model as TNG300 and at similar
resolution, but focus on high-mass halos \citep{TNGClusters}.
Hence, these zoom-in simulations have a very different halo mass function compared to the TNG300 volume;
in particular, while the TNG300 simulation only has a few halos with
$M_\text{200c} > 10^{14.5}\,h^{-1}M_\odot$, the majority of the zoom-in primary objects are more
massive than this.
The electron pressure PDF and CDF obtained in these highly biased training samples are plotted
as the dashed lines in Fig.~\ref{fig:PDFsparsity}.
It can be seen that about $70\,\%$ of the voxels
from the TNG-Cluster training set fall in the
``interesting" range, according to the conclusions derived from Fig.~\ref{fig:thresholds}.
We point out that for the zoom-in simulations we did not have gravity-only
simulations available. This implies that the mapping from input (which is gravity-only during
validation and testing) to output
is not exactly identical for the TNG-Cluster training samples
and the TNG300 validation and testing set.
We would expect this effect to be most prominent on small scales.

\subsection{Tailed distributions}

The second aspect of the sparsity problem is the remaining high dynamic range of the voxel values.
Even though we have argued that the network can essentially ignore any $P_e < 10^{-2}\sigma(P_e)$,
the remaining ``interesting" electron pressure values still span several orders of magnitude
(c.f. Fig.~\ref{fig:PDFsparsity}).
Likewise, the dark matter density which we take as the network input also has a tailed distribution.

The solution for the problem of the tailed distribution of input data is relatively standard
\citep[e.g.,][]{Troesteretal2019},
and consists in shrinking the dynamic range of the input data (this can be interpreted as a
generalization of the unit-variance zero-centred rescaling that is standard practice in many
machine learning applications).
We use the transformation
\begin{equation}
\tilde x_\text{DM} = a [\log(1 + b x_\text{DM}) - c]\,,
\label{eq:transformations}
\end{equation}
where $b$ is related to the standard deviation of the field, and $a$ and $c$ are obtained by
requiring zero mean and unit variance.
For reproducibility, we list the values of these parameters in Appendix~\ref{app:transformations}.

The tailed target distribution constitutes a challenge in two ways.
First, vanilla neural networks are not designed for the flow of high-dynamic-range data;
standard architectures and training procedures are calibrated for well behaved distributions of
the internal representation.
To address this problem, we take two steps:
(1) as mentioned before, we provide a semi-analytical model for the target,
so that the network only has to learn the difference between target and semi-analytical model\footnote{As can be
seen from the network architecture, Fig.~\ref{fig:network}, the network has the freedom
to rescale the semi-analytical model by a constant factor. In principle it could set this constant
to zero if it deems the semi-analytical model too inaccurate, but empirically we find it to be close to unity.
Thus, the word ``residual" is to be understood in a somewhat generalized sense.};
(2) before concatenating the network output with the semi-analytical model, we apply the sinh function to the
network output, as illustrated in Fig.~\ref{fig:network}.
We empirically found a sinh to work better than the exponential function,
presumably the network
prefers the possibility that some of
the output data fall into the approximately linear range of the sinh.
Both measures serve to decrease the dynamic range of the internal representation\footnote{In fact,
we examined the values of weights and biases after training and found them to be
approximately Gaussian distributed around mean zero with variance of order unity,
which confirms the effect we expect the mentioned techniques to have.}.

The second challenge stemming from the tailed target distribution is related to the choice of loss
function. It is clear that for good performance on summary statistics such as the power spectrum
we would like to use a loss function of the form $L(p,t) = \Sigma (p - t)^2$, where $p$ and $t$ are the
prediction and target respectively, and $\Sigma$ schematically indicates summation over voxels.
However, empirically it is found that starting training with this loss function leads to unstable
behaviour, presumably because large gradients are
backpropagated due to the high dynamic range of $t$.

Thus, in order to stabilize training, we adopt a time-dependent loss function,
\begin{align}
L_\tau(p,t) &= \sum [f_\tau(p) - f_\tau(t)]^2\,; \label{eq:timedeploss}\\
f_\tau(x)   &= e^{-\tau/\tau_0} \text{sgn}(x) \log(1+10 |x|) + (1 - e^{-\tau/\tau_0}) x\,,
\label{eq:targettransform}
\end{align}
where $p$ and $t$ have been divided by the standard deviation of the target field.
The parameter $\tau$ measures time in units of epochs, and we found that a good choice for $\tau_0$ is about
30-100, with lower values leading to more unstable training, while larger values prolong the training
process unnecessarily and increase the risk of overfitting.
The transformation $f_\tau$ smoothly interpolates between the more gentle $\log(1+x)$ and a
linear function, converging at the desired mean squared error (MSE)
loss as training progresses beyond $\tau_0$ epochs.
The signum function and absolute value are only required for the electron momentum density which can
have positive and negative values.
We always evaluate the validation loss with $\lim_{\tau\rightarrow\infty}L_\tau$,
i.e. the traditional MSE, since it is a good measure to quantify performance on the power spectrum
and other commonly used summary statistics.

\section{Results}
\label{sec:results}

Having explained the methods, we now proceed to present our results;
we will split this section into four parts:
In Section~\ref{subsec:Pe}, we present the results for the electron pressure (i.e. tSZ effect)
as target,
Section.~\ref{subsec:ne} and \ref{subsec:pe} are concerned with the electron density and momentum
density respectively,
and in Section~\ref{subsec:correlations} we consider cross-correlations between different fields.
We will give somewhat more detail on the electron pressure, with the results generalizing to a large
extent to the other two cases.
As usual in machine learning, the results here are evaluated on a testing set,
i.e. parts of the simulation the network has not seen before,
as described at the end of Appendix~\ref{app:trainingboxes}.
We will compare the performance of the network with that of the simple semi-analytical models we use
as a guess for the target. It should be emphasized that these semi-analytical models are not necessarily
the best we could possibly construct;
however, they should serve as useful reference points.

\newpage

\subsection{Electron pressure}
\label{subsec:Pe}

The network was trained with electron pressure as target on the mentioned zoom-in simulations.
The training set is relatively small (181 independent simulation boxes with enhanced resolution
centered around massive halos).
We observe the disadvantage of such a small training set in that the network starts overfitting
after 176 epochs, when we halt training.
Mixing samples from the TNG300 volume with the zoom-in simulations to enlarge the training set does not
seem to yield improved performance in our experiments, indicating that the network updates are driven by
the high-density regions in the extremely massive clusters.

\begin{figure}
\begin{center}
\begin{minipage}[t]{0.48\textwidth}
\vspace{0pt}
\includegraphics[width=\textwidth]{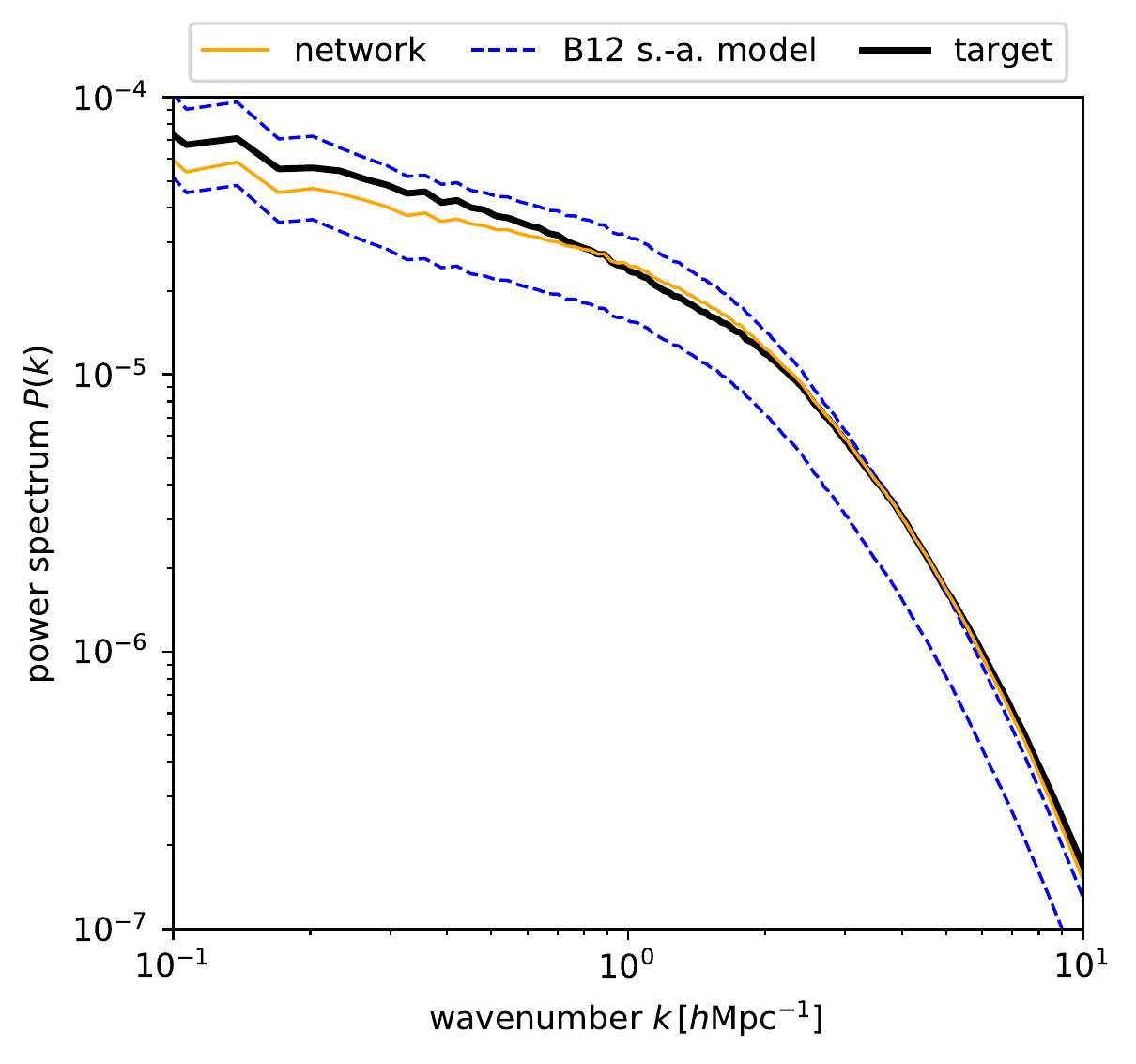}
\captionof{figure}
	{Electron pressure power spectrum.
	 We plot two versions of the B12 semi-analytical model: the lower curve is the original output,
         the upper curve is the same rescaled by a constant factor to make the comparison
	 fairer.
	 While we observe excellent agreement between network prediction
	 and target result (from the TNG300 hydrodynamical simulation),
	 there is a lack of power ($\sim 20\,\%$) on large scales.
	 This is likely related to errors in the prediction of low-pressure voxels,
	 as demonstrated in Fig.~\ref{fig:Petestprojpowerspectrum}.}
\label{fig:Petestpowerspectrum}
\includegraphics[width=\textwidth]{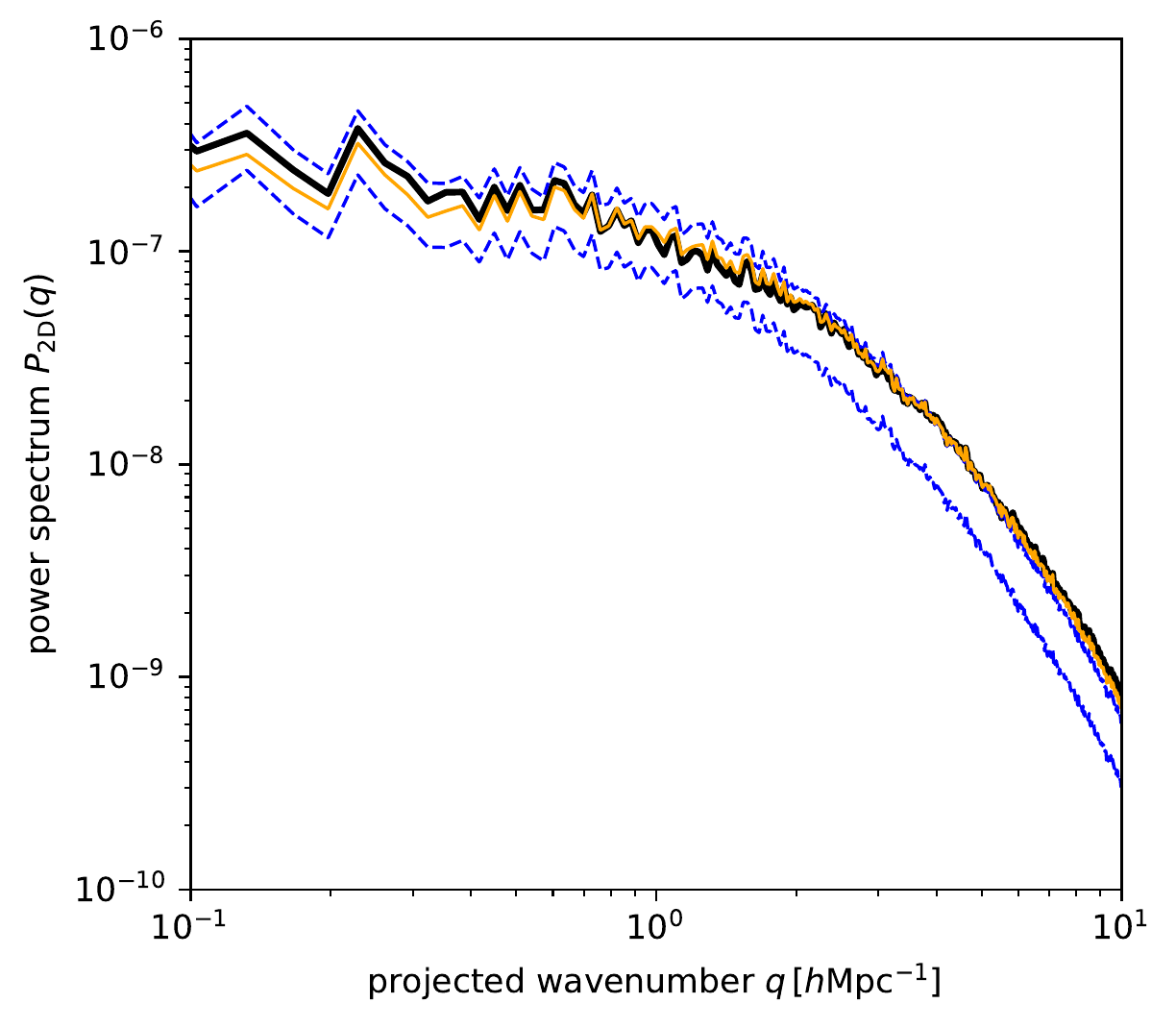}
\captionof{figure}
	{Electron pressure projected power spectrum,
         illustrating that upon projection along one axis the small electron pressure
	 values, which the network is unable to recover correctly, are of lesser importance,
	 compared to the 3-dimensional power spectrum in Fig.~\ref{fig:Petestpowerspectrum}.}
\label{fig:Petestprojpowerspectrum}
\end{minipage}\hspace{1em}%
\begin{minipage}[t]{0.48\textwidth}
\vspace{0pt}
\includegraphics[width=\textwidth]{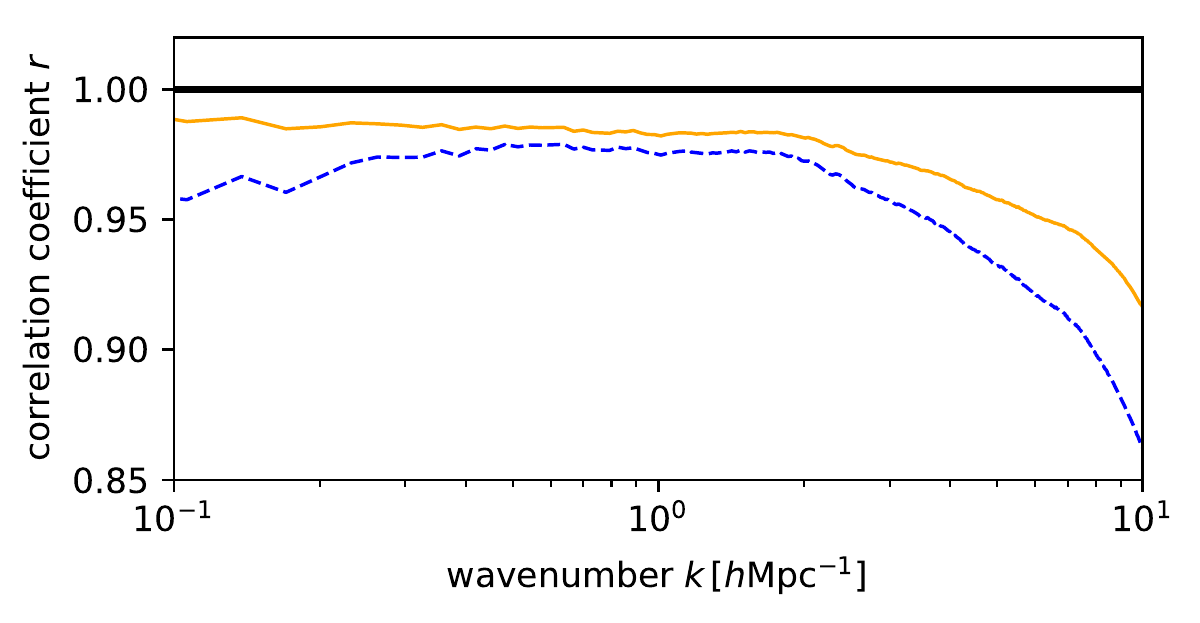}
\captionof{figure}
	{Electron pressure cross-correlation coefficient between network prediction/B12
	semi-analytical model
  	 output and the target simulation field.
	 Note that this measure is unaffected by the rescaling of the B12 semi-analytical model
	 employed in Fig.~\ref{fig:Petestpowerspectrum}.}
\label{fig:Petestcorrelation}
\includegraphics[width=\textwidth]{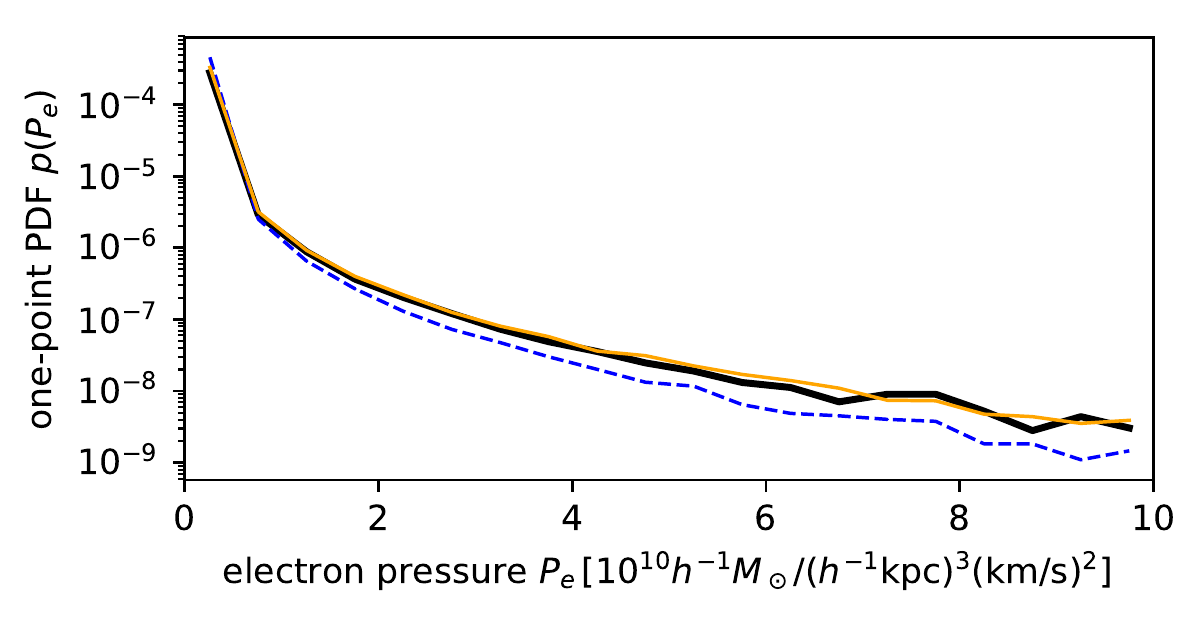}
\captionof{figure}
	{Electron pressure one-point PDF.
	 The network prediction matches the target simulation very well.
	 We only plot the original B12 semi-analytical model, not the rescaled version shown in
	 Fig.~\ref{fig:Petestpowerspectrum}.}
\label{fig:Petestonepoint}
\includegraphics[width=\textwidth]{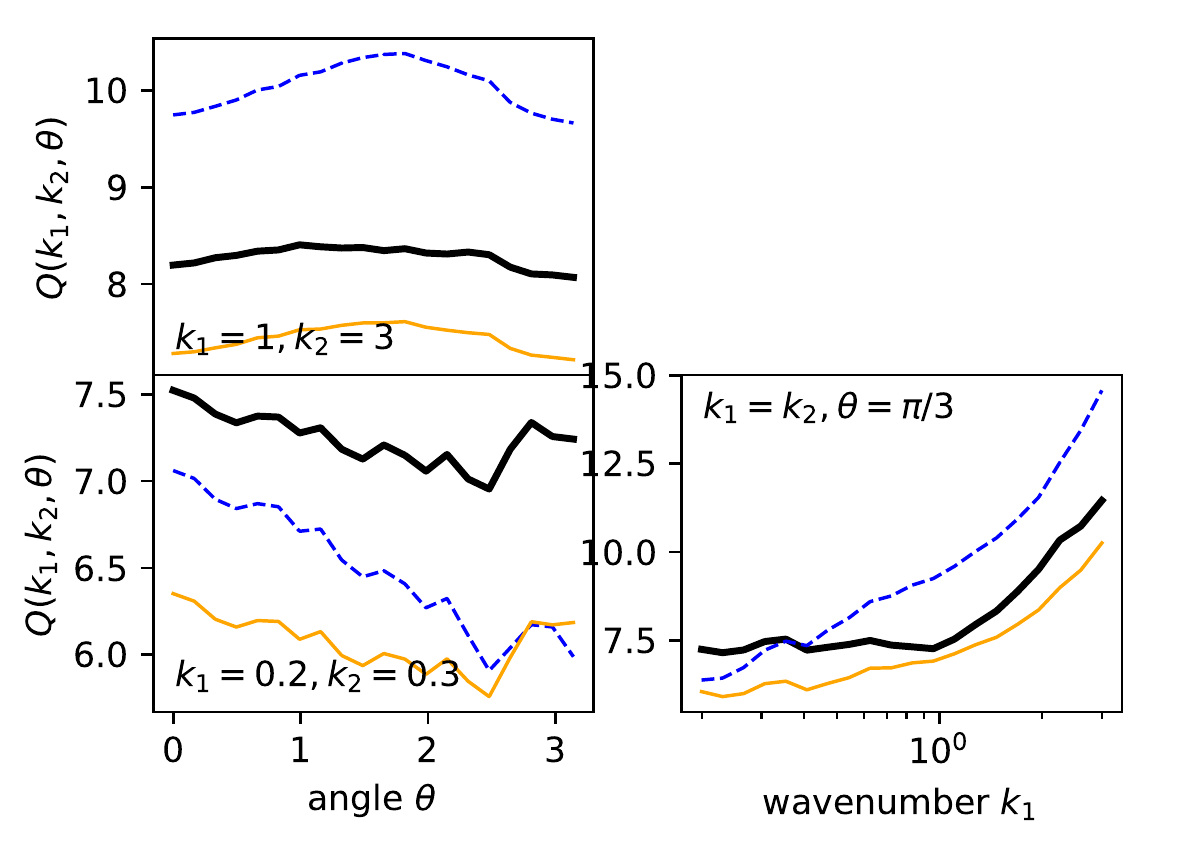}
\captionof{figure}
	{Some (reduced) bispectra for the electron pressure.
	 Wavenumbers are given in units of $h\text{Mpc}^{-1}$.
         \emph{Upper left}: small scales with varying triangle shapes,
	 \emph{lower left}: large scales with varying triangle shapes,
	 \emph{lower right}: varying scales with fixed equilateral triangle shape.}
\label{fig:Petestbispectra}
\end{minipage}
\end{center}
\end{figure}

In Fig.~\ref{fig:Petestpowerspectrum}, we plot the electron pressure power spectrum;
here, and elsewhere, computed using
\texttt{nbodykit}\footnote{\url{https://github.com/bccp/nbodykit}} \citep{nbodykit}.
We plot two versions of the B12 semi-analytical model, the lower curve being the original output
and the upper curve having the electron pressure values rescaled by a constant factor
to make for a fairer comparison.
In terms of the power spectrum, the network's prediction matches the TNG300 target
result better than the B12 semi-analytical model
on all scales and shows very good agreement with the
target on small scales ($k\gtrsim 1\,h\text{Mpc}^{-1}$).
It should be emphasized, however, that B12 was calibrated on a different simulation.

The deficit in power on larger scales ($k\lesssim 1\,h\text{Mpc}^{-1}$) is
consistent with the zero-order approximation that the network neglects to predict electron
pressures below a certain threshold (c.f. Fig.~\ref{fig:thresholds}).
We point out that if this explanation is correct, the lack of power is not particularly
worrisome since upon projection along the line of sight
(to obtain the Compton-$y$ observable Eq.~\ref{eq:Comptony})
the low pressure values will be further diminished in their contribution
to any relevant summary statistics.
In order to test this interpretation, we project the electron pressure along one cartesian axis and compute
the power spectrum of the resulting two-dimensional field;
this is plotted in Fig.~\ref{fig:Petestprojpowerspectrum}.
We observe that upon projection the agreement between the network prediction and the target
improves, consistent with our explanation for the lack of power.
It is quite conceivable that if one were to construct a complete tSZ map, i.e. project up to redshift
1100, the discrepancy would entirely disappear.

While the power spectrum measures the amplitudes of the different modes,
the cross-correlation coefficient between two fields $A$ and $B$
\begin{equation}
r_{AB}(k) \equiv \frac{P_{AB}(k)}{\sqrt{P_{AA}(k)P_{BB}(k)}}
\end{equation}
is a measure of the correlation between their phases.
We plot the correlation coefficent with respect to the target electron pressure in
Fig.~\ref{fig:Petestcorrelation}
(a perfect prediction would have unit cross-correlation with the target field).
Again, the network matches the reference simulation better than the semi-analytical model
and maintains a correlation of $98\,\%$ up to $k<2\,h\text{Mpc}^{-1}$.

While power spectrum and correlation coefficient are useful summary statistics,
non-Gaussian statistics are of great importance for non-linear fields such as the tSZ.
In Fig.~\ref{fig:Petestonepoint}, we plot the one-point PDF; again we observe very good agreement
between the network prediction and the target, while the semi-analytical model appears to lack
high-pressure voxels.
This is a strong indication that the simulation details (in particular the amount of feedback) are
sufficiently different between the IllustrisTNG simulations and the simulations the B12
semi-analytical model has
been calibrated on that a direct comparison between the two is not particularly meaningful.
Note that on linear $P_e$-scale the discrepancies at low values,
which explain large-scale differences in the power spectrum,
are not visible.

Another non-Gaussian statistic is the bispectrum, which we plot in Fig.~\ref{fig:Petestbispectra}
for several different triangle configurations;
the computation was performed using
\texttt{Pylians}\footnote{\url{https://github.com/franciscovillaescusa/Pylians}} \citep{Pylians}.
The network shows agreement with the target at the $20\,\%$ level.
It should be pointed out that we did not use the bispectrum during cross-validation;
one could imagine a training procedure where it is taken into consideration as well if more accurate
predictions are needed.
Another alternative would be to use a more general loss function
if bispectrum predictions are desired.
The large-scale bispectrum (lower left corner of Fig.~\ref{fig:Petestbispectra})
is the only summary statistics in this section for which the B12 semi-analytical model matches the simulation
better than the network's prediction.

\begin{center}
\begin{figure}
\begin{minipage}[b]{0.7\textwidth}
\includegraphics[width=\textwidth]{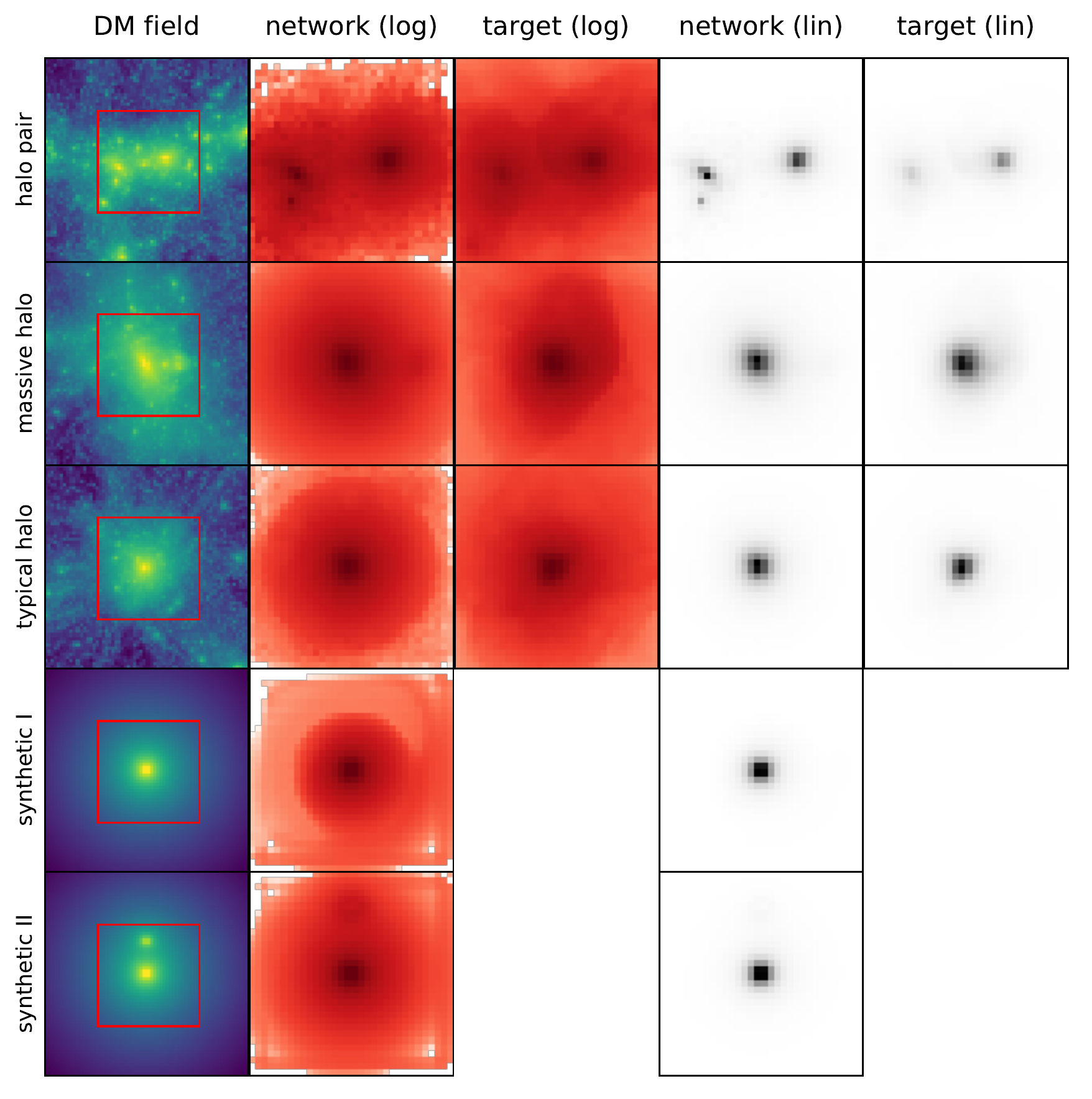}
\end{minipage}%
\begin{minipage}[b]{0.3\textwidth}
\captionof{figure}
	{Some individual halos.
	 We plot the matter density and electron pressure
	 (both network prediction and simulation target),
	 projected across $6.4$ and $3.2\,h^{-1}\text{Mpc}$ respectively.
	 For the electron pressure, both logarithmic and linear scales are provided.
	 The red boxes in the left column indicate the extent of the target boxes
	 in the dark matter boxes
	 (c.f. the discussion of padding in Appendix~\ref{app:trainingboxes}).
	 The first three rows are taken from the TNG300 box,
	 while the last two are synthetic pure NFW dark matter halos
	 for which we do not have a hydrodynamical simulation target available.
	 Note that the color scales are different in each row in order to make
	 interpretation easier.}
\label{fig:interrogation}
\end{minipage}
\end{figure}
\end{center}

In order to gain more intuition, we present images of the electron pressure in several
different halos in Fig.~\ref{fig:interrogation}.
The first three rows are taken from the TNG300 box; for the ones labeled ``massive halo"
and ``typical halo" (with total halo masses of $M_\text{200c} = 8.8\times10^{14}$ and
$2.6\times10^{14}\,h^{-1}M_\odot$, respectively); we observe very good agreement between network prediction and target.
On the other hand, the network predicts too high electron pressure values for the ``halo pair",
possibly it is unable to take the long-range interaction between the two halos into account,
the number of such situations in the training set being very small.
The last two rows are synthetic pure NFW dark matter profiles.
For these we choose a halo mass of the primary object as $M_\text{200c} = 10^{14}\,h^{-1}M_\odot$;
the last row also contains a smaller halo with $2\times 10^{13}\,h^{-1}M_\odot$.
Given these masses (and redshift $z=0$), we use the concentration-mass relation from
\cite{Duffy2008} and the density profile from \cite{NFW1997} to create the ``synthetic" input data.
We observe that the network has
learned spherical symmetry quite well (the small deviations near the edges of the box are completely
irrelevant for the loss function, as evidenced by the plots in linear color scale). 

The network performance for electron pressure is remarkable given that it was trained
on the TNG-Cluster sample, which is unusually small (for U-Net standards) and highly biased,
and which furthermore is expected to show some small-scale differences to the test set
of TNG300 halos.
We expect these small-scale differences because
we do not have gravity-only zoom-in simulations available, so that the input data
will be slightly different due to feedback and baryon-baryon interactions.
The fact that these small-scale differences seem to be of minor importance indicates that the
network does not assign much importance to features at the voxel scale;
this fact gives us confidence in generalizability.
The good performance despite strong biasing of the training set is interesting as well:
we have mentioned that most of the halos in the training set are more massive
than the most massive halos in the testing set.
We interpret the fact that our procedure still works as an indication that the mapping between dark
matter and electron gas may be dominated by sub-halo-scale features.


\subsection{Electron density}
\label{subsec:ne}

\begin{figure}
\begin{center}
\begin{minipage}[t]{0.48\textwidth}
\vspace{0pt}
\includegraphics[width=\textwidth]{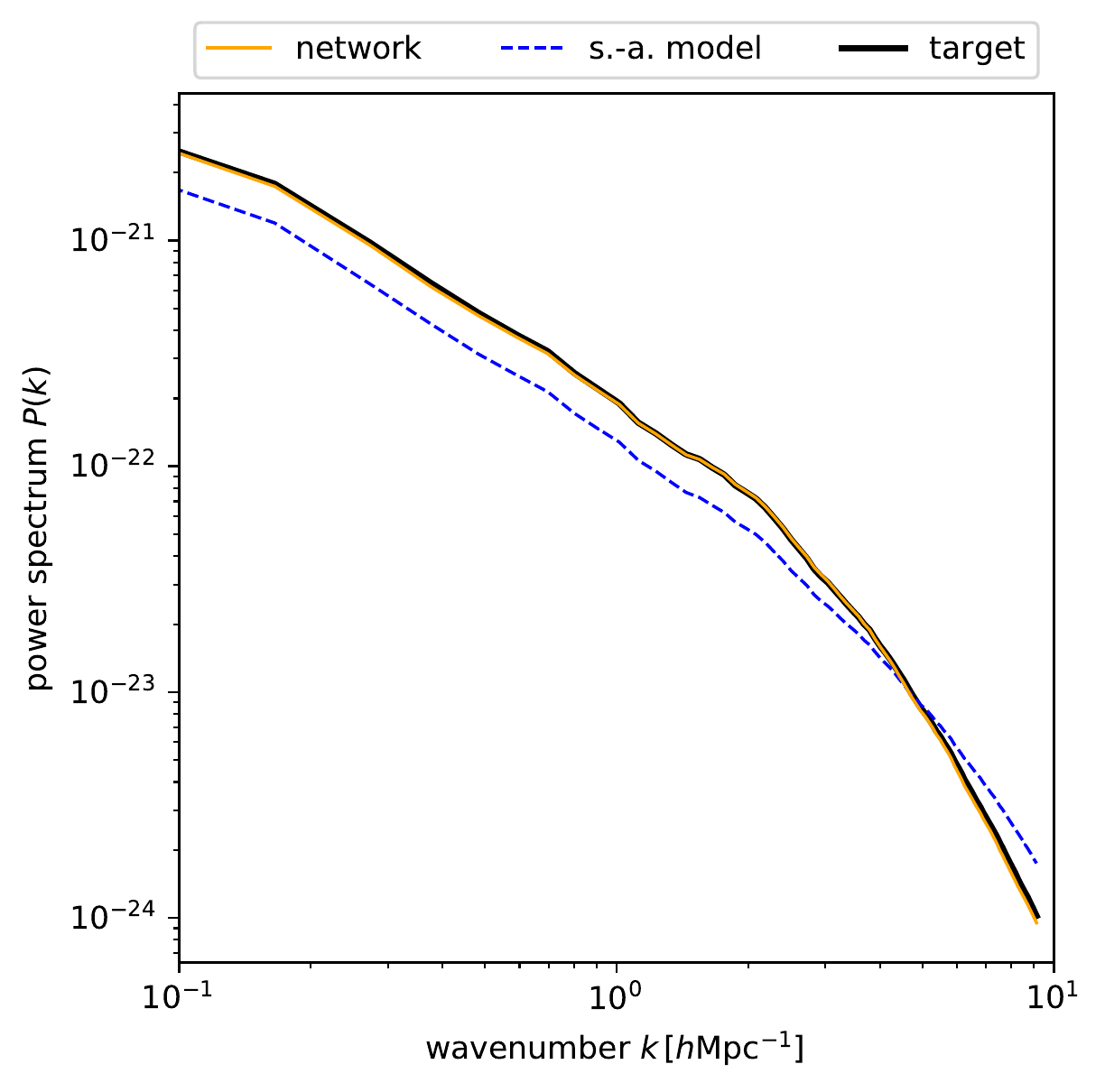}
\captionof{figure}
	{Electron density power spectrum.
	 The network achieves excellent agreement with the target simulation,
	 which is an indication that electron density is an easier target than both
	 pressure and momentum density.}
\label{fig:Netestpowerspectrum}
\end{minipage}\hspace{1em}%
\begin{minipage}[t]{0.48\textwidth}
\vspace{0pt}
\includegraphics[width=\textwidth]{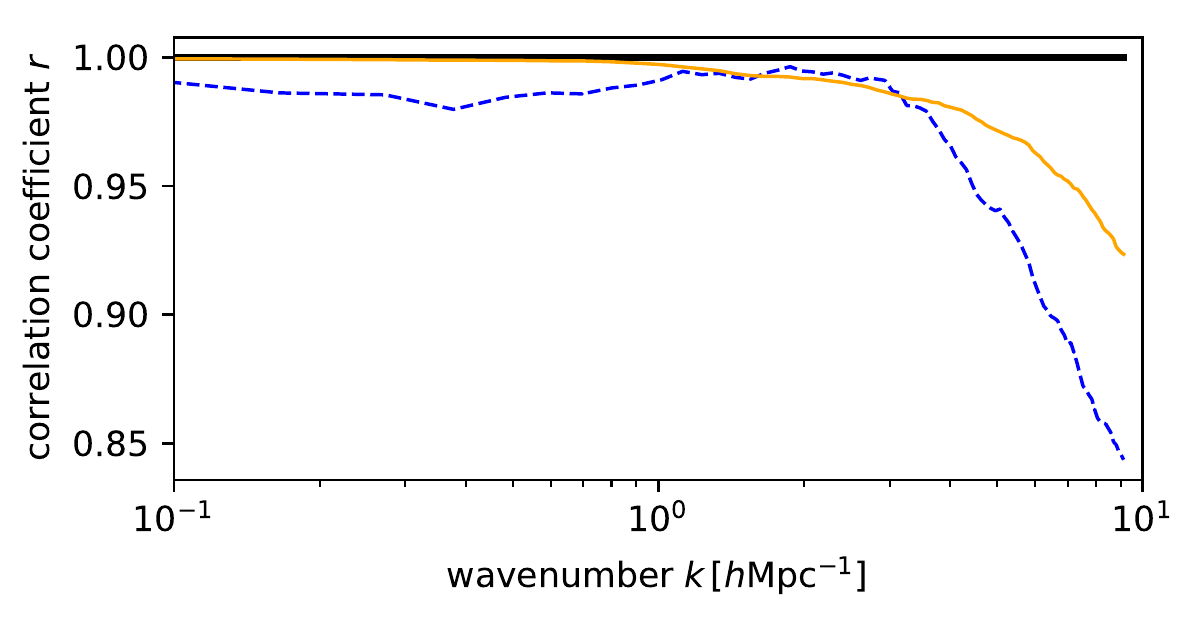}
\captionof{figure}{Electron density cross-correlation coefficient with the target simulation.}
\label{fig:Netestcorrelation}
\includegraphics[width=\textwidth]{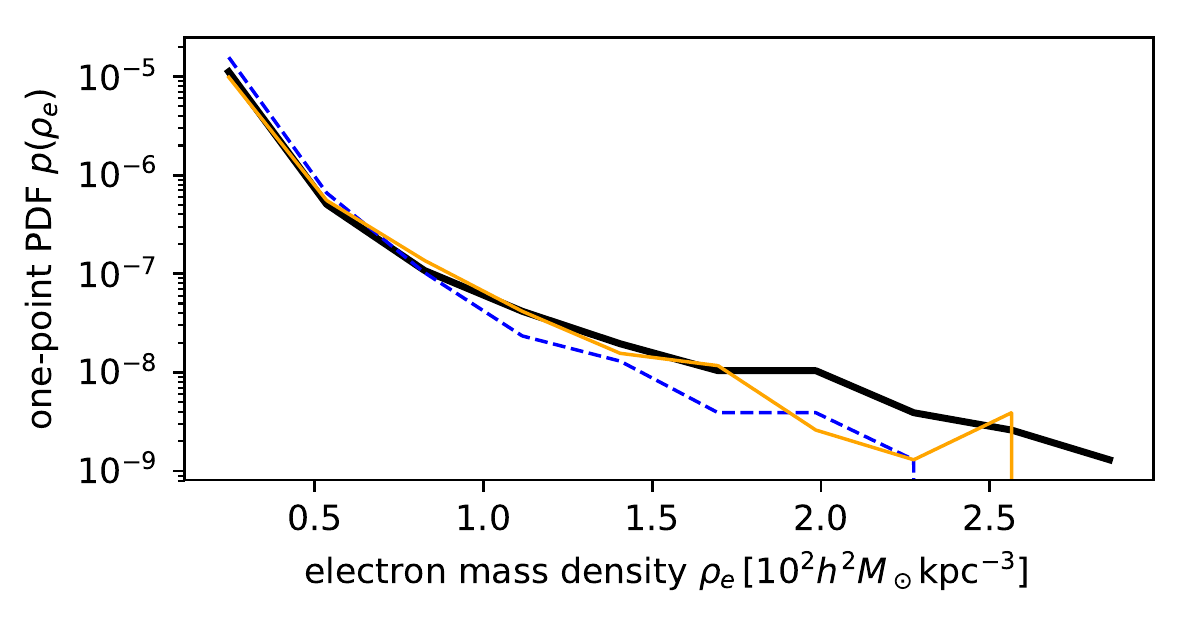}
\captionof{figure}
	{Electron density one-point PDF.
	 Note that the test box is only $10\,\%$ the size used in Fig.~\ref{fig:Petestonepoint},
	 which explains the larger scatter in the tail.}
\label{fig:Netestonepoint}
\end{minipage}
\end{center}
\end{figure}

As mentioned in Section~\ref{subsec:fewinterestingvoxels}, learning the mapping from matter to electron density does not
require such a strongly biased selection of the training samples as it did for the electron pressure.
Thus, we can use sub-volumes of the TNG300 box as training samples, giving us a
substantially larger training set. In this case, we do not observe any problems with overfitting.
Since we expect the network to utilize quite similar features in its prediction of both electron
pressure and density, we use the same network architecture and 
start the training for
electron density with the best trained version for the electron pressure.
Confirming our expectation, the network quickly (within about 20 epochs) adapts to the new target
and converges slowly afterwards. Since no overfitting occurs, we train up to 210 epochs, when the
validation loss shows no significant improvement anymore.

In Fig.~\ref{fig:Netestpowerspectrum}, we plot the electron density power spectrum, which shows
remarkable agreement between network prediction and target.
Similarly, the cross-correlation coefficient with the target simulation, plotted in Fig.~\ref{fig:Netestcorrelation}, exceeds
$98\,\%$ up to $k < 4\,h\text{Mpc}^{-1}$.
Finally, the one-point PDF, plotted in Fig.~\ref{fig:Netestonepoint}, also shows good agreement. The
relatively noisy behaviour in the high-density tail (compared to Fig.~\ref{fig:Petestonepoint}) occurs because
the volume of the testing box is only $10\,\%$ of the volume in the electron pressure testing set.

In summary, the electron density is an easier target compared to the electron pressure.
Due to the better behaved distribution of target values, we are able to use
a much larger training set\footnote{While we have 181 primary objects in the TNG-Cluster zoom-in simulations
available, there are 1416 halos in the TNG300 with masses larger than
the cut-offs in Eqs.~\eqref{eq:rhoesampling},~\eqref{eq:pesampling}.},
preventing any problems with overfitting and leading to very good performance on the testing
set. The fact that, apart from the change in training set, no modification to the network architecture
and training procedure was necessary, indicates that the techniques to confront the sparsity problem,
explained in Section~\ref{sec:sparsity}, generalize well to different datasets\footnote{This is not obvious,
since many hyperparameters were tuned during extensive trials on the electron pressure, while the
only tuning necessary for the electron density was the biasing of the sample selection function,
Eq.~\ref{eq:rhoesampling}.}.

\subsection{Electron momentum density}
\label{subsec:pe}

As in the case of the electron density, we find that we do not require the additional zoom-in
simulations to learn the mapping to the momentum density;
thus, the training set is identical to the one used in the previous section
(the sample selection function is slightly different though, c.f.
Section~\ref{subsec:fewinterestingvoxels}).

The electron momentum density is somewhat special in that $\mathbf{p}_e$ is a
three-component vector (while $P_e$ and $\rho_e$ are positive scalars).
As mentioned in Section~\ref{subsec:dataprep}, one input channel is the dark matter density,
and we include information on the dark matter momentum density in the input as well.
Then we are left with three possible input-output configurations:
\begin{enumerate}
	\item three directions as input, three directions as output; \label{io:tt}
	\item three directions as input, one direction as output; \label{io:to}
	\item one direction as input, one direction as output. \label{io:oo}
\end{enumerate}
The shape of the semi-analytical model would be chosen equal to the desired output shape.
Empirically, we find options \ref{io:tt} and \ref{io:to} not to work well with network architectures analogous
to the one described in Section~\ref{subsec:nn}.
Presumably, the network finds it difficult to learn that it has to establish three separate paths,
with small coupling between them.
This problem could possibly be solved by modifying the network architecture more fundamentally,
however, in the interest of consistency, we choose not to do this and opt for the input-output
configuration \ref{io:oo}.
This necessitates only minor modifications to the network architecture, as described in the caption
to Fig.~\ref{fig:network}.
It should be pointed out that this choice has some drawbacks:
First, it is probably inefficient, since many features can be assumed to be shared between the three
different spatial directions.
Second, we could imagine situations in which the dark matter velocity in a particular voxel satisfies
inequalities like $v_\text{DM}^1 \ll v_\text{DM}^2$
(with the superscripts representing different cartesian coordinate directions)
and we would like to predict $p_e^2$.
In such a situation the network would be unable to incorporate small rotations
relating the dark matter and electron velocity directions
(which would make $v_e^2$ drastically different from $v_\text{DM}^2$),
leading to inaccurate predictions of $p_e^2$.

Besides the described modifications to the network architecture, we use the tuned sample selection
function of Eq.~\eqref{eq:pesampling}, and find that increasing the noise to $20\,\%$
(c.f. Section~\ref{subsec:training}) yields better performance.
We observe slight overfitting and halt training after 188 epochs.

\begin{figure}
\begin{center}
\begin{minipage}[t]{0.48\textwidth}
\includegraphics[width=\textwidth]{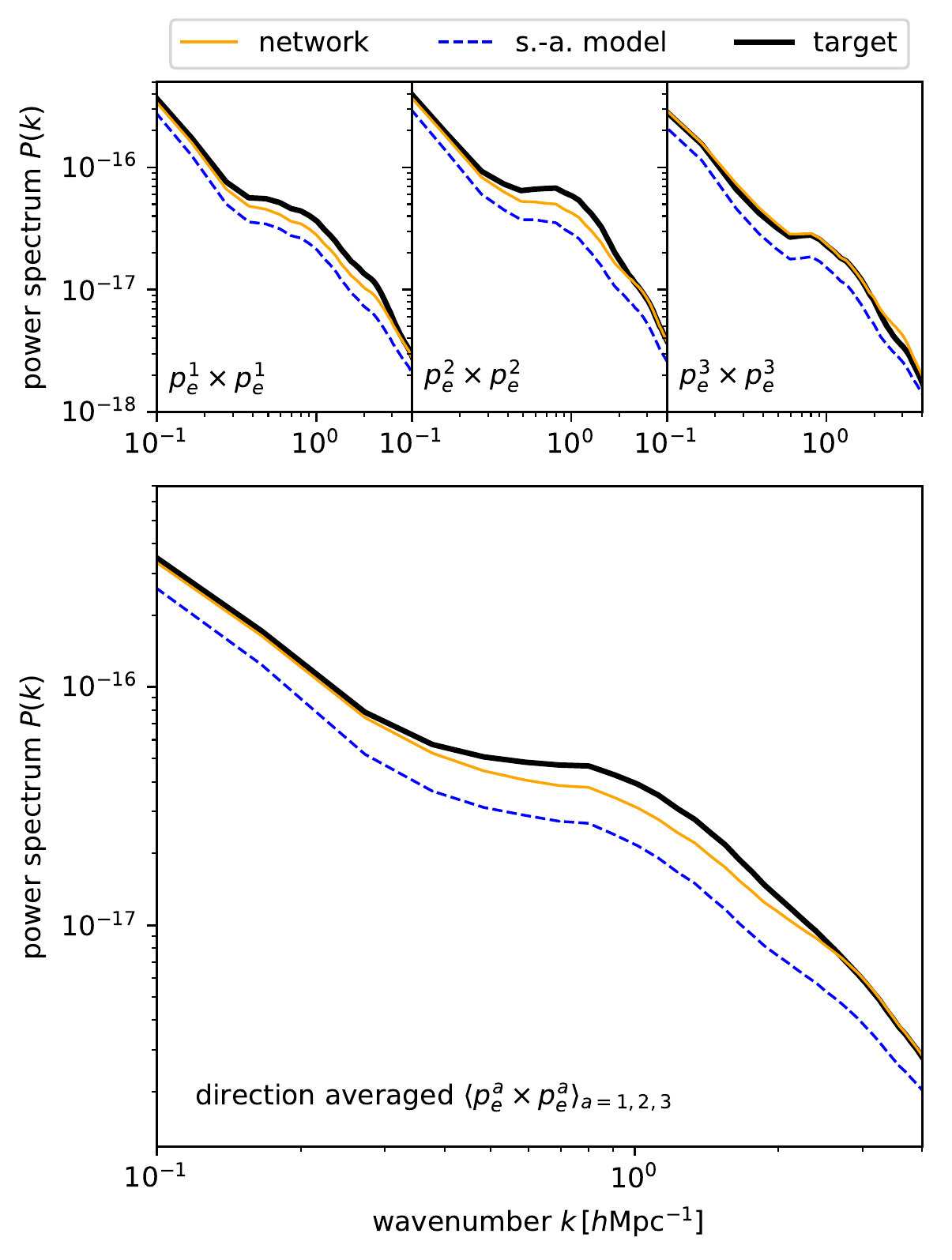}
\captionof{figure}
	{Electron momentum density power spectrum.
	 While the upper panels show the power spectra of the momentum components
	 in three cartesian directions, the lower panel is the average.
	 Although the network performs much better than the semi-analytical model,
	 some discrepancies ($\sim 20\,\%$) are clearly visible.
	 They are likely related to a combination of overfitting and a sub-optimal
	 network architecture.}
\label{fig:MOMtestpowerspectrum}
\includegraphics[width=\textwidth]{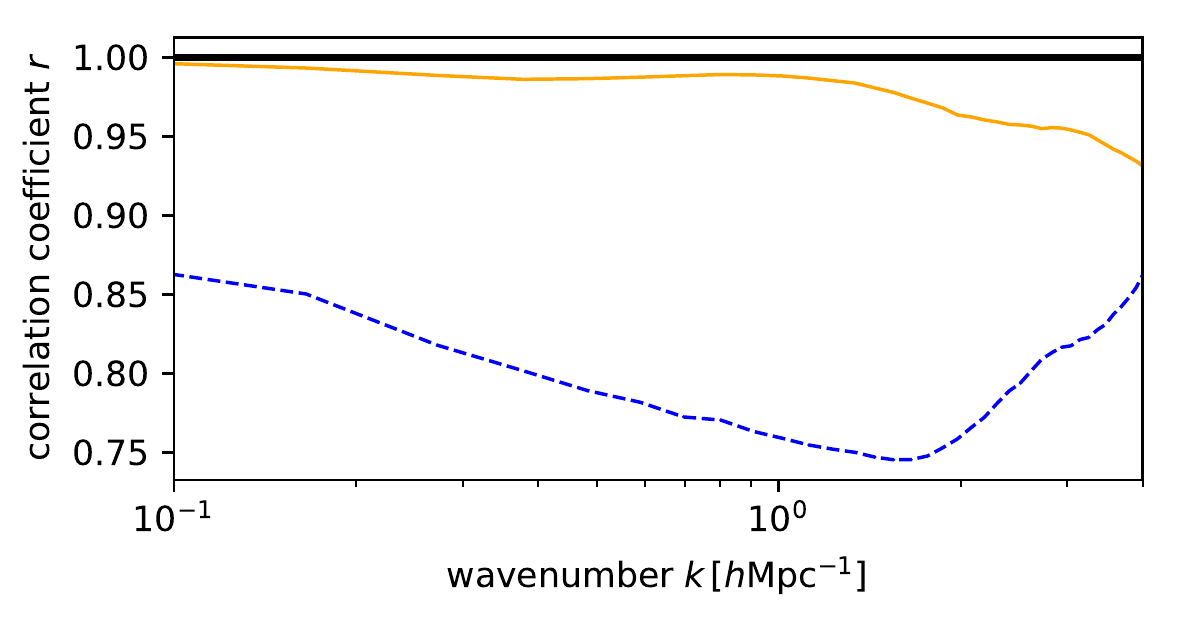}
\captionof{figure}
	{Electron momentum density cross-correlation coefficient
	 with the target simulation,
	 averaged over the three cartesian directions.
	 The scatter between different directions is small.}
\label{fig:MOMtestcorrelation}
\end{minipage}\hspace{1em}%
\begin{minipage}[t]{0.48\textwidth}
\includegraphics[width=\textwidth]{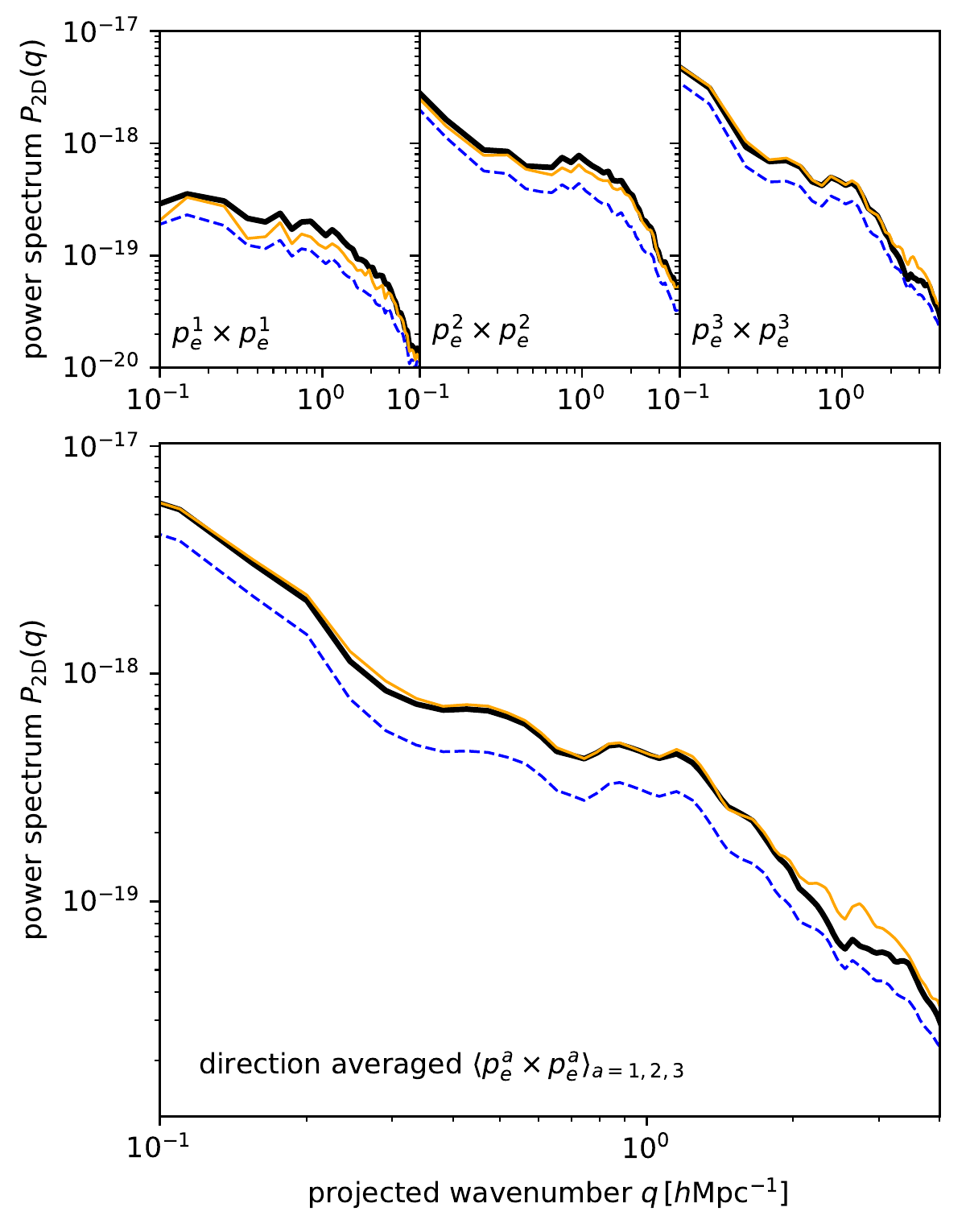}
\captionof{figure}
	{Electron momentum density projected power spectrum.
	 The format is the same as in Fig.~\ref{fig:MOMtestpowerspectrum}.}
\label{fig:MOMtestprojpowerspectrum}
\includegraphics[width=\textwidth]{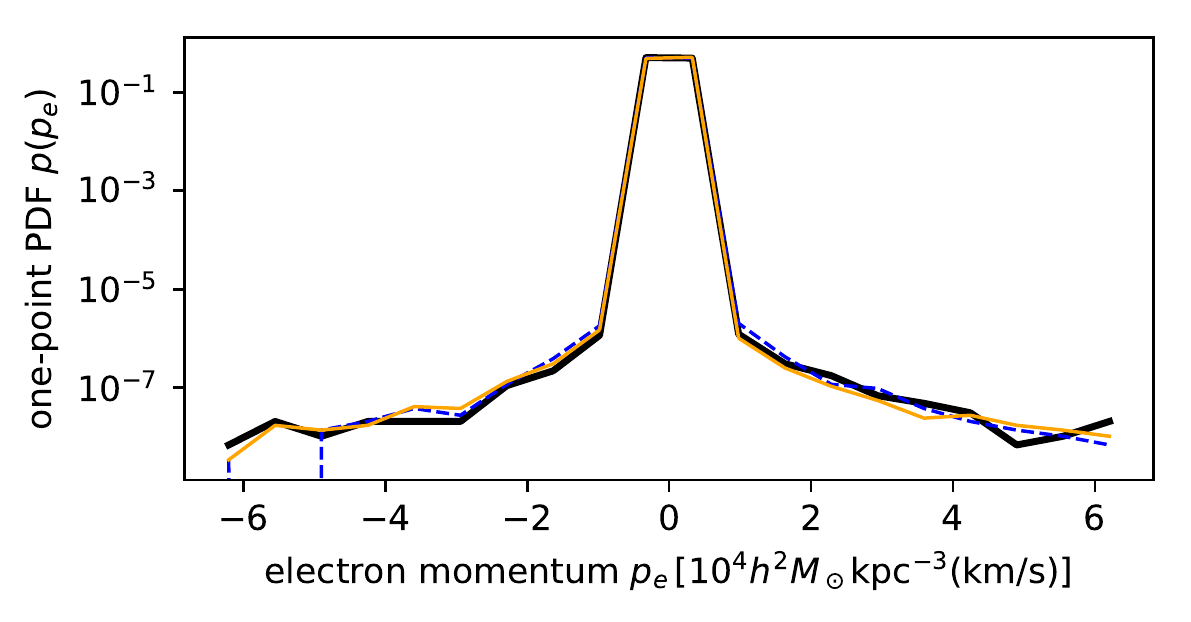}
\captionof{figure}
	{Electron momentum density one-point PDF.
	 Again, all three cartesian directions are included.}
\label{fig:MOMtestonepoint}
\end{minipage}
\end{center}
\end{figure}

In Fig.~\ref{fig:MOMtestpowerspectrum}, we plot electron momentum density power spectra.
The three plots in the top row represent the three directions, while the larger plot in the bottom
row is simply the average of these.
Of course, taking this average is not a particularly meaningful operation,
given that the momentum components are correlated, but it should serve as a less noisy
representation of the network performance. 
We see that although the network outperforms the simple semi-analytical model,
it shows a power deficit of up to $20\,\%$ in the intermediate $k$-range.
Interestingly, we actually observed a slight power excess when evaluating the network on the
validation set, indicating that we overfit on the validation set.
As a useful proxy for the actual kSZ-observable, we plot the projected power spectra
in Fig.~\ref{fig:MOMtestprojpowerspectrum} (projections are along the individual momentum
directions, which is the physically relevant case).
We observe a large variability in the power spectra for the three different directions (top row),
making the averaged version less meaningful.
It can be seen that the power deficit observed in the 3D power spectra is still present in the
projected versions, although upon averaging it is less prominent.

The correlation coefficient shown in Fig.~\ref{fig:MOMtestcorrelation} is an average over the three
directions, we did not observe significant variability between the directions.
We recover correlations exceeding $96\,\%$ up to $k<2\,h\text{Mpc}^{-1}$,
indicating that the predicted momentum field has somewhat lower fidelity than in the cases of
electron pressure and density.
However, we are comparing correlations at an excellent level here.

It should be noted that the semi-analytical model performs significantly worse
than in the other two cases (compare Figs.~\ref{fig:Petestcorrelation}, \ref{fig:Netestcorrelation}).
Since the network builds on the semi-analytical  model, worse performance should be expected.
From the one-point PDF in Fig.~\ref{fig:MOMtestonepoint} we see that network and semi-analytical model perform
about equally well on this summary statistics.
One conclusion to be drawn from this is that the deficit in power exhibited by the semi-analytical model cannot be
solved by simply rescaling it.

\newpage

\subsection{Cross-correlations between different fields}
\label{subsec:correlations}

As the last part of the section on results, we present cross-correlations between different fields
evaluated on the testing set.
The purpose of this is twofold.
First, correlating different cosmological observables can be less susceptible to systematic effects,
which makes them useful summary statistics
\citep[e.g., for weak lensing cross tSZ,][]{HillSpergel2014, vanWaerbekeHinshawMurray2014,
Hojjatietal2017}.
Second, these cross-correlation functions were not considered during training,
making them a ``double blind" way of evaluating the network performance
(unseen data and new types of summary statistics)
and therefore they represent a useful diagnostic.

\begin{figure}
\begin{center}
\includegraphics[width=\textwidth]{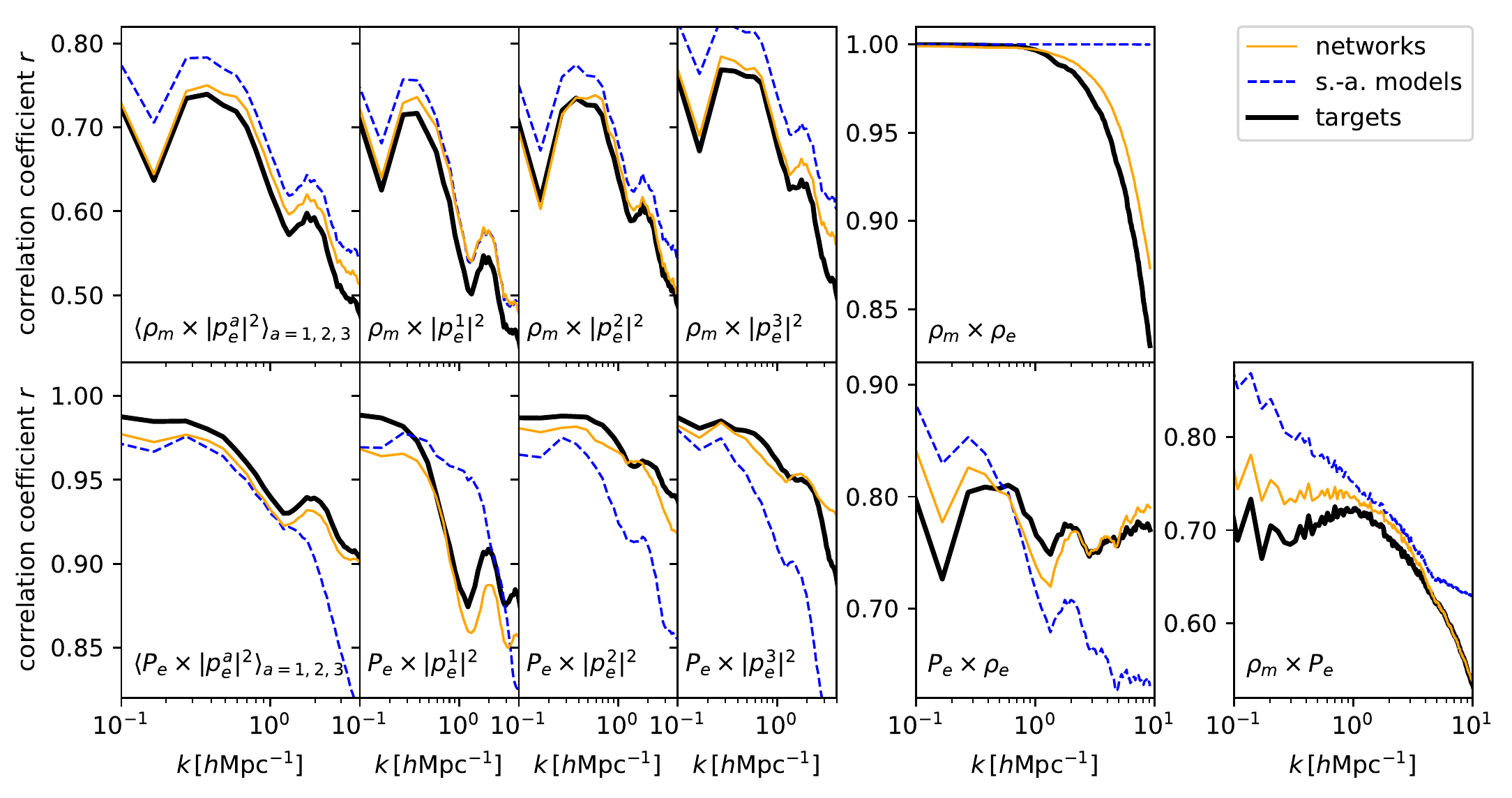}
\caption{Cross-correlation coefficients between different fields.
         We observe relatively good agreement between network prediction and target simulation
	 for a range of observables and scales,
	 but discrepancies for others.
	 The networks consistently outperform the semi-analytical models,
	 but is is likely that the network prediction quality depends significantly on the
	 semi-analytical model's
	 quality.}
\label{fig:testcorrelations}
\end{center}
\end{figure}

Fig.~\ref{fig:testcorrelations} shows the 3-dimensional cross-correlations.
Note that the $\rho_m$-field is the same for all correlations including it,
since it comes from the gravity-only simulation.
\begin{itemize}
\item $\rho_m\times |\mathbf{p}_e|^2$ (element-wise squared):
The top left panel in Fig.~\ref{fig:testcorrelations}
is direction averaged (with the caveats discussed in Section~\ref{subsec:pe}),
while the narrower panels to its right are for the individual momentum directions.
This type of correlation function can be seen as a proxy for
$\text{weak~lensing}\times\text{kSZ}^2$.
The network vastly outperforms the semi-analytical model on larger scales ($k\lesssim 0.5\,h\text{Mpc}^{-1}$),
while on smaller scales its performance is about a factor two better.
\item $P_e\times |\mathbf{p}_e|^2$ (element-wise squared):
The panels are arranged analogously to the previous case.
This correlation function emulates $\text{tSZ}\times\text{kSZ}^2$ correlations.
We observe that the semi-analytical models struggle to generate remotely correct predictions for this
correlation function (although the average, perhaps by chance, happens to look reasonable for
$k\lesssim 1\,h\text{Mpc}^{-1}$).
The networks have problems with this correlation as well, presumably because it combines two predicted
fields (while $\rho_m$ in the previous case was correct by design).
\item $\rho_m\times\rho_e$:
This correlation is mostly a diagnostic and not related to any real-world observable.
As expected from the design of the semi-analytical model, Eq.~\eqref{eq:rhoemodel}, the
semi-analytical model's
prediction for $\rho_e$ is perfectly correlated with the matter field.
The network's prediction is closer to the target correlation function.
\item $P_e\times\rho_e$:
Again, this plot is a diagnostic.
As observed before, correlating two generated fields proves to be a challenge for the
semi-analytical models.
The networks' prediction is about a factor two better than the semi-analytical models' for $k\lesssim
1.5\,h\text{Mpc}^{-1}$ and much better on smaller scales.
We observe the trend already remarked in Section~\ref{subsec:Pe} that the network prediction for
$P_e$ is less accurate on large scales.
However, unlike the $P_e\times P_e$ auto-power, projecting the field does not seem to
substantially improve the agreement between network prediction and simulation target.
\item $\rho_m\times P_e$:
This correlation is a proxy for $\text{weak~lensing}\times\text{tSZ}$ statistics.
Again, the network struggles on large scales with a correlation function involving $P_e$,
while the small-scale predictions are excellent.
\end{itemize}
We said at the beginning of this section that considering the cross-correlations has two purposes.
With respect to real-world data analyses, we observe high-fidelity predictions
for a subset of the correlations and not necessarily in the whole range of wavenumbers considered.
However, we point to the fact that, as observed before, projecting the fields can help to bring the
networks' predictions closer to the truth.

With respect to the diagnostic purpose of these plots, we first note that the discrepancies observed
here are roughly in line with those seen in different Gaussian summary statistics before,
indicating that the we did not strongly overfit on the type of summary statistics.
A final comment is that we find the discrepancies between network predictions and truth to be mostly
in the direction of the discrepancies between semi-analytical models and truth.
This is an indication that the network performance depends substantially on the semi-analytical
model's quality.

\section{Conclusions}
\label{sec:conclusions}

As part of an effort to quickly generate maps of non-linear observables,
we have developed machine learning techniques that allow us to map the gravity-only matter field
to three electron gas properties, namely pressure, density, and momentum density.
This is the first step towards the generation of Sunyaev-Zel'dovich maps solely
from gravity-only simulations, or even from high-redshift initial conditions.

In this work, we have focused on the three-dimensional fields at redshift $z = 0$.
We have trained a deep convolutional neural network with a U-Net architecture
using the output of cosmological gravity-only and full-physics hydrodynamical simulations
from the IllustrisTNG project, specifically the TNG300 box and TNG-Cluster, a set of zoom-in
cluster simulations.
We have shown that the network outputs match the reference target simulation better than
the predictions from simple semi-analytical models, according to a range of summary statistics.
To achieve this, we have encountered the problem of sparse data sets,
and explained a number of techniques that enabled us to solve it.

While we cannot, at present, exactly quantify the speed-up gained by our technique,
it is clear that it will enable us to generate much larger sky areas
than would be possible with computationally expensive hydrodynamical simulations.
As point of reference, running the reference simulation used in this work, TNG300,
took 34.9M CPU-hours,
while evaluating the neural network on the same volume took 16 GPU-hours.
In order to construct a light-cone, we would probably require evaluation on a few dozen
redshift slices.

Some interesting findings made in the course of this work, and explained in more detail
in previous sections, are:
(1) the networks do not assign much weight to voxel-scale features;
(2) the mapping between dark matter and electron pressure is dominated by sub-halo-scale features;
(3) electron density is a much easier target than pressure,
    indicating that the degree of sparsity influences the network prediction quality.

Sub-optimal network performance in some cases (kSZ, some cross-correlations)
can be explained by three factors:
(1) scarcity of training data (leading to overfitting in some cases);
(2) sub-optimal network architecture for the kSZ effect;
(3) sub-optimal quality of the semi-analytical models.
We believe that these problems are technical in nature and do not present
a fundamental obstacle.
For example, one could re-calibrate the B12 semi-analytical model for different simulations.

This work is only the first step in a longer program.
While we only work at redshift $z=0$, this is the prerequisite for the prediction of light cones.
A further obvious and easy generalization would be to the relativistic tSZ effect
\citep{Nozawaetal2006, RemazeillesChluba2020, Leeetal2020}.
Another extension would be to X-ray maps,
for which excellent observational data will be available from ROSAT and eROSITA.
Due to the signal's dependence on the square of the electron density the sparsity problem
is likely quite severe;
the data augmentation with zoom-in simulations
which we used for the tSZ effect will be essential in that case.

In terms of future work, a clear goal is the generalization to redshifts $z \neq 0$,
which will ultimately enable us to generate maps of actual observables.
There is a good chance that the dependence on redshift is a relatively simple function;
this would imply that we would effectively enlarge the training data set
and mitigate the overfitting problem.
While we have worked with a single simulation (TNG300) so far,
there is a need to compare networks trained on different simulations
(one could also imagine mixing samples with different sub-grid physics,
although the outcome of such a training strategy should be validated thoroughly).
A possible route could be to use
the CAMEL simulations \citep{CAMEL}, which contain state-of-the-art hydrodynamical
simulations spanning thousands of different cosmologies and astrophysics models run with two
different hydrodynamical solvers and subgrid implementations.

As mentioned in the Introduction, our approach does not capture stochasticity
in the mapping from gravity-only matter to electron gas.
The probabilistic nature stems from unresolved features
that are coupled to the scales of interest, for example,
from the stochasticity in the efficiency of AGN feedback.
Possible extensions to this work which would take this effect into account
could involve GANs \citep{Goodfellowetal2014} or Bayesian neural networks
\citep{TishbyLevinSolla1989}.
We note that with GANs it would be challenging to generate large enough volumes
such that the two-halo term is captured correctly.
A further difficulty in both approaches would be to create consistent predictions
of several different observables,
as is required for the cross-correlation functions.
However, we believe that stochasticity does not present a fundamental obstacle to our approach,
since it could accurately be modeled as additional white noise in any parameter inference.

Our work opens up the avenue towards several possible applications.
Given the ability to quickly generate maps of the Sunyaev-Zel'dovich effects
and other observables,
we could create covariance matrices and perform parameter inference from summary statistics.
In order to utilize the non-Gaussian information in maps,
likelihood-free inference \citep{Alsing2018} could be a promising approach;
for example applied to the tSZ one-point PDF
\citep{Hilletal2014, ThieleHillSmith2019}.
Furthermore, we could imagine interpreting the network output to learn more
about the effect of sub-grid astrophysical models.

\acknowledgments

We would like to thank J.C. Hill and E. Komatsu for useful suggestions.
The training of the neural networks has been carried out on the Tiger supercomputer of Princeton University.
We thank their system administrators for their help 
at several stages of this work.  FVN acknowledges funding from the WFIRST program through NNG26PJ30C and NNN12AA01C.
The Flatiron Institute is supported by the Simons Foundation.
The IllustrisTNG simulations including TNG300 were undertaken with compute time awarded by the
Gauss Centre for Supercomputing (GCS) under GCS Large-Scale Projects GCS-ILLU and GCS-DWAR on the
Hazel Hen supercomputer at the High Performance Computing Center Stuttgart (HLRS). The TNG-Cluster
simulation has been executed on Draco and Cobra machines of the Max Planck Computing and Data
Facility (MPCDF) in Garching, Germany.


\begin{thebibliography}{}
\expandafter\ifx\csname natexlab\endcsname\relax\def\natexlab#1{#1}\fi
\providecommand{\url}[1]{\href{#1}{#1}}
\providecommand{\dodoi}[1]{doi:~\href{http://doi.org/#1}{\nolinkurl{#1}}}
\providecommand{\doeprint}[1]{\href{http://ascl.net/#1}{\nolinkurl{http://ascl.net/#1}}}
\providecommand{\doarXiv}[1]{\href{https://arxiv.org/abs/#1}{\nolinkurl{https://arxiv.org/abs/#1}}}

\bibitem[{{Afshordi} {et~al.}(2007){Afshordi}, {Lin}, {Nagai}, \& {Sand
  erson}}]{Afshordietal2007}
{Afshordi}, N., {Lin}, Y.-T., {Nagai}, D., \& {Sand erson}, A. J.~R. 2007,
  \mnras, 378, 293, \dodoi{10.1111/j.1365-2966.2007.11776.x}

\bibitem[{{Agarwal} {et~al.}(2018){Agarwal}, {Dav{\'e}}, \&
  {Bassett}}]{Agarwaletal2018}
{Agarwal}, S., {Dav{\'e}}, R., \& {Bassett}, B.~A. 2018, \mnras, 478, 3410,
  \dodoi{10.1093/mnras/sty1169}

\bibitem[{{Allison} {et~al.}(2011){Allison}, {Taylor}, {Jones}, {Rawlings}, \&
  {Kay}}]{Allisonetal2011}
{Allison}, J.~R., {Taylor}, A.~C., {Jones}, M.~E., {Rawlings}, S., \& {Kay},
  S.~T. 2011, \mnras, 410, 341, \dodoi{10.1111/j.1365-2966.2010.17447.x}

\bibitem[{{Alsing} {et~al.}(2018){Alsing}, {Wandelt}, \& {Feeney}}]{Alsing2018}
{Alsing}, J., {Wandelt}, B., \& {Feeney}, S. 2018, \mnras, 477, 2874,
  \dodoi{10.1093/mnras/sty819}

\bibitem[{{Alvarez}(2016)}]{Alvarez2016}
{Alvarez}, M.~A. 2016, \apj, 824, 118, \dodoi{10.3847/0004-637X/824/2/118}

\bibitem[{{Arnaud} {et~al.}(2010){Arnaud}, {Pratt}, {Piffaretti},
  {B{\"o}hringer}, {Croston}, \& {Pointecouteau}}]{Arnaudetal2010}
{Arnaud}, M., {Pratt}, G.~W., {Piffaretti}, R., {et~al.} 2010, \aap, 517, A92,
  \dodoi{10.1051/0004-6361/200913416}

\bibitem[{{Atrio-Barandela} {et~al.}(2008){Atrio-Barandela}, {Kashlinsky},
  {Kocevski}, \& {Ebeling}}]{AtrioBarandelaetal2008}
{Atrio-Barandela}, F., {Kashlinsky}, A., {Kocevski}, D., \& {Ebeling}, H. 2008,
  \apjl, 675, L57, \dodoi{10.1086/533437}

\bibitem[{{Battaglia}(2016)}]{Battaglia2016}
{Battaglia}, N. 2016, \jcap, 2016, 058, \dodoi{10.1088/1475-7516/2016/08/058}

\bibitem[{{Battaglia} {et~al.}(2012){Battaglia}, {Bond}, {Pfrommer}, \&
  {Sievers}}]{Battagliaetal2012}
{Battaglia}, N., {Bond}, J.~R., {Pfrommer}, C., \& {Sievers}, J.~L. 2012, \apj,
  758, 75, \dodoi{10.1088/0004-637X/758/2/75}

\bibitem[{{Battaglia} {et~al.}(2010){Battaglia}, {Bond}, {Pfrommer}, {Sievers},
  \& {Sijacki}}]{Battagliaetal2010}
{Battaglia}, N., {Bond}, J.~R., {Pfrommer}, C., {Sievers}, J.~L., \& {Sijacki},
  D. 2010, \apj, 725, 91, \dodoi{10.1088/0004-637X/725/1/91}

\bibitem[{{Bode} {et~al.}(2009){Bode}, {Ostriker}, \&
  {Vikhlinin}}]{Bodeetal2009}
{Bode}, P., {Ostriker}, J.~P., \& {Vikhlinin}, A. 2009, \apj, 700, 989,
  \dodoi{10.1088/0004-637X/700/2/989}

\bibitem[{{Bode} {et~al.}(2007){Bode}, {Ostriker}, {Weller}, \&
  {Shaw}}]{Bodeetal2007}
{Bode}, P., {Ostriker}, J.~P., {Weller}, J., \& {Shaw}, L. 2007, \apj, 663,
  139, \dodoi{10.1086/518432}

\bibitem[{{Capelo} {et~al.}(2012){Capelo}, {Coppi}, \&
  {Natarajan}}]{CapeloCoppiNatarajan2012}
{Capelo}, P.~R., {Coppi}, P.~S., \& {Natarajan}, P. 2012, \mnras, 422, 686,
  \dodoi{10.1111/j.1365-2966.2012.20648.x}

\bibitem[{{Chaudhuri} \& {Majumdar}(2011)}]{ChaudhuriMajumdar2011}
{Chaudhuri}, A., \& {Majumdar}, S. 2011, \apjl, 728, L41,
  \dodoi{10.1088/2041-8205/728/2/L41}

\bibitem[{{da Silva} {et~al.}(2000){da Silva}, {Barbosa}, {Liddle}, \&
  {Thomas}}]{daSilvaetal2000}
{da Silva}, A.~C., {Barbosa}, D., {Liddle}, A.~R., \& {Thomas}, P.~A. 2000,
  \mnras, 317, 37, \dodoi{10.1046/j.1365-8711.2000.03553.x}

\bibitem[{{da Silva} {et~al.}(2001){da Silva}, {Barbosa}, {Liddle}, \&
  {Thomas}}]{daSilvaetal2001}
---. 2001, \mnras, 326, 155, \dodoi{10.1046/j.1365-8711.2001.04580.x}

\bibitem[{{Dolag} {et~al.}(2016){Dolag}, {Komatsu}, \&
  {Sunyaev}}]{DolagKomatsuSunyaev2016}
{Dolag}, K., {Komatsu}, E., \& {Sunyaev}, R. 2016, \mnras, 463, 1797,
  \dodoi{10.1093/mnras/stw2035}

\bibitem[{{Duffy} {et~al.}(2008){Duffy}, {Schaye}, {Kay}, \& {Dalla
  Vecchia}}]{Duffy2008}
{Duffy}, A.~R., {Schaye}, J., {Kay}, S.~T., \& {Dalla Vecchia}, C. 2008,
  \mnras, 390, L64, \dodoi{10.1111/j.1745-3933.2008.00537.x}

\bibitem[{{Efstathiou} \& {Migliaccio}(2012)}]{EfstathiouMigliaccio2012}
{Efstathiou}, G., \& {Migliaccio}, M. 2012, \mnras, 423, 2492,
  \dodoi{10.1111/j.1365-2966.2012.21059.x}

\bibitem[{{Goodfellow} {et~al.}(2014){Goodfellow}, {Pouget-Abadie}, {Mirza},
  {Xu}, {Warde-Farley}, {Ozair}, {Courville}, \& {Bengio}}]{Goodfellowetal2014}
{Goodfellow}, I.~J., {Pouget-Abadie}, J., {Mirza}, M., {et~al.} 2014, arXiv
  e-prints, arXiv:1406.2661.
\newblock \doarXiv{1406.2661}

\bibitem[{{Gupta} {et~al.}(2017){Gupta}, {Saro}, {Mohr}, {Dolag}, \&
  {Liu}}]{Guptaetal2017}
{Gupta}, N., {Saro}, A., {Mohr}, J.~J., {Dolag}, K., \& {Liu}, J. 2017, \mnras,
  469, 3069, \dodoi{10.1093/mnras/stx715}

\bibitem[{{Hallman} {et~al.}(2007{\natexlab{a}}){Hallman}, {Burns}, {Motl}, \&
  {Norman}}]{Hallmanetal2007betamodel}
{Hallman}, E.~J., {Burns}, J.~O., {Motl}, P.~M., \& {Norman}, M.~L.
  2007{\natexlab{a}}, \apj, 665, 911, \dodoi{10.1086/519447}

\bibitem[{{Hallman} {et~al.}(2007{\natexlab{b}}){Hallman}, {O'Shea}, {Norman},
  {Wagner}, \& {Burns}}]{Hallmanetal2007maps}
{Hallman}, E.~J., {O'Shea}, B.~W., {Norman}, M.~L., {Wagner}, R., \& {Burns},
  J.~O. 2007{\natexlab{b}}, in Heating versus Cooling in Galaxies and Clusters
  of Galaxies, ed. H.~{B{\"o}hringer}, G.~W. {Pratt}, A.~{Finoguenov}, \&
  P.~{Schuecker}, 355, \dodoi{10.1007/978-3-540-73484-0_64}

\bibitem[{{Hand} {et~al.}(2018){Hand}, {Feng}, {Beutler}, {Li}, {Modi},
  {Seljak}, \& {Slepian}}]{nbodykit}
{Hand}, N., {Feng}, Y., {Beutler}, F., {et~al.} 2018, \aj, 156, 160,
  \dodoi{10.3847/1538-3881/aadae0}

\bibitem[{{He} {et~al.}(2019){He}, {Li}, {Feng}, {Ho}, {Ravanbakhsh}, {Chen},
  \& {P{\'o}czos}}]{Heetal2019}
{He}, S., {Li}, Y., {Feng}, Y., {et~al.} 2019, Proceedings of the National
  Academy of Science, 116, 13825, \dodoi{10.1073/pnas.1821458116}

\bibitem[{{Hill} \& {Spergel}(2014)}]{HillSpergel2014}
{Hill}, J.~C., \& {Spergel}, D.~N. 2014, \jcap, 2014, 030,
  \dodoi{10.1088/1475-7516/2014/02/030}

\bibitem[{{Hill} {et~al.}(2014){Hill}, {Sherwin}, {Smith}, {Addison},
  {Battaglia}, {Battistelli}, {Bond}, {Calabrese}, {Devlin}, {Dunkley},
  {Dunner}, {Essinger-Hileman}, {Gralla}, {Hajian}, {Hasselfield}, {Hincks},
  {Hlozek}, {Hughes}, {Kosowsky}, {Louis}, {Marsden}, {Moodley}, {Niemack},
  {Page}, {Partridge}, {Schmitt}, {Sehgal}, {Sievers}, {Spergel}, {Staggs},
  {Swetz}, {Thornton}, {Trac}, \& {Wollack}}]{Hilletal2014}
{Hill}, J.~C., {Sherwin}, B.~D., {Smith}, K.~M., {et~al.} 2014, arXiv e-prints,
  arXiv:1411.8004.
\newblock \doarXiv{1411.8004}

\bibitem[{{Hojjati} {et~al.}(2017){Hojjati}, {Tr{\"o}ster},
  {Harnois-D{\'e}raps}, {McCarthy}, {van Waerbeke}, {Choi}, {Erben}, {Heymans},
  {Hildebrandt}, {Hinshaw}, {Ma}, {Miller}, {Viola}, \&
  {Tanimura}}]{Hojjatietal2017}
{Hojjati}, A., {Tr{\"o}ster}, T., {Harnois-D{\'e}raps}, J., {et~al.} 2017,
  \mnras, 471, 1565, \dodoi{10.1093/mnras/stx1659}

\bibitem[{{Jaffe} \& {Kamionkowski}(1998)}]{JaffeKamionkowski1998}
{Jaffe}, A.~H., \& {Kamionkowski}, M. 1998, \prd, 58, 043001,
  \dodoi{10.1103/PhysRevD.58.043001}

\bibitem[{{Jo} \& {Kim}(2019)}]{JoKim2019}
{Jo}, Y., \& {Kim}, J.-h. 2019, \mnras, 489, 3565,
  \dodoi{10.1093/mnras/stz2304}

\bibitem[{{Kay} {et~al.}(2012){Kay}, {Peel}, {Short}, {Thomas}, {Young},
  {Battye}, {Liddle}, \& {Pearce}}]{Kayetal2012}
{Kay}, S.~T., {Peel}, M.~W., {Short}, C.~J., {et~al.} 2012, \mnras, 422, 1999,
  \dodoi{10.1111/j.1365-2966.2012.20623.x}

\bibitem[{{Komatsu} \& {Kitayama}(1999)}]{KomatsuKitayama1999}
{Komatsu}, E., \& {Kitayama}, T. 1999, \apjl, 526, L1, \dodoi{10.1086/312364}

\bibitem[{{Komatsu} \& {Seljak}(2001)}]{KomatsuSeljak2001}
{Komatsu}, E., \& {Seljak}, U. 2001, \mnras, 327, 1353,
  \dodoi{10.1046/j.1365-8711.2001.04838.x}

\bibitem[{{Komatsu} \& {Seljak}(2002)}]{KomatsuSeljak2002}
---. 2002, \mnras, 336, 1256, \dodoi{10.1046/j.1365-8711.2002.05889.x}

\bibitem[{{Le Brun} {et~al.}(2014){Le Brun}, {McCarthy}, {Schaye}, \&
  {Ponman}}]{LeBrunetal2014}
{Le Brun}, A. M.~C., {McCarthy}, I.~G., {Schaye}, J., \& {Ponman}, T.~J. 2014,
  \mnras, 441, 1270, \dodoi{10.1093/mnras/stu608}

\bibitem[{{Le Brun} {et~al.}(2017){Le Brun}, {McCarthy}, {Schaye}, \&
  {Ponman}}]{LeBrunetal2017}
---. 2017, \mnras, 466, 4442, \dodoi{10.1093/mnras/stw3361}

\bibitem[{{Lee} {et~al.}(2020){Lee}, {Chluba}, {Kay}, \&
  {Barnes}}]{Leeetal2020}
{Lee}, E., {Chluba}, J., {Kay}, S.~T., \& {Barnes}, D.~J. 2020, \mnras, 493,
  3274, \dodoi{10.1093/mnras/staa450}

\bibitem[{{Lee} \& {Suto}(2003)}]{LeeSuto2003}
{Lee}, J., \& {Suto}, Y. 2003, \apj, 585, 151, \dodoi{10.1086/345931}

\bibitem[{{Ma} \& {Fry}(2002)}]{MaFry2002}
{Ma}, C.-P., \& {Fry}, J.~N. 2002, \prl, 88, 211301,
  \dodoi{10.1103/PhysRevLett.88.211301}

\bibitem[{{Marinacci} {et~al.}(2018){Marinacci}, {Vogelsberger}, {Pakmor},
  {Torrey}, {Springel}, {Hernquist}, {Nelson}, {Weinberger}, {Pillepich},
  {Naiman}, \& {Genel}}]{IllustrisTNG4}
{Marinacci}, F., {Vogelsberger}, M., {Pakmor}, R., {et~al.} 2018, \mnras, 480,
  5113, \dodoi{10.1093/mnras/sty2206}

\bibitem[{{Mead} {et~al.}(2020){Mead}, {Tr{\"o}ster}, {Heymans}, {Van
  Waerbeke}, \& {McCarthy}}]{Meadetal2020}
{Mead}, A.~J., {Tr{\"o}ster}, T., {Heymans}, C., {Van Waerbeke}, L., \&
  {McCarthy}, I.~G. 2020, arXiv e-prints, arXiv:2005.00009.
\newblock \doarXiv{2005.00009}

\bibitem[{{Moster} {et~al.}(2020){Moster}, {Naab}, {Lindstr{\"o}m}, \&
  {O'Leary}}]{Mosteretal2020}
{Moster}, B.~P., {Naab}, T., {Lindstr{\"o}m}, M., \& {O'Leary}, J.~A. 2020,
  arXiv e-prints, arXiv:2005.12276.
\newblock \doarXiv{2005.12276}

\bibitem[{{Nagai} {et~al.}(2007){Nagai}, {Kravtsov}, \&
  {Vikhlinin}}]{Nagaietal2007}
{Nagai}, D., {Kravtsov}, A.~V., \& {Vikhlinin}, A. 2007, \apj, 668, 1,
  \dodoi{10.1086/521328}

\bibitem[{{Naiman} {et~al.}(2018){Naiman}, {Pillepich}, {Springel},
  {Ramirez-Ruiz}, {Torrey}, {Vogelsberger}, {Pakmor}, {Nelson}, {Marinacci},
  {Hernquist}, {Weinberger}, \& {Genel}}]{IllustrisTNG2}
{Naiman}, J.~P., {Pillepich}, A., {Springel}, V., {et~al.} 2018, \mnras, 477,
  1206, \dodoi{10.1093/mnras/sty618}

\bibitem[{{Navarro} {et~al.}(1997){Navarro}, {Frenk}, \& {White}}]{NFW1997}
{Navarro}, J.~F., {Frenk}, C.~S., \& {White}, S.~D.~M. 1997, \apj, 490, 493,
  \dodoi{10.1086/304888}

\bibitem[{{Nelson} {et~al.}(2018){Nelson}, {Pillepich}, {Springel},
  {Weinberger}, {Hernquist}, {Pakmor}, {Genel}, {Torrey}, {Vogelsberger},
  {Kauffmann}, {Marinacci}, \& {Naiman}}]{IllustrisTNG3}
{Nelson}, D., {Pillepich}, A., {Springel}, V., {et~al.} 2018, \mnras, 475, 624,
  \dodoi{10.1093/mnras/stx3040}

\bibitem[{{Nelson} {et~al.}(2019){Nelson}, {Springel}, {Pillepich},
  {Rodriguez-Gomez}, {Torrey}, {Genel}, {Vogelsberger}, {Pakmor}, {Marinacci},
  {Weinberger}, {Kelley}, {Lovell}, {Diemer}, \& {Hernquist}}]{Nelsonetal2019}
{Nelson}, D., {Springel}, V., {Pillepich}, A., {et~al.} 2019, Computational
  Astrophysics and Cosmology, 6, 2, \dodoi{10.1186/s40668-019-0028-x}

\bibitem[{{Nelson} {et~al.}(2014){Nelson}, {Lau}, \&
  {Nagai}}]{NelsonLauNagai2014}
{Nelson}, K., {Lau}, E.~T., \& {Nagai}, D. 2014, \apj, 792, 25,
  \dodoi{10.1088/0004-637X/792/1/25}

\bibitem[{{Nelson et al.}({in prep.})}]{TNGClusters}
{Nelson et al.} {in prep.}

\bibitem[{{Nozawa} {et~al.}(2006){Nozawa}, {Itoh}, {Suda}, \&
  {Ohhata}}]{Nozawaetal2006}
{Nozawa}, S., {Itoh}, N., {Suda}, Y., \& {Ohhata}, Y. 2006, Nuovo Cimento B
  Serie, 121, 487, \dodoi{10.1393/ncb/i2005-10223-0}

\bibitem[{{Ostriker} {et~al.}(2005){Ostriker}, {Bode}, \&
  {Babul}}]{OstrikerBodeBabul2005}
{Ostriker}, J.~P., {Bode}, P., \& {Babul}, A. 2005, \apj, 634, 964,
  \dodoi{10.1086/497122}

\bibitem[{{Ostriker} \& {Vishniac}(1986)}]{OstrikerVishniac1986}
{Ostriker}, J.~P., \& {Vishniac}, E.~T. 1986, \apjl, 306, L51,
  \dodoi{10.1086/184704}

\bibitem[{{Park} {et~al.}(2018){Park}, {Alvarez}, \& {Bond}}]{Parketal2018}
{Park}, H., {Alvarez}, M.~A., \& {Bond}, J.~R. 2018, \apj, 853, 121,
  \dodoi{10.3847/1538-4357/aaa0da}

\bibitem[{{Park} {et~al.}(2016){Park}, {Komatsu}, {Shapiro}, {Koda}, \&
  {Mao}}]{Parketal2016}
{Park}, H., {Komatsu}, E., {Shapiro}, P.~R., {Koda}, J., \& {Mao}, Y. 2016,
  \apj, 818, 37, \dodoi{10.3847/0004-637X/818/1/37}

\bibitem[{{Persi} {et~al.}(1995){Persi}, {Spergel}, {Cen}, \&
  {Ostriker}}]{Persietal1995}
{Persi}, F.~M., {Spergel}, D.~N., {Cen}, R., \& {Ostriker}, J.~P. 1995, \apj,
  442, 1, \dodoi{10.1086/175416}

\bibitem[{{Pfrommer} {et~al.}(2007){Pfrommer}, {En{\ss}lin}, {Springel},
  {Jubelgas}, \& {Dolag}}]{Pfrommeretal2007}
{Pfrommer}, C., {En{\ss}lin}, T.~A., {Springel}, V., {Jubelgas}, M., \&
  {Dolag}, K. 2007, \mnras, 378, 385, \dodoi{10.1111/j.1365-2966.2007.11732.x}

\bibitem[{{Pillepich} {et~al.}(2018{\natexlab{a}}){Pillepich}, {Nelson},
  {Hernquist}, {Springel}, {Pakmor}, {Torrey}, {Weinberger}, {Genel}, {Naiman},
  {Marinacci}, \& {Vogelsberger}}]{IllustrisTNG5}
{Pillepich}, A., {Nelson}, D., {Hernquist}, L., {et~al.} 2018{\natexlab{a}},
  \mnras, 475, 648, \dodoi{10.1093/mnras/stx3112}

\bibitem[{{Pillepich} {et~al.}(2018{\natexlab{b}}){Pillepich}, {Springel},
  {Nelson}, {Genel}, {Naiman}, {Pakmor}, {Hernquist}, {Torrey}, {Vogelsberger},
  {Weinberger}, \& {Marinacci}}]{Pillepichetal2018}
{Pillepich}, A., {Springel}, V., {Nelson}, D., {et~al.} 2018{\natexlab{b}},
  \mnras, 473, 4077, \dodoi{10.1093/mnras/stx2656}

\bibitem[{{Planck Collaboration} {et~al.}(2013){Planck Collaboration}, {Ade},
  {Aghanim}, {Arnaud}, {Ashdown}, {Atrio-Barandela}, {Aumont}, {Baccigalupi},
  {Balbi}, {Banday}, {Barreiro}, {Bartlett}, {Battaner}, {Benabed},
  {Beno{\^\i}t}, {Bernard}, {Bersanelli}, {Bhatia}, {Bikmaev}, {Bobin},
  {B{\"o}hringer}, {Bonaldi}, {Bond}, {Borgani}, {Borrill}, {Bouchet},
  {Bourdin}, {Brown}, {Burenin}, {Burigana}, {Cabella}, {Cardoso}, {Carvalho},
  {Castex}, {Catalano}, {Cay{\'o}n}, {Chamballu}, {Chiang}, {Chon},
  {Christensen}, {Churazov}, {Clements}, {Colafrancesco}, {Colombi}, {Colombo},
  {Comis}, {Coulais}, {Crill}, {Cuttaia}, {Da Silva}, {Dahle}, {Danese},
  {Davis}, {de Bernardis}, {de Gasperis}, {de Zotti}, {Delabrouille},
  {D{\'e}mocl{\`e}s}, {D{\'e}sert}, {Diego}, {Dolag}, {Dole}, {Donzelli},
  {Dor{\'e}}, {D{\"o}rl}, {Douspis}, {Dupac}, {Efstathiou}, {En{\ss}lin},
  {Eriksen}, {Finelli}, {Flores-Cacho}, {Forni}, {Fosalba}, {Frailis},
  {Franceschi}, {Frommert}, {Galeotta}, {Ganga}, {G{\'e}nova-Santos}, {Giard},
  {Giraud-H{\'e}raud}, {Gonz{\'a}lez-Nuevo}, {G{\'o}rski}, {Gregorio},
  {Gruppuso}, {Hansen}, {Harrison}, {Hempel}, {Henrot-Versill{\'e}},
  {Hern{\'a}ndez-Monteagudo}, {Herranz}, {Hildebrandt}, {Hivon}, {Hobson},
  {Holmes}, {Hurier}, {Jaffe}, {Jaffe}, {Jagemann}, {Jones}, {Juvela},
  {Keih{\"a}nen}, {Khamitov}, {Kisner}, {Kneissl}, {Knoche}, {Knox}, {Kunz},
  {Kurki-Suonio}, {Lagache}, {L{\"a}hteenm{\"a}ki}, {Lamarre}, {Lasenby},
  {Lawrence}, {Le Jeune}, {Leonardi}, {Liddle}, {Lilje}, {L{\'o}pez-Caniego},
  {Luzzi}, {Mac{\'\i}as-P{\'e}rez}, {Maino}, {Mandolesi}, {Maris}, {Marleau},
  {Marshall}, {Mart{\'\i}nez-Gonz{\'a}lez}, {Masi}, {Massardi}, {Matarrese},
  {Mazzotta}, {Mei}, {Melchiorri}, {Melin}, {Mendes}, {Mennella}, {Mitra},
  {Miville-Desch{\^e}nes}, {Moneti}, {Montier}, {Morgante}, {Mortlock},
  {Munshi}, {Murphy}, {Naselsky}, {Nati}, {Natoli}, {N{\o}rgaard-Nielsen},
  {Noviello}, {Novikov}, {Novikov}, {Osborne}, {Pajot}, {Paoletti}, {Pasian},
  {Patanchon}, {Perdereau}, {Perotto}, {Perrotta}, {Piacentini}, {Piat},
  {Pierpaoli}, {Piffaretti}, {Plaszczynski}, {Pointecouteau}, {Polenta},
  {Ponthieu}, {Popa}, {Poutanen}, {Pratt}, {Prunet}, {Puget}, {Rachen},
  {Reach}, {Rebolo}, {Reinecke}, {Remazeilles}, {Renault}, {Ricciardi},
  {Riller}, {Ristorcelli}, {Rocha}, {Roman}, {Rosset}, {Rossetti},
  {Rubi{\~n}o-Mart{\'\i}n}, {Rusholme}, {Sandri}, {Savini}, {Scott}, {Smoot},
  {Starck}, {Sudiwala}, {Sunyaev}, {Sutton}, {Suur-Uski}, {Sygnet}, {Tauber},
  {Terenzi}, {Toffolatti}, {Tomasi}, {Tristram}, {Tuovinen}, {Valenziano}, {Van
  Tent}, {Varis}, {Vielva}, {Villa}, {Vittorio}, {Wade}, {Wandelt}, {Welikala},
  {White}, {White}, {Yvon}, {Zacchei}, \& {Zonca}}]{Planck2013TSZProfiles}
{Planck Collaboration}, {Ade}, P.~A.~R., {Aghanim}, N., {et~al.} 2013, \aap,
  550, A131, \dodoi{10.1051/0004-6361/201220040}

\bibitem[{{Planelles} {et~al.}(2017){Planelles}, {Fabjan}, {Borgani},
  {Murante}, {Rasia}, {Biffi}, {Truong}, {Ragone-Figueroa}, {Granato}, {Dolag},
  {Pierpaoli}, {Beck}, {Steinborn}, \& {Gaspari}}]{Planellesetal2017}
{Planelles}, S., {Fabjan}, D., {Borgani}, S., {et~al.} 2017, \mnras, 467, 3827,
  \dodoi{10.1093/mnras/stx318}

\bibitem[{{Ramos-Ceja} {et~al.}(2015){Ramos-Ceja}, {Basu}, {Pacaud}, \&
  {Bertoldi}}]{RamosCejaetal2015}
{Ramos-Ceja}, M.~E., {Basu}, K., {Pacaud}, F., \& {Bertoldi}, F. 2015, \aap,
  583, A111, \dodoi{10.1051/0004-6361/201425534}

\bibitem[{{Refregier} {et~al.}(2000){Refregier}, {Komatsu}, {Spergel}, \&
  {Pen}}]{Refregieretal2000}
{Refregier}, A., {Komatsu}, E., {Spergel}, D.~N., \& {Pen}, U.-L. 2000, \prd,
  61, 123001, \dodoi{10.1103/PhysRevD.61.123001}

\bibitem[{{Remazeilles} \& {Chluba}(2020)}]{RemazeillesChluba2020}
{Remazeilles}, M., \& {Chluba}, J. 2020, \mnras, 494, 5734,
  \dodoi{10.1093/mnras/staa1135}

\bibitem[{{Scaramella} {et~al.}(1993){Scaramella}, {Cen}, \&
  {Ostriker}}]{Scaramellaetal1993}
{Scaramella}, R., {Cen}, R., \& {Ostriker}, J.~P. 1993, \apj, 416, 399,
  \dodoi{10.1086/173245}

\bibitem[{{Sch{\"a}fer} {et~al.}(2006){Sch{\"a}fer}, {Pfrommer}, {Bartelmann},
  {Springel}, \& {Hernquist}}]{Schaeferetal2006}
{Sch{\"a}fer}, B.~M., {Pfrommer}, C., {Bartelmann}, M., {Springel}, V., \&
  {Hernquist}, L. 2006, \mnras, 370, 1309,
  \dodoi{10.1111/j.1365-2966.2006.10552.x}

\bibitem[{{Schaye} {et~al.}(2015){Schaye}, {Crain}, {Bower}, {Furlong},
  {Schaller}, {Theuns}, {Dalla Vecchia}, {Frenk}, {McCarthy}, {Helly},
  {Jenkins}, {Rosas-Guevara}, {White}, {Baes}, {Booth}, {Camps}, {Navarro},
  {Qu}, {Rahmati}, {Sawala}, {Thomas}, \& {Trayford}}]{Schayeetal2015}
{Schaye}, J., {Crain}, R.~A., {Bower}, R.~G., {et~al.} 2015, \mnras, 446, 521,
  \dodoi{10.1093/mnras/stu2058}

\bibitem[{{Shaw} {et~al.}(2010){Shaw}, {Nagai}, {Bhattacharya}, \&
  {Lau}}]{Shawetal2010}
{Shaw}, L.~D., {Nagai}, D., {Bhattacharya}, S., \& {Lau}, E.~T. 2010, \apj,
  725, 1452, \dodoi{10.1088/0004-637X/725/2/1452}

\bibitem[{{Shaw} {et~al.}(2012){Shaw}, {Rudd}, \& {Nagai}}]{ShawRuddNagai2012}
{Shaw}, L.~D., {Rudd}, D.~H., \& {Nagai}, D. 2012, \apj, 756, 15,
  \dodoi{10.1088/0004-637X/756/1/15}

\bibitem[{{Shi} \& {Komatsu}(2014)}]{ShiKomatsu2014}
{Shi}, X., \& {Komatsu}, E. 2014, \mnras, 442, 521,
  \dodoi{10.1093/mnras/stu858}

\bibitem[{{Sijacki} {et~al.}(2008){Sijacki}, {Pfrommer}, {Springel}, \&
  {En{\ss}lin}}]{Sijackietal2008}
{Sijacki}, D., {Pfrommer}, C., {Springel}, V., \& {En{\ss}lin}, T.~A. 2008,
  \mnras, 387, 1403, \dodoi{10.1111/j.1365-2966.2008.13310.x}

\bibitem[{{Spacek} {et~al.}(2018){Spacek}, {Richardson}, {Scannapieco},
  {Devriendt}, {Dubois}, {Peirani}, \& {Pichon}}]{Spaceketal2018}
{Spacek}, A., {Richardson}, M. L.~A., {Scannapieco}, E., {et~al.} 2018, \apj,
  865, 109, \dodoi{10.3847/1538-4357/aada01}

\bibitem[{{Springel} {et~al.}(2001){Springel}, {White}, \&
  {Hernquist}}]{SpringelWhiteHernquist2001}
{Springel}, V., {White}, M., \& {Hernquist}, L. 2001, \apj, 549, 681,
  \dodoi{10.1086/319473}

\bibitem[{{Springel} {et~al.}(2018){Springel}, {Pakmor}, {Pillepich},
  {Weinberger}, {Nelson}, {Hernquist}, {Vogelsberger}, {Genel}, {Torrey},
  {Marinacci}, \& {Naiman}}]{IllustrisTNG1}
{Springel}, V., {Pakmor}, R., {Pillepich}, A., {et~al.} 2018, \mnras, 475, 676,
  \dodoi{10.1093/mnras/stx3304}

\bibitem[{Strobl {et~al.}(2016)Strobl, Formella, \& Pöschel}]{Overlap}
Strobl, S., Formella, A., \& Pöschel, T. 2016, Journal of Computational
  Physics, 311, 158 , \dodoi{https://doi.org/10.1016/j.jcp.2016.02.003}

\bibitem[{{Sunyaev} \& {Zeldovich}(1970)}]{SZ1970}
{Sunyaev}, R.~A., \& {Zeldovich}, Y.~B. 1970, \apss, 7, 3,
  \dodoi{10.1007/BF00653471}

\bibitem[{{Thiele} {et~al.}(2019){Thiele}, {Hill}, \&
  {Smith}}]{ThieleHillSmith2019}
{Thiele}, L., {Hill}, J.~C., \& {Smith}, K.~M. 2019, \prd, 99, 103511,
  \dodoi{10.1103/PhysRevD.99.103511}

\bibitem[{Tishby {et~al.}(1989)Tishby, Levin, \& Solla}]{TishbyLevinSolla1989}
Tishby, N., Levin, E., \& Solla, S. 1989, in IJCNN Int Jt Conf Neural Network,
  ed. Anon (Publ by IEEE), 403--409

\bibitem[{{Trac} {et~al.}(2011){Trac}, {Bode}, \&
  {Ostriker}}]{TracBodeOstriker2011}
{Trac}, H., {Bode}, P., \& {Ostriker}, J.~P. 2011, \apj, 727, 94,
  \dodoi{10.1088/0004-637X/727/2/94}

\bibitem[{{Tr{\"o}ster} {et~al.}(2019){Tr{\"o}ster}, {Ferguson},
  {Harnois-D{\'e}raps}, \& {McCarthy}}]{Troesteretal2019}
{Tr{\"o}ster}, T., {Ferguson}, C., {Harnois-D{\'e}raps}, J., \& {McCarthy},
  I.~G. 2019, \mnras, 487, L24, \dodoi{10.1093/mnrasl/slz075}

\bibitem[{{Valageas} {et~al.}(2001){Valageas}, {Balbi}, \&
  {Silk}}]{ValageasBalbiSilk2001}
{Valageas}, P., {Balbi}, A., \& {Silk}, J. 2001, \aap, 367, 1,
  \dodoi{10.1051/0004-6361:20000403}

\bibitem[{{Van Waerbeke} {et~al.}(2014){Van Waerbeke}, {Hinshaw}, \&
  {Murray}}]{vanWaerbekeHinshawMurray2014}
{Van Waerbeke}, L., {Hinshaw}, G., \& {Murray}, N. 2014, \prd, 89, 023508,
  \dodoi{10.1103/PhysRevD.89.023508}

\bibitem[{{Villaescusa-Navarro}(2018)}]{Pylians}
{Villaescusa-Navarro}, F. 2018, {Pylians: Python libraries for the analysis of
  numerical simulations}.
\newblock \doeprint{1811.008}

\bibitem[{{Villaescusa-Navarro et al.}({in prep.})}]{CAMEL}
{Villaescusa-Navarro et al.} {in prep.}

\bibitem[{{Vishniac}(1987)}]{Vishniac1987}
{Vishniac}, E.~T. 1987, \apj, 322, 597, \dodoi{10.1086/165755}

\bibitem[{{Weinberger} {et~al.}(2017){Weinberger}, {Springel}, {Hernquist},
  {Pillepich}, {Marinacci}, {Pakmor}, {Nelson}, {Genel}, {Vogelsberger},
  {Naiman}, \& {Torrey}}]{Weinbergeretal2017}
{Weinberger}, R., {Springel}, V., {Hernquist}, L., {et~al.} 2017, \mnras, 465,
  3291, \dodoi{10.1093/mnras/stw2944}

\bibitem[{{White} {et~al.}(2002){White}, {Hernquist}, \&
  {Springel}}]{WhiteHernquistSpringel2002}
{White}, M., {Hernquist}, L., \& {Springel}, V. 2002, \apj, 579, 16,
  \dodoi{10.1086/342756}

\bibitem[{{Xu} {et~al.}(2013){Xu}, {Ho}, {Trac}, {Schneider}, {Poczos}, \&
  {Ntampaka}}]{Xuetal2013}
{Xu}, X., {Ho}, S., {Trac}, H., {et~al.} 2013, \apj, 772, 147,
  \dodoi{10.1088/0004-637X/772/2/147}

\bibitem[{{Zandanel} {et~al.}(2014){Zandanel}, {Pfrommer}, \&
  {Prada}}]{ZandanelPfrommerPrada2014}
{Zandanel}, F., {Pfrommer}, C., \& {Prada}, F. 2014, \mnras, 438, 116,
  \dodoi{10.1093/mnras/stt2196}

\bibitem[{{Zeldovich} \& {Sunyaev}(1969)}]{ZS1969}
{Zeldovich}, Y.~B., \& {Sunyaev}, R.~A. 1969, \apss, 4, 301,
  \dodoi{10.1007/BF00661821}

\bibitem[{{Zhang} {et~al.}(2004){Zhang}, {Pen}, \& {Trac}}]{ZhangPenTrac2004}
{Zhang}, P., {Pen}, U.-L., \& {Trac}, H. 2004, \mnras, 347, 1224,
  \dodoi{10.1111/j.1365-2966.2004.07298.x}

\bibitem[{{Zhang} {et~al.}(2019){Zhang}, {Wang}, {Zhang}, {Sun}, {He},
  {Contardo}, {Villaescusa-Navarro}, \& {Ho}}]{Zhangetal2019}
{Zhang}, X., {Wang}, Y., {Zhang}, W., {et~al.} 2019, arXiv e-prints,
  arXiv:1902.05965.
\newblock \doarXiv{1902.05965}

\end{thebibliography}
\bibliographystyle{aasjournal}

\appendix

\section{Pixelization}
\label{app:pixelization}

Our primary input data are properties of the dark matter field from the gravity-only simulations.
For prediction of the electron pressure and density, we use the dark matter density,
while for the electron momentum density we also use the dark matter momentum density. 
We make use of the \texttt{subfind\_hsml} fields in the snapshot to obtain spheres of varying size in
which we assume the dark matter properties to be uniform;
then we use the \texttt{Overlap} library\footnote{\url{https://github.com/severinstrobl/overlap}} \citep{Overlap}
to grid these spheres onto cubical voxels.
The target data (electron pressure, density, and momentum density) are taken from the
hydrodynamical run.
Here, we approximate the Voronoi cells as spheres, the radius being inferred from the volume given
in the simulation snapshot, and perform the gridding to voxels as described.
These gridding procedures are approximations whose accuracy improves as the number of
simulation elements in the individual voxels increases. Thus, these approximations are 
not worrisome in the high-density regions we are primarily focused on in this work.
We note that the electron pressure is not directly measured in the simulation;
however, under the assumption of complete ionization a linear relationship between thermal and
electron pressure allows us to infer the latter.
For reference, $P_\text{th}/P_e = (5X_\text{H}+3)/2(X_\text{H}+1) = 1.932$, with $X_\text{H}$
the primordial hydrogen mass fraction.

\section{Training boxes}
\label{app:trainingboxes}

As justified in Section~\ref{subsec:sampling},
we restrict the target volumes to cubical boxes of size $(3.2\,h^{-1}\text{Mpc})^3$;
the semi-analytical models are evaluated on the same volume.
The dark matter samples serving as input are taken as boxes of twice the
sidelength, i.e. volume $(6.4\,h^{-1}\text{Mpc})^3$; the target volume is centred in these boxes.
The additional padding of $1.6\,h^{-1}\text{Mpc}$ on each side of the target volume allows for
inclusion of long-range effects.
For electron pressure and density, we use voxels of sidelength $100.1\,h^{-1}\text{kpc}$, such that
the dark matter (electron gas) boxes are of size $64^3$ ($32^3$) voxels.
For the electron momentum density we use half this resolution (voxel sidelength
$200.2\,h^{-1}\text{kpc}$).
We chose the resolution by two considerations:
(1) we do not expect structure below length-scales of a few $100\,h^{-1}\text{kpc}$ to be important
in the mapping from dark matter to electron gas, while observationally smaller scales should also be
of minor importance;
(2) higher resolutions would be difficult to implement on the hardware available to us, the limiting
factor is GPU memory.

For electron density and momentum density as targets, we split the TNG300 box into training (70\%),
validation (20\%), and testing (10\%) cuboidal sub-boxes.
For the electron pressure we use external
data for training (as discussed in Section~\ref{subsec:fewinterestingvoxels}),
allowing us to use the whole simulation box
for testing (we also use 20\% of the box for validation, leading to an overlap between validation
and testing data; however, the small volume in comparison to the testing box ensures that this does
not invalidate the testing results).

\section{Transformations}
\label{app:transformations}

The numerical values $a$, $b$, $c$ introduced in Eq.~\eqref{eq:transformations}
are as follows:

\begin{tabular}{cllll}
& target = $P_e$ & target = $\rho_e$ & \multicolumn{2}{c}{target = $\mathbf{p}_e$}\\
\cline{1-5}
     & $x_\text{DM} = \rho_m$ & $x_\text{DM} = \rho_m$ & $x_\text{DM} = \rho_m$ & $x_\text{DM} = \mathbf{p}_m$ \\
\cline{1-5}
 $a$ & $2.2939$               & $2.2160$               & $8.4592$               & $1.8813$                     \\
 $b$ & $2.7984\times 10^6$    & $2.7984\times 10^6$    & $3.9381\times 10^6$    & $1.2871\times 10^4$          \\
 $c$ & $3.8477\times 10^{-1}$ & $3.8232\times 10^{-1}$ & $2.6845\times 10^{-2}$ & $0.0000$                     \\
\end{tabular}

We remind the reader of the unit conventions given in footnote~\ref{fn:units}.
The calculations for $a$, $b$, and $c$ were performed
by restricting to ``interesting" voxels in the sense of Section~\ref{subsec:fewinterestingvoxels},
these definitions were slightly different for different targets which explains the different
values.

\end{document}